\documentclass[twocolumn,showpacs,preprintnumbers,amsmath,amssymb,nofootinbib]{revtex4}
\usepackage{graphicx}% Include figure files
\usepackage{dcolumn}% Align table columns on decimal point
\usepackage{bm}% bold math
\usepackage{amssymb}
\usepackage{amsmath}
\usepackage{amsthm}
\begin{document}

\preprint{}

\title{Multiphoton effects in coherent radiation spectra}

\author{M.V.~Bondarenco}
 \email{bon@kipt.kharkov.ua}
 \affiliation{NSC Kharkov Institute of Physics and Technology, 1 Academic St.,
61108 Kharkov, Ukraine}

\date{\today}% It is always \today, today,

\begin{abstract}
At measurements of gamma-radiation spectra from ultra-relativistic
electrons in periodic structures, pileup of events in the
calorimeter may cause significant deviation of the detector signal
from the classically evaluated spectrum. That requires appropriate
resummation of multiphoton contributions. We describe the
resummation procedure for the photon spectral intensity and for the
photon multiplicity spectrum, and apply it to the study of spectra
of coherent radiation with an admixture of incoherent component.
Impact of multiphoton effects on the shape of the radiation spectrum
is investigated. The limit of high photon multiplicity for coherent
radiation is explored. A method for reconstruction of the underlying
single-photon spectrum from the multiphoton one is proposed.

%Dimuon registration efficiency is evaluated in analytic form.

\end{abstract}




%\end{keyword}

\maketitle

\section{Introduction}

Many efficient sources of quasi-monochromatic hard radiation exploit
transmission of ultrarelativistic electrons through periodic
structures (crystals or undulators) \cite{books-on-coh-sources}. For
such so-called coherent sources, high radiation brightness is
relatively easy to achieve by increasing the periodic structure
length. The price to pay, however, is that as the photon emission
probability reaches the order of unity, the description of the
source performance must regard the possibility of creation of a few
photons and electron-positron pairs per passing electron, i.e.,
essentially electromagnetic cascading \cite{Rossi}.

The proper procedure for calculation of electromagnetic multiple
particle production at ultra-relativistic energies is via a system
of kinetic equations for sequential photon and $e^+e^-$ pair
creation, allowing for energy redistribution at each branching. In a
non-trivial field of the radiator, this complete procedure is
involved, and generally requires numerical simulation. Fortunately,
in a quite typical case when typical photon energies are inferior to
the incident electron energy, the calculations may appreciably
simplify. First of all, the probability of pair production is not
enhanced so strongly as that of radiation \cite{noe+e-}, and thus
can be neglected in the first approximation. Secondly, the
negligibility of photon recoils allows treating the electron current
as fully determined by the electron's initial conditions, entailing
statistical independence of multiple photon emission acts
\cite{phot-stat-indep}. Altogether, that opens the possibility for
semiclassical description of the cascading process.

However, an extra impediment is that when emitted photon energies
belong to gamma-range\footnote{In this paper, by gamma-radiation we
will mean photon energies well above the electron-positron pair
creation threshold. The best-established coherent gamma-ray sources
nowadays are based on relativistic particle interactions with
crystals, though for forthcoming high-energy undulators, gamma-range
is within reach, as well (see Sec.~\ref{sec:high-intensity} and
\cite{Moortgat-Pick}).},
%an additional complication arises, since besides the particle dynamics and kinetics in the radiator,
the measured radiation spectra begin to depend on the photon
detection method. If it was feasible experimentally, in spite of
high radiation intensity, to distinguish arrivals of individual
photons, the detector signal under soft photon emission condition
would be described merely by formula from classical electrodynamics
for the radiation energy spectrum. But narrow beaming of the
radiation from ultra-relativistic electrons causes pileup of
reactions from different photons in the downstream detector volume,
thus affecting the detector signal. Spectrometers for gamma-quanta
\cite{high-energy-detectors} today are predominantly based on
calorimetry (measurement of the total energy deposition in a
detector) \cite{EMcal}, but to capture most of the energy of
electromagnetic shower created by the hard photon, the calorimeter
transverse size must be in excess of the typical shower lateral
spread (Moli\`{e}re radius), which amounts a few centimeters
\cite{PDG}. That in turn impedes angular resolution of the radiation
within its Lorentz-contracted emission cone, which might otherwise
be employed to eliminate photon pileups\footnote{The detector size
restrictions may be circumvented in several ways \cite{Bochek}, but
any of them requires installation of an additional bulky
equipment.}.

\begin{figure}
\includegraphics{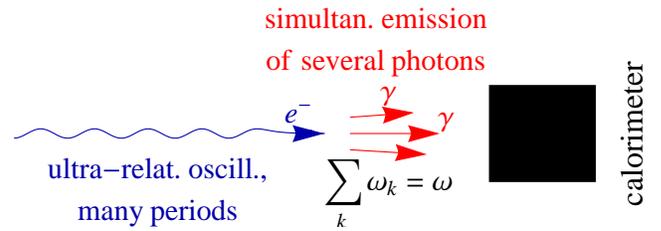}
\caption{\label{fig:calorim-scheme} Scheme of gamma-radiation
spectrum measurement yielding a multiphoton spectrum. The spent
electron is deflected by a magnet, so only the forward-flying
photons hit the calorimeter.}
\end{figure}

In view of appreciable difficulties with eliminating photon pileup
effects, a common attitude at laboratory measurements of
gamma-radiation spectra is not to pursue the objective of photon
counting at all. Then, a single electromagnetic calorimeter is used
to measure the \emph{total} energy deposited by \emph{all} the
$\gamma$-quanta per event of electron passage through the radiator
(see Fig.~\ref{fig:calorim-scheme}). Clearly, the latter method is
equivalent to measuring the electron's radiative energy loss
spectrum\footnote{Assuming the dominance of radiative losses,
holding for ultra-relativistic electrons.}. Furthermore, sometimes a
simplified procedure of radiation spectrum measurement is adopted,
when spent electron energies are measured, which then truly yields
the spectrum of full energy losses by electrons in the radiator.

Even though the calorimetric method for spectrum measurement hinders
applicability of formulas of classical electrodynamics, those can
still be useful for the spectrum calculation. Indeed, the theory
predicts that multiple soft photon emission probabilities factorize,
expressing through the same classically calculated spectral
intensity. To obtain the multiphoton spectrum, one thus needs to
resum all $n$-photon probabilities while holding the total
irradiated energy at a preselected value. Technically, the problem
is similar to that of fast charged particle energy loss under
statistically independent atom ionization acts in the medium
\cite{Landau}, and some other problems
\cite{Bloch-Nordsieck,EW-boson-resum,Feller}, whose solutions were
attained by integral transform methods. For generic energy loss
spectrum by a fast particle in the medium, a Laplace transform
solution was known since 1944 \cite{Landau}. In application to
radiation, it was later utilized in \cite{Akh-Shul-shower-ZhETF},
with the purpose of investigating soft (photon-dominated)
electromagnetic showers in crystals and polycrystals, and checking
the possibility of the shower length reduction due to coherent
radiation effects. Relatively recently, in \cite{BK} this method was
used for studying the shape of the soft part of the radiation
spectrum from electrons in an amorphous medium. Model solutions for
shapes of synchrotron-like radiation spectra in crystals were
discussed in \cite{Khokonov}. There were also several investigations
of shapes of radiation spectra in crystals, based on numerical
simulation (usually by Monte-Carlo), with
\cite{coh-rad-MCgenerators-negle+e-,BKS-simul,Artru-simul,Kononets-Ryabov-simul,Maisheev-simul}
or without \cite{coh-rad-MCgenerators-full} neglect of $e^+e^-$ pair
production as a first approximation.
%\footnote{Note
%that the radiation spectrum is generated globally, i.e. from the
%particle trajectory as a whole, which often necessitates limitation
%of the simulation statistics, consequently obscuring the fine
%spectral detail. An exception may be synchrotron-like radiation in
%intense fields at ultra-high energies, or coherent bremsstrahlung in
%bent crystals, which is generated within a well-defined spatial
%domain depending on the photon momentum \cite{Bondarenco-CBBC}.}.

Notwithstanding the pretty long history of the problem, there still
remain many open questions concerning multiphoton effects on
coherent radiation spectrum shapes. For example, since coherent
radiation spectra are typically sharply peaked and discontinuous,
the prime task is to investigate how do such features modify under
multiphoton emission conditions. In so doing, it is also important
to incorporate an incoherent bremsstrahlung component, which may
require some special treatment, because of its divergent total
probability. Next, it is important to give proper study for the
behavior of the resummed coherent radiation spectrum in the high
intensity (or photon multiplicity) limit. There, one generally
expects the spectrum to tend to a Gaussian form, but corrections to
this form may be significant at practice, and demand evaluation.

Besides the abovementioned most urgent problems, there is a number
of developments which are desirable to make in the theory. First, it
is expedient to supplement the notion of the resummed spectrum by
that of photon multiplicity spectrum, which was proven measurable in
experiments at CERN \cite{multiphot-CERN-Kirsebom}. Secondly, it
would be valuable for various purposes to find a possibility to
reconstruct the single-photon spectrum from a calorimetrically
measured one. Other issues include correspondence with the
perturbation theory, separation of energetically ordered and
anti-ordered contributions, etc.

The goal of the present paper is to provide a comprehensive
formulation for the theory of resummed soft coherent radiation
spectra, and with its aid, to tackle problems raised above. Although
conditions of coherent radiation in different sources may vary, we
concentrate on one of the most typical cases -- dipole radiation
(for overview of its conditions see, e.g., \cite{dipole-rad}), under
the simplifying assumption of approximately harmonic transverse
oscillation of the radiating electron (the neglect of higher
harmonics). We incorporate also an incoherent radiation component,
which, fortunately, has a simple structure, free from model
assumptions. The interference between coherent and incoherent
spectral components may be neglected in the first approximation.
Therewith, there emerges a fairly model-independent description of
radiation, which can embrace diverse practical situations.

This paper is laid out as follows. In Sec.~\ref{sec:2} we establish
main relations of the theory of soft multiphoton spectra. Along with
the resummed spectrum itself, the photon multiplicity spectrum is
defined. In Sec.~\ref{sec:coh+incoh}, the developed generic
techniques are applied to the study of coherent and incoherent
contributions to the radiation. Modification of various spectral
features due to multiphoton effects, as well as interplay of
coherent and incoherent radiation components, is investigated.
Sec.~\ref{sec:high-intensity} explores the high-intensity limit,
benefiting from the analogy between the multiphoton energy loss and
a random walk. Having calculated the multiphoton spectrum in the
central region, we also examine the spectrum behavior beyond it, and
assess long-range effects due to the incoherent bremsstrahlung
component. Sec.~\ref{sec:reconstr} works out the principles of
reconstruction of the single-photon radiation spectrum from the
calorimetrically measured one. The summary and some outlook is given
in Sec.~\ref{sec:summary}. A few technical issues are relegated to
the Appendices.

\section{Generic resummation procedure}\label{sec:2}

As was mentioned in the Introduction, for radiation emitted by an
ultra-relativistic electron (or positron), the radiation angles are
typically small (inversely proportional to the electron's Lorentz
factor), wherefore they often remain experimentally unresolved. In
the latter case, only the radiation spectrum is measured. For
coherent radiation, the spectrum is concentrated on a characteristic
energy scale $\omega_0$, which is often much smaller than the
initial electron energy $E_e$:
\begin{equation}\label{omega0llEe}
\omega_0\ll E_e.
\end{equation}
Under those circumstances, the radiation recoils may be neglected,
and the radiation process be treated as statistically independent
emissions of photons by a classically moving charged particle. The
description of the multiphoton spectrum then proceeds as
follows\footnote{An alternative formulation based on kinetic
equation will be given in Sec.~\ref{subsec:random-walk}.}.

\subsection{Multiphoton spectra under statistically independent photon emission}\label{subsec:Stat-indep-emission}

Let the classically evaluated radiation energy spectrum, integrated
over relativistically small emission angles, equal
$\frac{dE_1}{d\omega}$. The corresponding photon number density
could be derived as
\begin{equation}\label{primary}
\frac{dw_1}{d\omega}=\frac1\omega \frac{dE_1}{d\omega}.
\end{equation}
That might as well be interpreted as the probability distribution
for emission of a photon, but only provided its integral
\begin{equation}\label{w1-def}
w_1=\int_0^\infty d\omega\frac{dw_1}{d\omega},
\end{equation}
which is to correspond to the total emission probability, is $\ll1$.
Sometimes the latter condition is fulfilled thanks to the smallness
of electron coupling with the electromagnetic field,
$\alpha=\frac1{137}$.

On the other hand, in a long radiator, $w_1$ can accumulate and get
arbitrarily large. As it becomes sizable, proper probabilistic
treatment must take into account emission of an arbitrary number of
photons. Assuming that the radiation process remains soft and
intrinsically semiclassical, photon emission acts can be proven to
be statistically independent \cite{phot-stat-indep,Poisson-rad},
thereby obeying Poisson statistics\footnote{\label{exp-before-aver}A
word of caution must be sound that Poisson statistics holds only for
a completely prescribed classical electromagnetic current. At
practice, the radiation spectrum can yet depend on the electron
impact parameters and momentum dispersion within the initial beam,
fluctuations of the intra-crystal field due to thermal vibrations
and lattice defects, etc. Therefore, the Poisson distributions must
be averaged over the beam and target ensembles, but only at the last
step of the calculation (cf., e.g., \cite{Poisson-rad}). In the
present paper, we will refrain from implementation of the averaging
procedures, just indicating which of them are most relevant for
specific coherent radiation sources. Fortunately, in a number of
cases we consider, the ensemble averaging effects are negligible.}:
\begin{equation}\label{Poisson}
\frac{dW_n}{d\omega_1\ldots
d\omega_n}=W_0\frac1{n!}\frac{dw_1}{d\omega_1}\ldots\frac{dw_1}{d\omega_n}.
\end{equation}
Here factors $W_0$ and $1/n!$ are consequences of probability
conservation during photon generation\footnote{At derivation of the
Poisson distribution from Feynman diagrams \cite{phot-stat-indep},
each diagram involves no factorial factors (which cancel with the
number of pairings between creation and annihilation operators due
to the Wick theorem). In this framework, factor $1/n!$ is
conventionally attributed to photon equivalence.}. From the
normalization of the total probability to unity,
\begin{equation}\label{norm}
W_0+\sum_{n=1}^{\infty}\int_0^\infty d\omega_1\ldots\int_0^\infty
d\omega_n\frac{dW_n}{d\omega_1\ldots d\omega_n}=1,
\end{equation}
one readily derives a relation of $W_0$ with the single-photon
spectrum:
\begin{equation}\label{W0-def}
W_0=e^{-w_1}.
\end{equation}
Since $W_0$ may be regarded as the 0-th term of the series, it is to
be interpreted as the probability of photon non-emission. Quantity
$w_1$ also admits an independent physical meaning, being equal to
the mean number of emitted photons (the photon multiplicity):
\begin{equation}\label{mean-n-def}
\bar n=e^{-w_1}\sum^\infty_{n=1}\frac{n}{n!}w_1^n=w_1.
\end{equation}

From the experimental point of view, completely exclusive spectrum
$\frac{dW_n}{d\omega_1\ldots d\omega_n}$ is arguably too detailed.
If we (idealistically) limit ourselves to a single-inclusive
spectrum, where the energy of only one of the photons is measured,
the corresponding observable would equal
\begin{eqnarray}\label{class}
W_0\sum_{n=1}^\infty\frac1{n!} \int_0^\infty\!  d\omega_1
\frac{dw_1}{d\omega_1}\ldots\int_0^\infty\!  d\omega_n
\frac{dw_1}{d\omega_n}
\sum_{k=1}^n\delta\left(\omega-\omega_k\right)\nonumber\\
=\frac{dw_1}{d\omega},\qquad\qquad\qquad\qquad\qquad\qquad\qquad\qquad
\end{eqnarray}
thus reproducing the photon number spectrum, connected with the
classical energy spectrum via Eq.~(\ref{primary}).

\begin{figure}
\includegraphics{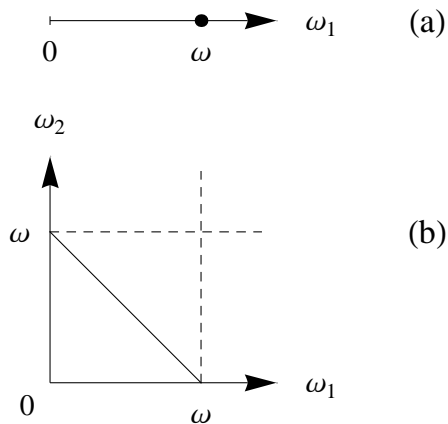}
\caption{\label{fig:integr-domain} Integration domains for partial
probabilities in Eqs.~(\ref{class}) and (\ref{generic-sum-int}). (a)
Component $n=1$. Only the point $\omega_1=\omega$ is selected by
perfect detector of either kind. (b) Component $n=2$. Dashed lines
mark the region selected by a perfect photon counter and
spectrometer. Solid line, the region selected by a perfect
calorimeter.}
\end{figure}

For the case of calorimetric measurement, however, the situation is
different. There, the probability of emission of an arbitrary number
of photons with definite aggregate energy $\omega$ is measured,
taking the form\footnote{In principle, all detectors have finite
energy resolution. To take this into account, the $\delta$-functions
in Eqs.~(\ref{class}) or (\ref{generic-sum-int}) must be replaced by
a function of the detector response (see, e.g., \cite{Kolch-Uch}).
That would be equivalent to ensemble averaging over the final state,
which is left beyond the scope of this paper (see
Footnote~\ref{exp-before-aver}). At practice, though, the energy
resolution of calorimeters is usually high enough for the
$\delta$-function approximation in (\ref{generic-sum-int}) to work
satisfactorily.}
\begin{eqnarray}\label{generic-sum-int}
\frac{dw}{d\omega}&=&W_0\sum_{n=1}^\infty\frac1{n!} \int_0^\infty  d\omega_1 \frac{dw_1}{d\omega_1}\ldots\nonumber\\
&\,&\qquad\qquad\times \int_0^\infty  d\omega_n
\frac{dw_1}{d\omega_n}
\delta\left(\omega-\sum_{k=1}^n\omega_k\right).
\end{eqnarray}
The latter distribution\footnote{In the probability theory,
distributions of the form (\ref{generic-sum-int}) are known as
compound (or generalized) Poisson distributions \cite{Feller}. They
are usually defined in terms of self-convolutions of the initial
probability density
\[
f*f(z)=\int dy f(z-y)f(y) .
\]
It is easy to see that in Eq.~(\ref{generic-sum-int}),
\[
\int_0^\infty \! d\omega_1 \frac{dw_1}{d\omega_1}\ldots\!
\int_0^\infty \! d\omega_n \frac{dw_1}{d\omega_n}
\delta\left(\omega-\sum_{k=1}^n\omega_k\right)=\left(\frac{dw_1}{d\omega}\right)^{n*}
\]
is an $n$-fold self-convolution of the single-photon spectrum.} is
already properly normalized:
\begin{equation}\label{total-probab-1}
\int_0^\infty d\omega \frac{dw}{d\omega}=1-W_0.
\end{equation}
As illustrated in Fig.~\ref{fig:integr-domain}, in the limit of low
radiation intensity, distributions (\ref{class}) and
(\ref{generic-sum-int}) match, since terms $n=1$ in their series are
identical (Fig.~\ref{fig:integr-domain}a). But the rest of the
terms, with $n\geq2$, do differ (cf. Fig.~\ref{fig:integr-domain}b).
Distribution (\ref{generic-sum-int}) is often called multiphoton
spectrum, and in contrast to that, (\ref{primary}) is called
single-photon spectrum.

Another way of representing the results is via a multiphoton energy
spectrum
\begin{equation}\label{energy-spectrum}
\frac{dE_{\text{rad}}}{d\omega}:=\omega\frac{dw}{d\omega}\neq\frac{dE_1}{d\omega}.
\end{equation}
In this paper, though, it will be advantageous for us to work with
the probability spectral density. So, in what follows, the term
`spectrum' will mostly be used in relation to quantities such as
$\frac{dw_1}{d\omega_1}$ and $\frac{dw}{d\omega}$.

Before proceeding with the resummation of series
(\ref{generic-sum-int}), it is expedient to inspect its structure
more closely. First of all, in (\ref{generic-sum-int}), upper
integration limits for individual photon energies may actually be
lowered down to $\omega$, granted that if energy of any of the
emitted photons exceeds $\omega$, the energy-conserving
$\delta$-function will definitely yield zero. At $n=1$, one yet
needs to ensure that the $\delta$-function falls into the
integration interval completely, so the integration upper limit
should be written more precisely as $\omega+0$. With the forthcoming
resummation procedure in mind, it is convenient to let the upper
integration limit be the same everywhere, writing:
\begin{eqnarray}\label{generic-sum-int-upper-omega}
\frac{dw}{d\omega}=W_0\sum_{n=1}^\infty\frac1{n!} \int_0^{\omega+0}
d\omega_1 \frac{dw_1}{d\omega_1}\ldots\qquad\qquad\qquad\nonumber\\
\times\int_0^{\omega+0} d\omega_n \frac{dw_1}{d\omega_n}
\delta\left(\omega-\sum_{k=1}^n\omega_k\right).
\end{eqnarray}

Next, even though in Eq.~(\ref{generic-sum-int-upper-omega}) upper
integration limits for partial probabilities are rendered finite, in
integral (\ref{w1-def}) entering $W_0$ through Eq.~(\ref{W0-def}),
the integration extends over an infinite range of $\omega_1$. That
fact needs care, since at $\omega\sim E_e$, quantum corrections due
to photon recoils invalidate the statistical independence of photon
emission acts. For pure coherent radiation, whose spectrum is
strongly suppressed beyond a sufficiently low energy $\omega_0$,
such a problem may be absent. However, there often exists an
incoherent bremsstrahlung contribution to $\frac{dw_1}{d\omega_1}$,
scaling as $\frac{dw_1}{d\omega_1}\propto\omega_1^{-1}$. Therewith,
integral $w_1$ logarithmically diverges on the upper limit, whereby
factor $e^{-w_1}$ would nullify the multiphoton radiation spectrum
for any finite $\omega$, if there was no physical end of the
bremsstrahlung spectrum, situated at $\omega=E_e$. Within the
leading logarithmic (LL) accuracy, the upper integration limit could
be merely replaced by $E_e$:
\begin{equation}\label{w1-regEe}
w_1\simeq\int_0^{E_e}d\omega_1 \frac{dw_1}{d\omega_1}\qquad(\text{LL
accuracy}).
\end{equation}
However, the usage of the leading-log accuracy in
Eq.~(\ref{Poisson}), where $w_1$ is exponentiated, may be
unsatisfactory, as long as it introduces an indeterminate overall
factor. Thence, it is worth promoting it to the next-to-leading
logarithmic (NLL) accuracy. That is equivalent to effectively
replacing Eq.~(\ref{w1-regEe}) by
\begin{equation}\label{w1-NLL}
w_1\simeq\int_0^{E}d\omega_1 \frac{dw_1}{d\omega_1}\qquad(\text{NLL
accuracy}),
\end{equation}
where parameter $E=\kappa E_e\sim1$ must be ab initio calculable in
NLL. A study delivered in Appendix~\ref{app:kappa} (though still
under the neglect of pair production) suggests that
$\kappa\approx0.5$. Greater accuracy may be unnecessary, since
$\kappa$ only enters in terms a power $\kappa^a$ with $a\ll 1$ (see
Sec.~\ref{subsec:incoh-bremsstr}).

The incoherent bremsstrahlung contribution also dominates in the
opposite extreme $\omega_1\to0$, where its behavior
$\frac{dw_1}{d\omega_1}\sim\omega_1^{-1}$ makes all the integrals
present in Eq.~(\ref{generic-sum-int-upper-omega}) diverge on the
lower limit. Fortunately, though, the infrared (IR) divergence
cancels for energy-resummed distributions by virtue of the
Bloch-Nordsieck theorem \cite{phot-stat-indep,Bloch-Nordsieck}, as
will be demonstrated in the next subsection.

%\newpage

\subsection{Contour integral representations}\label{subsec:cont-int-repr}

Resummation of series of type (\ref{generic-sum-int}) proceeds by
applying a Laplace transform\footnote{Spoken mathematically --
resorting to characteristic functions of probability distributions
 \cite{Feller}.}, which reduces the
multiple convolution to a simple product of equal single integrals,
and the resulting power series resums to an exponential:
%\begin{subequations}\label{generic-exponentiation}
\begin{eqnarray}\label{generic-exponentiation}
\int_0^\infty d\omega \frac{dw}{d\omega}e^{-s\omega}&=&e^{-w_1}\sum_{n=1}^\infty\frac1{n!} \prod_{k=1}^n \int_0^{E} d\omega_k \frac{dw_1}{d\omega_k} e^{-s\omega_k}  \nonumber\\
&=&e^{\int_0^{E} d\omega_1
\frac{dw_1}{d\omega_1}\left(e^{-s\omega_1}-1\right) } -W_0.
\end{eqnarray}
[Prior to the resummation, we implemented a proper cutoff parameter
$E$, in the manner of Eq.~(\ref{w1-NLL})]. An immediate observation
from Eq.~(\ref{generic-exponentiation}) is that by virtue of factor
$e^{-s\omega_1}-1$ tending to zero as $\omega_1\to0$, the infrared
singularity brought by $\frac{dw_1}{d\omega_1}$ is suppressed,
providing the integral convergence on the lower limit if
$\frac{dw_1}{d\omega_1}=\mathcal{O}(\omega_1^{-1})$. If $s$ is put
to zero, Eq.~(\ref{generic-exponentiation}) reproduces normalization
condition (\ref{total-probab-1}).

%Integrating in the exponent of (\ref{generic-exponentiation-01-Ee})
%by parts, one can as well recast it form
%\begin{equation}\label{generic-exponentiation-03}
%\int_0^\infty d\omega \frac{dw}{d\omega}e^{-s\omega}=e^{-s\int_0^{E}
%d\omega_1
%e^{-s\omega_1}\int^{E}_{\omega_1}d\omega'_1\frac{dw_1}{d\omega'_1}}
%-W_0.
%\end{equation}
%\end{subequations}
%Convenience of the latter transformation will become clear somewhat later.

The Laplace transform is further inverted by computing a Mellin
integral in the complex $s$-plane:
\begin{equation}\label{generic-Mellin-int}
\frac{dw}{d\omega}=\frac1{2\pi i}\int_{c-i\infty}^{c+i\infty}ds
e^{s\omega}\int_0^\infty d\omega_1
\frac{dw}{d\omega_1}e^{-s\omega_1}.
\end{equation}
Here constant $c$ is required to be greater than real parts of all
singularities of the integrand, otherwise being arbitrary. Combining
Eqs.~(\ref{generic-Mellin-int}) and (\ref{generic-exponentiation})
yields the generic integral representation for the multiphoton
radiation spectrum:
%\begin{subequations}\label{10}
\begin{equation}\label{generic-eq}
\frac{dw}{d\omega}=\frac1{2\pi i}\int_{c-i\infty}^{c+i\infty}ds e^{s\omega+\int_0^{E} d\omega_1 \frac{dw_1}{d\omega_1}\left(e^{-s\omega_1}-1\right)}%\nonumber\\&\,&
-W_0\delta(\omega).%\nonumber\\
%\\
%&\,&\qquad\qquad (0<\omega\ll E_e),\nonumber
\end{equation}
The appearance in Eq.~(\ref{generic-eq}) of a negative term
proportional to Dirac $\delta$-function does not imply negativity of
the resummed spectrum at $\omega=0$. It just cancels the
corresponding positive singularity in the contour integral, arising
because its integrand tends to $W_0 e^{s\omega}$ as $s\to\pm
i\infty$. At $\omega>0$, the $\delta$-term does not contribute,
anyway, but it proves important at derivation of various integral
relations involving $\frac{dw}{d\omega}$ [in particular, maintaining
the correct normalization (\ref{total-probab-1})]. The integral on
the r.h.s. of (\ref{generic-eq}),
\begin{equation}\label{=Pi}
\frac{dw}{d\omega}+W_0\delta(\omega)=\Pi(E_e-\omega),
\end{equation}
has an independent physical meaning, representing the distribution
function for the radiating electrons, which is singular at
$\omega=0$, but is normalized to unity:\footnote{It is essential
that in Eq.~(\ref{Pi-norm}), the integral over $\omega$ extends to
infinity, because even if the single-photon spectrum terminates at
$\omega_1=E_e$, the spectrum resummed without account of electron
energy degradation after photon emission will actually extend beyond
the limit $\omega=E_e$ (see discussion in
Sec.~\ref{subsec:incoh-bremsstr}).}
\begin{equation}\label{Pi-norm}
\int_0^\infty
d\omega\left\{\frac{dw}{d\omega}+W_0\delta(\omega)\right\}=1.
\end{equation}
Representation (\ref{generic-eq}) without the $\delta$-term, i.e.
holding for function $\Pi(E_e-\omega)$ [or pertaining to case
$w_1=\infty$, $W_0\equiv0$], was derived by Landau \cite{Landau} on
the basis of a kinetic equation.

%where in the r.h.s. we dropped term  vanishing at $\omega>0$. The appearance of the $\delta(\omega)$ term can be avoided if instead of a rectilinear integration path in Eqs.~(\ref{10}) one chooses a contour turning at infinity to the left, i.e. $\mathfrak{Re}s\to-\infty$. Such a contour is depicted in Fig.~\ref{fig:contours} as $\mathcal{C}_2$. (It should be noted, however, that in Eqs.~(\ref{generic-exponentiation}) the term $-e^{-w_1}$ remains necessary to guarantee vanishing of the r.h.s. at $s\to+\infty$.)

Next, let us note that contour integral representation
(\ref{generic-eq}) needs not be the only possible one. The example
of representation (\ref{generic-sum-int-upper-omega}) shows that in
the exponentiated Laplace transform of $\frac{dw_1}{d\omega_1}$, the
upper integration limit may be arbitrary, provided it exceeds
$\omega$. For instance, lowering it right down to $\omega$ leads to
the integral representation
\begin{eqnarray}\label{generic-eq-upperlimitomega}
\frac{dw}{d\omega}=e^{-\!\int^{E}_{\omega}\!d\omega_1\!\frac{dw_1}{d\omega_1}}\frac1{2\pi i}\int_{c-i\infty}^{c+i\infty}\!ds e^{s\omega+\int_0^{\omega+0} \!d\omega_1\! \frac{dw_1}{d\omega_1}\!\left(\! e^{-s\omega_1}-1\!\right)}\nonumber\\
-W_0\delta(\omega).\qquad\qquad\qquad\qquad\qquad\qquad\qquad%\\
%\equiv e^{-\!\int^{E}_{\omega}\!d\omega_1\frac{dw_1}{d\omega_1}}\frac1{2\pi i}\int_{c-i\infty}^{c+i\infty}\!ds e^{s\omega-s\!\int_0^{\omega+0} \!d\omega_1 e^{-s\omega_1}\!\int^{\omega}_{\omega_1}\!d\omega'_1\frac{dw_1}{d\omega'_1}}\nonumber\\
%-W_0\delta(\omega).\qquad\qquad\qquad\qquad\qquad\qquad\qquad\label{generic-eq-upperlimitomega-byparts}
%\\
%&\,&\qquad\qquad (0<\omega\ll E_e).\nonumber
\end{eqnarray}
%\end{subequations}
Indeed, expanding here $e^{\int_0^{\omega+0} d\omega_1
\frac{dw_1}{d\omega_1}e^{-s\omega_1}}$ to Taylor series and
integrating over $s$ termwise leads to decomposition
(\ref{generic-sum-int-upper-omega}). Relation
(\ref{generic-eq-upperlimitomega}) was first noticed in \cite{BK},
wherein it was regarded as only an \emph{approximate} consequence of
representation (\ref{generic-eq}) (in the sense that in the integral
over $s$, the contributing $s$ are $\sim\omega^{-1}$, and hence in
$\int_0^{E}d\omega_1 \frac{dw_1}{d\omega_1}e^{-s\omega_1}$ the
contributing $\omega_1$ are $\lesssim s^{-1}\sim\omega$). But here,
by deriving both representations (\ref{generic-eq}) and
(\ref{generic-eq-upperlimitomega}) from the same series, we reveal
their exact equivalence.

Although the derivation of representation
(\ref{generic-eq-upperlimitomega}) given above is illuminating, it
may be desirable sometimes to avoid the use of power series, which
may be beset by IR divergences. In that case, one may utilize the
following generic statement:
\newtheorem{theorem}{Lemma}
\begin{theorem}\label{lemma1}
Let $\omega\geq\epsilon\geq0$ be real parameters, and $f(\omega_1)$
be a positive function of real variable, integrable on interval
$\epsilon<\omega_1<\infty$. Also let $\mathcal{F}(z)$ be an analytic
function of complex variable $z$ in domain $\mathfrak{Re}z>0$, and
such that $\left|\mathcal{F}'\left(\int_{\epsilon}^{\omega+\varpi}
d\omega_1 e^{-s\omega_1}f(\omega_1)\right)\right|$ grows with
$\mathfrak{Re} s\to+\infty$ slower than exponentially. Then, contour
integral
\begin{equation}\label{lemma}
\int_{c-i\infty}^{c+i\infty}ds e^{s\omega}
\mathcal{F}\left(\int_{\epsilon}^{\omega\!+\!\varpi} d\omega_1
e^{-s\omega_1}f(\omega_1)\!\right)\!,\,
\end{equation}
where $\mathfrak{Re} c$ is greater than real parts of all
singularities of the integrand in the $s$-plane, is independent of
$\varpi$ on the semiaxis $\varpi>0$.
\end{theorem}
%The proof is trivial:

\begin{proof}
Differentiation of Eq.~(\ref{lemma}) by parameter $\varpi$ gives:
\begin{eqnarray}\label{lemma-diff}
\frac{\partial}{\partial\varpi}\int_{c-i\infty}^{c+i\infty}\!ds e^{s\omega}\mathcal{F}\left(\int_{\epsilon}^{\omega+\varpi} d\omega_1 e^{-s\omega_1}f(\omega_1)\!\right)\qquad\qquad\nonumber\\
=f(\omega+\varpi)\!\int_{c-i\infty}^{c+i\infty}\!ds e^{-\varpi
s}\mathcal{F}'\!\left(\int_{\epsilon}^{\omega+\varpi}\!\! d\omega_1
e^{-s\omega_1}f(\omega_1)\!\right).\nonumber\\
\end{eqnarray}
The condition that function
$\left|\mathcal{F}'\left(\int_{\epsilon}^{\omega+\varpi} d\omega_1
e^{-s\omega_1}f(\omega_1)\right)\right|$ grows with $\mathfrak{Re}s$
slower than exponentially guarantees the exponential decrease of the
integrand of (\ref{lemma-diff}) when $\varpi>0$. Thus, if the
integration contour, defined to pass on the right of all
singularities of the integrand, is withdrawn rightwards to infinity,
the whole integral exponentially vanishes. Therewith, derivative
(\ref{lemma-diff}) is identically zero, which entails independence
of integral (\ref{lemma}) on $\varpi$ under the conditions assumed.
\end{proof}

In our case (\ref{generic-eq}), (\ref{generic-eq-upperlimitomega}),
function $\mathcal{F}(z)$ has to be identified with an exponential
\begin{equation}\label{FFF}
\mathcal{F}(z)=e^{z-\int_{\epsilon}^{E} d\omega_1
\frac{dw_1}{d\omega_1}}\equiv \mathcal{F}'(z)
\end{equation}
(where $\epsilon$ furnishes an IR cutoff, if necessary). Function
(\ref{FFF}) satisfies the conditions of our lemma liberally: since
$z(s)=\int^E_{\epsilon}d\omega_1
e^{s\omega_1}\frac{dw_1}{d\omega_1}$ increases at $\mathfrak{Re}
s\to+\infty$ at most logarithmically, $|\mathcal{F}'[z(s)]|$ in this
limit can only increase as some power of $s$. As a whole,
$\mathcal{F}(z)$ is actually IR-finite, so after unifying
$\omega_1$-integrals in the exponent of $\mathcal{F}$, it is safe to
put $\epsilon=0$. By choosing in expression (\ref{lemma}) $\varpi$
arbitrarily small or large, one can deduce
Eq.~(\ref{generic-eq-upperlimitomega}) from Eq.~(\ref{generic-eq}),
or vice versa. We will also employ Lemma \ref{lemma1} for
simplification of the exponentiated integrals with specific profiles
of $dw_1/d\omega_1$ in the next section.

%All 4 representations (\ref{generic-eq}-\ref{generic-eq-upperlimitomega-byparts}) will be helpful in our following calculations.

Eq. (\ref{generic-eq-upperlimitomega}) has an appealing property
that apart from the factor
\begin{equation}\label{W0omega-def}
e^{-\int^{E}_{\omega}d\omega_1\frac{dw_1}{d\omega_1}}=W_0(\omega),
\end{equation}
the multiphoton spectrum involves only contributions from the
single-photon spectrum with $\omega_1<\omega$. That looks natural in
view of the positivity of contributing photon energies. In this
sense, it seems natural to term Eq.
(\ref{generic-eq-upperlimitomega}) `energetically ordered' form [to
distinguish it from `non-ordered' form (\ref{generic-eq})]. Still,
one should mind the existence of factor (\ref{W0omega-def}), which
depends on contributions from $\omega_1>\omega$, and is nothing but
the probability of non-emission of any photon with energy greater
than $\omega$ \cite{BK} (check:
\begin{eqnarray*}
W_0+\sum_{n=1}^\infty\int_0^\infty d\omega_1\ldots\int_0^\infty
d\omega_n\frac{dW_n}{d\omega_1\ldots
d\omega_n}\prod_{k=1}^n\theta(\omega-\omega_k)\nonumber\\
=W_0(\omega),\qquad\qquad\qquad\qquad\qquad\qquad\qquad\qquad\qquad
\end{eqnarray*}
with $\theta$ standing for the Heaviside unit step function). So,
the effect of contributions from $\omega_1>\omega$ is
`non-dynamical', resulting in a rather trivial explicit factor, but
nonetheless, due to this factor, the energetic ordering does not
hold strictly for multiphoton probability spectrum. Remarkably, in
Sec.~\ref{sec:reconstr} we will encounter representations which may
be regarded even as energetically anti-ordered.
%Later on we will also analyze the origin of factor $W_0(\omega)$ from the viewpoint of a kinetic equation.

\subsection{Perturbation series}\label{subsec:Ln}

The manifestation of multiphoton effects may be studied in terms of
non-linearity of the radiation spectrum dependence on the radiator
length (or the crystalline target thickness), $L$ (cf.
\cite{relevance-of-multiphot,relevance-of-multiphot2}). In the
simplest approximation (reasonable for long radiators), one may
assume that $\frac{dw_1}{d\omega_1}\propto L$. Then, $L$-dependence
of $\frac{dw}{d\omega}$ may be studied based on expansion
(\ref{generic-sum-int}), where factor $W_0$ is yet a non-linear
function of $w_1$. Expanding the latter factor to series in $w_1$
leads to a double series for $\frac{dw}{d\omega}$.

In fact, a single power series in $L$ can be obtained using the
following trick. Issuing from contour integral representation
(\ref{generic-eq}),  integrate in the exponent by parts:
\begin{eqnarray}
\frac{dw}{d\omega}&=&\frac1{2\pi i}\int_{c-i\infty}^{c+i\infty}ds e^{s\omega-s\int_0^{E} d\omega_1 e^{-s\omega_1}\int^{E}_{\omega_1}d\omega'_1\frac{dw_1}{d\omega'_1}}\nonumber\\
&\,&-W_0\delta(\omega).%\nonumber\\
\label{generic-eq-2}
%\\
%&\,&\qquad\qquad (0<\omega\ll E_e),\nonumber
\end{eqnarray}
Next, expand the $\frac{dw_1}{d\omega_1}$-dependent part of the
exponential to power series:
\begin{equation}\label{sum-wn}
\frac{dw}{d\omega}=\sum^\infty_{n=1}\frac{dw_n}{d\omega},
\end{equation}
with
\begin{eqnarray*}\label{}
\frac{dw_n}{d\omega}&=&\frac{(-1)^n}{n!}\frac{1}{2\pi
i}\int_{c-i\infty}^{c+i\infty} ds s^n
e^{s\omega}\nonumber\\
&\,&\times\prod^n_{k=1}\int_0^E d\omega_k
e^{-s\omega_k}\int_{\omega_k}^{E}d\omega'_k\frac{dw_1}{d\omega'_k}.
\end{eqnarray*}
Now, it is straightforward to integrate over $s$ termwise with the
use of the identity
\begin{equation}\label{}
\frac1{2\pi i}\int_{c-i\infty}^{c+i\infty}ds
e^{s\left(\omega-\sum_{k=1}^n\omega_k\right)}s^n=\frac{\partial^n}{\partial\omega^n}\delta\left(\omega-\sum_{k=1}^n\omega_k\right),
\end{equation}
which yields
\begin{eqnarray}\label{dwn-pert-ser}
\frac{dw_n}{d\omega}=\frac{(-1)^n}{n!}\frac{\partial^n}{\partial\omega^n}\int_0^{E}d\omega_1\ldots \int_0^{E}d\omega_n\qquad\qquad\quad\nonumber\\
\times\delta\left(\omega-\sum_{k=1}^n\omega_k\right)\int^{E}_{\omega_1}d\omega'_1\frac{dw_1}{d\omega'_1}\ldots
\int^{E}_{\omega_n}d\omega'_n\frac{dw_1}{d\omega'_n}.\quad
\end{eqnarray}
The latter multiple integral converges even if the  mean photon
number $w_1$ diverges logarithmically; its multiple derivatives
always exist, as well.
%It may be amusing to note that Eq.~(\ref{dwn-pert-ser}) realizes a unitary perturbation theory on the probability level.

%From the practical viewpoint, if $\frac{dw_1}{d\omega_1}$ is
%proportional to the radiator length $L$, series (\ref{sum-wn},
%\ref{dwn-pert-ser}) may be viewed as a power expansion in $L$:
%$\frac{dw_n}{d\omega_1}\propto L^n$, describing successive nonlinear
%corrections in the target thickness dependence. Such nonlinearities
%are observed for coherent radiation in even not too thick ($\sim
%10^{-1}$ mm) crystals
%.

In contrast to Eq.~(\ref{generic-sum-int}), however, terms of series
(\ref{dwn-pert-ser}) need not be everywhere positive. That is
chained to the fact that multiple photon emission leads to a
redistribution of the spectrum, not only to pileup of events. From
the standpoint of QED, negative terms in the perturbative expansion
typically arise from interference of loop (photon reabsorption)
diagrams with lower-order ones \cite{phot-stat-indep}.
Eq.~(\ref{dwn-pert-ser}) indicates that for an observable such as
the calorimetrically measured spectrum, the structure of those
corrections assumes a simple generic form.

Correspondence of (\ref{dwn-pert-ser}) with expansion
(\ref{generic-sum-int}) may be established by differentiating in
(\ref{dwn-pert-ser}) the delta-function under the integral sign by
formula
\begin{equation}\label{dwn-def}
(-1)^n\frac{\partial^n}{\partial\omega^n}\delta\left(\!\omega-\sum_{k=1}^n\omega_k\!\right)=\frac{\partial^n}{\partial\omega_1\ldots\partial\omega_n}\delta\left(\!\omega-\sum_{k=1}^n\omega_k\!\right)\!,
\end{equation}
and subsequently integrating by parts over all $\omega_k$. It should
be minded thereat that the endpoint terms are non-vanishing,
producing the powers of the total single-photon emission probability
$w_1$ present in Eq.~(\ref{generic-sum-int}) in the exponent.
However, if $w_1$ diverges, the integration by parts is no longer
possible, wherewith series (\ref{generic-sum-int}) does not exist,
but Eq.~(\ref{dwn-pert-ser}) remains valid, anyway.
%At practice, though, if there exists an additive representation for primary spectra like (\ref{coh+inc}), it may actually be easier to evaluate corrections to $\frac{dw_{1\text{c}}}{d\omega_1}$ by Eq.~(\ref{generic-sum-int}) not requiring differentiation, and then convolve with $\frac{dw_{\text{i}}}{d(\omega-\omega_1)}$ (see Sec.~\ref{subsec:coh+incoh}).

\subsection{Spectral moments}\label{subsec:moments}

Characterization of compact probability distributions is often
conducted in terms of their moments. Spectrum $\frac{dw}{d\omega}$,
however, is not normalized to unity, so it can not be used
straightforwardly as a weighting distribution. In capacity of a
normalized probability distribution, one should either use electron
distribution function (\ref{=Pi}), or the rescaled radiation
spectrum $\frac1{1-W_0}\frac{dw}{d\omega}$. We will adopt the first
option, which leads to simpler final results.

The mean emitted photon energy for our resummed spectrum is
computed, e.g., by termwise integration in
Eq.~(\ref{generic-sum-int}):
\begin{eqnarray}\label{mean-omega}
\overline \omega&=&\int_0^\infty d\omega \omega\left[\frac{dw}{d\omega}+W_0\delta(\omega)\right]\equiv\int_0^\infty d\omega \omega\frac{dw}{d\omega}\nonumber\\
&=&e^{-w_1}\sum_{n=1}^\infty\frac{n}{n!}w_1^{n-1}\overline{\omega_1\!}=\overline{\omega_1\!}.
\end{eqnarray}
Thus, it coincides with that for the single-photon spectrum,
\begin{equation}\label{}
\overline{\omega_1\!}=\int_0^\infty d\omega_1
\omega_1\frac{dw_1}{d\omega_1}.
\end{equation}
That is quite natural, inasmuch as once photon energies in all the
events are summed over, it no longer matters whether they were
measured by photon counters or calorimeters. [In particular, that
justifies the use of classical formulas for the rate of radiative
energy losses (see, e.g., \cite{mean-energy-loss}) under conditions
of multiple photon emission.]

To further characterize the spectrum width, asymmetry, and so on,
one can refer to higher moments about the mean value (central
moments), presuming their existence. Those are conveniently
calculated based on representation (\ref{generic-eq}):
\begin{eqnarray}\label{higher-moments}
\overline{(\omega-\overline{\omega})^n}
&=&\int_0^\infty d\omega (\omega-\overline{\omega_1\!})^n\left[\frac{dw}{d\omega}+W_0\delta(\omega)\right]\qquad\qquad\nonumber\\
&=&\left(-\frac{\partial}{\partial s}-\overline{\omega_1\!}\right)^n e^{\int_0^{E} d\omega_1 \frac{dw_1}{d\omega_1}\left(e^{-s\omega_1}-1\right) }\bigg|_{s=0}\nonumber\\
&\equiv&(-1)^n\frac{\partial^n}{\partial s^n}e^{\int_0^{E} d\omega_1
\frac{dw_1}{d\omega_1}\left(e^{-s\omega_1}-1+s\omega_1\right)
}\bigg|_{s=0}.\,\,
\end{eqnarray}
The above representation implies that exponential $e^{\int_0^{E}
d\omega_1
\frac{dw_1}{d\omega_1}\left(e^{-s\omega_1}-1+s\omega_1\right) }$
serves as a generating function \cite{Gnedenko-Kolmogorov,Feller}
for central moments, while $e^{\int_0^{E} d\omega_1
\frac{dw_1}{d\omega_1}\left(e^{-s\omega_1}-1\right) }$ offers a
generating function for ordinary moments. Applying
Eq.~(\ref{higher-moments}) for lowest $n$, one finds\footnote{The
exponential form of generating functions suggests that instead of
moments for resummed spectra, it could be easier to calculate the
corresponding cumulants (semi-invariants)
\cite{Gnedenko-Kolmogorov}, which for Poisson distribution coincide
with moments for the single-photon spectrum through all orders. But
for our purposes, we will not need moments of order higher than 4,
anyway.}
\begin{equation}\label{mean-omega2-alt}
\overline{\left(\omega-\overline{\omega}\right)^2
}=\overline{\omega_1^2},
\end{equation}
\begin{equation}\label{mean-omega3}
\overline{\left(\omega-\overline{\omega}\right)^3
}=\overline{\omega_1^3},
\end{equation}
\begin{equation}\label{mean-omega4}
\overline{\left(\omega-\overline{\omega}\right)^4
}=3\big(\overline{\omega_1^2}\big)^2+\overline{\omega_1^4},
\end{equation}
etc. Relations (\ref{mean-omega}), (\ref{mean-omega2-alt})
(historically known as Campbell's theorem \cite{Feller}) were
discussed in \cite{Kolchuzhkin} with the objective of extracting
information about the single-photon distribution from the resummed
one. Eqs.~(\ref{mean-omega3}), (\ref{mean-omega4}) may serve for the
same purpose. In addition, in Sec.~\ref{sec:reconstr} we will
formulate a reconstruction procedure for the complete spectrum, not
resorting to the notion of moments.

%It is worth emphasizing that the moments defined above are
%\emph{unnormalized}. They are proportional to the radiation
%intensity, and for example, may be vanishing if the intensity is
%small. So, in order to characterize the spectrum shape, one yet has
%to adjust the normalization.

%\begin{widetext}
\begin{figure*}
\includegraphics{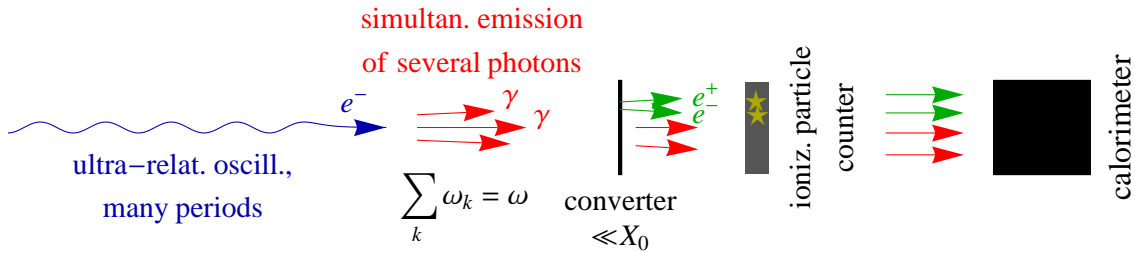}
\caption{\label{fig:multty-measur-scheme} Scheme of measurement of
the photon multiplicity spectrum. Due to the converter thinness, the
number of $e^+e^-$ pairs created in it is proportional to the number
of incident photons. The knowledge of the proportionality
coefficient then allows assessing the photon number in a radiative
event. The total energy emitted per event is measured by the final
calorimeter, as in Fig.~\ref{fig:calorim-scheme}.}
\end{figure*}
%\end{widetext}

A predicament with the use of the above quoted moments is that with
the account of an incoherent bremsstrahlung component
(Sec.~\ref{sec:coh+incoh}), they all diverge in the ultraviolet.
Therefore, they may seem to be only useful in the context of pure
coherent radiation. Nonetheless, later on we will show that
treatments of coherent and incoherent radiation components can
always be separated, permitting the usage of spectral moments for
pure coherent component regardless of the presence of an incoherent
one.

\subsection{Photon multiplicity spectrum}

As was mentioned in the Introduction, it is rather difficult in a
multiphoton event to practically pin down individual energies of all
the photons, i.e., to measure joint probability distribution
$\frac{dW_n}{d\omega_1\ldots d\omega_n}$. But there exist relatively
compact experimental methods allowing to receive partial information
about the photon number content. For instance, placement of a thin
converter and an ionizing particle counter upstream the calorimeter
allows measuring the mean number of photons (the photon
multiplicity) as a function of the total energy $\omega$ deposited
in the calorimeter -- see Fig.~\ref{fig:multty-measur-scheme} and,
e.g., \cite{multiphot-CERN-Kirsebom}. In this subsection, we will
extend our resummation procedure to description of the photon
multiplicity spectrum.

In the spirit of Eqs.~(\ref{mean-n-def}) and
(\ref{generic-sum-int}), to construct the photon multiplicity at a
given $\omega$, one must incorporate in (\ref{generic-sum-int}) a
weighting factor $n$ equal to the number of emitted photons,
whereupon divide the resulting sum by the unweighted expression:
\begin{eqnarray}\label{mean-n-omega-def}
\bar n(\omega)=\frac{W_0}{dw/d\omega}\sum_{n=1}^\infty\frac{n}{n!} \int_0^\infty  d\omega_1 \frac{dw_1}{d\omega_1}\ldots\qquad\nonumber\\
\times\int_0^\infty  d\omega_n \frac{dw_1}{d\omega_n}
\delta\left(\omega-\sum_{k=1}^n\omega_k\right).
\end{eqnarray}
Since all the weights $n$ in this sum are greater than unity, so
must be their mean number:
\begin{equation}\label{mean-n-gtr1}
\bar n(\omega)\geq 1.
\end{equation}
In particular, when $\omega\to0$, integration phase space volumes
for $n\geq2$ terms shrink to zero, wherefore only term $n=1$
survives. This term cancels with the correspondent term in the
denominator, yielding
\begin{equation}\label{mean-n-to1}
\bar n(0)=1.
\end{equation}

The sum entering Eq.~(\ref{mean-n-omega-def}) can again be evaluated
by the Laplace transform method, as follows:
\begin{subequations}\label{mean-n-omega-contour}
\begin{eqnarray}
\bar n(\omega)\frac{dw}{d\omega}&=&\frac1{2\pi i}\int_{c-i\infty}^{c+i\infty}ds  e^{s\omega+\int_0^{E} d\omega_1 \frac{dw_1}{d\omega_1} \left(e^{-s\omega_1}-1\right)}\quad\nonumber\\
&\,&\qquad\qquad\quad\times\int_0^{E} d\omega'_1 \frac{dw_1}{d\omega'_1} e^{-s\omega'_1}\label{mean-n-omega-contour-non-casual}\\
&\equiv&\frac{W_0(\omega)}{2\pi i}\int_{c-i\infty}^{c+i\infty}ds  e^{s\omega+\int_0^{\omega} d\omega_1 \frac{dw_1}{d\omega_1} \left(e^{-s\omega_1}-1\right)}\quad\nonumber\\
&\,&\qquad\qquad\quad\times\int_0^{\omega} d\omega'_1
\frac{dw_1}{d\omega'_1}
e^{-s\omega'_1}.\label{mean-n-omega-contour-casual}
\end{eqnarray}
\end{subequations}
[The latter equality may be inferred from Lemma~\ref{lemma1} with
$\mathcal{F}(z)=z e^{z-\int_0^{E} d\omega_1
\frac{dw_1}{d\omega_1}}$]. This may also be expressed as a
convolution of the single-photon spectrum with the resummed one:
\begin{equation}\label{mean-n-omega-convol}
\bar
n(\omega)\frac{dw}{d\omega}=\int_0^{\omega}d\omega_1\frac{dw_1}{d\omega_1}\frac{dw}{d\omega'}\bigg|_{\omega'=\omega-\omega_1}+W_0\frac{dw_1}{d\omega}.
\end{equation}
Here the last term stems from the $\delta$-function term of
Eq.~(\ref{generic-eq-upperlimitomega}).

Unlike $\frac{dw}{d\omega}$, however, the multiplicity spectrum is
not an IR-safe quantity. But fortunately, the IR divergence of $w_1$
affects only the constant, $\omega$-independent part of $\bar
n(\omega)$. Indeed, the contribution from vicinity of the lower
integration limit in the convolution term in
(\ref{mean-n-omega-convol}) is $\propto dw/d\omega$, and so is the
l.h.s. Hence, at $W_0\to0$, $\omega$-dependent factors in both sides
of Eq.~(\ref{mean-n-omega-convol}) cancel.

To isolate the IR divergence of $\bar n(\omega)$ explicitly, one can
rewrite Eq.~(\ref{mean-n-omega-convol}), e.g., as
\begin{eqnarray}\label{mean-n-omega-convol-regul}
\bar n(\omega)-\!\int_{\epsilon}^{\omega}\!d\omega_1\frac{dw_1}{d\omega_1}=\int_0^{\omega}\!d\omega_1\frac{dw_1}{d\omega_1}\!\left(\frac{\frac{dw}{d\omega'}\big|_{\omega'=\omega-\omega_1}}{{dw}/{d\omega}}-1\!\right)\nonumber\\
\\
(w_1\to\infty).\qquad\qquad\qquad\qquad\qquad\nonumber
\end{eqnarray}
Here the r.h.s. is IR-safe (thence we put there $\epsilon=0$), while
in the l.h.s., the dependence of $\bar n(\omega)$ on the IR cutoff
$\epsilon$ must be additive, to cancel with the additive dependence
on IR cutoff of the integral term. On the other hand, the dependence
of on $E$ for $\bar n(\omega)$ must disappear, as will be
demonstrated in the next section.

%Indeed, sending in (\ref{mean-n-omega-convol}) $W_0$, and regulating
%the integral by a cutoff $\epsilon$ on the lower limit,
%\begin{eqnarray}\label{mean-n-omega-convol-regul}
%\frac{dw}{d\omega}\bar n(\omega)&=&\int_{\epsilon}^{\omega}d\omega_1\frac{dw_1}{d\omega_1}\frac{dw}{d\omega'}\bigg|_{\omega'=\omega-\omega_1}\nonumber\\
%&=&-\int_{\epsilon}^{\omega}d\int^{\omega}_{\omega_1}d\omega'_1\frac{dw_1}{d\omega'_1}\frac{dw}{d\omega'}\bigg|_{\omega'=\omega-\omega_1}\nonumber\\
%&=&\frac{dw}{d\omega}\int^{\omega}_{\epsilon}d\omega'_1\frac{dw_1}{d\omega'_1}\nonumber\\
%&\,&+\int_0^{\omega}d\omega_1\int^{\omega}_{\omega_1}d\omega'_1\frac{dw_1}{d\omega'_1}\frac{d}{d\omega_1}\frac{dw}{d\omega'}\bigg|_{\omega'=\omega-\omega_1}.
%\end{eqnarray}
%The last integral in (\ref{mean-n-omega-convol-regul}) is already IR safe, so its lower limit is again replaced by 0. Dividing by $\frac{dw}{d\omega}$, we get
%\begin{eqnarray}\label{mean-n-omega-convol-regul}
%\bar n(\omega)&=&\int^{\omega}_{\epsilon}d\omega'_1\frac{dw_1}{d\omega'_1}\nonumber\\
%&\,&+\frac1{{dw}/{d\omega}}\int_0^{\omega}d\omega_1\int^{\omega}_{\omega_1}d\omega'_1\frac{dw_1}{d\omega'_1}\frac{d}{d\omega_1}\frac{dw}{d\omega'}\bigg|_{\omega'=\omega-\omega_1}.
%\end{eqnarray}
%Differs from $w_1(\omega)$, which is the first term of (\ref{mean-n-omega-convol-regul}).
%\newpage

\section{Properties of incoherent and coherent multiphoton radiation spectra}\label{sec:coh+incoh}

Generic resummation formulas established in the previous section are
valid for radiation characterized by an arbitrary single-photon
profile $dw_1/d\omega_1$. In what follows, we will mostly
concentrate on a category of coherent radiation spectra involving an
admixture of incoherent radiation. In that case, the salient
features shared by the single-photon spectra are:
\begin{itemize}
\item spectral discontinuities at finite $\omega$ in the coherent radiation component (coherent emission edges);
\item a `tail' towards large $\omega$ brought by the incoherent bremsstrahlung component;
\item an infrared divergence, originating from the incoherent bremsstrahlung, or from non-zero net deflection by coherent fields (see, e.g., \cite{Bondarenco-CBBC}). For simplicity, in this paper we will be only considering the first case.
\end{itemize}

Our first task is to investigate how do the above mentioned features
modify under multiphoton emission conditions. Since multiphoton
effects interplay nonlinearly, it is expedient first to analyze them
for pure coherent and incoherent components separately. Combining
them, we can thereupon distinguish effects of their mutual
influence.

\subsection{Pure incoherent bremsstrahlung}\label{subsec:incoh-bremsstr}

Let us first scrutinize the case of purely incoherent
bremsstrahlung. Its single-photon spectrum must be proportional to
Bethe-Heitler's spectrum of radiation at a single
collision\footnote{\label{foot:no-medium-effects}This assumption
admittedly neglects dense-medium effects such as transition
radiation due to the change of the dielectric density, and LPM
radiation suppression by multiple scattering \cite{Ter-Mik}. Their
neglect is justified under typical conditions $L\sim0.1\div1$~mm,
$E_e<100$~GeV, $\omega\gtrsim1$~MeV, although some of those
conditions can be relaxed at the expense of the others. The
influence of forward transition radiation (noticeable when $L<0.1$
mm), and of LPM suppression on resummed radiation spectra was
studied in \cite{BK}.} \cite{Rossi}. Moreover, in the soft radiation
limit we are restricted to, the single-photon spectrum reduces to
the well-known semiclassical form \cite{1/omega-spectrum}
\begin{equation}\label{dw1inc}
\frac{dw_{1\text{i}}}{d\omega_1}\simeq\frac{a}{\omega_1}\qquad
(\omega_1\ll E_e),
\end{equation}
where $a$ is proportional to the target thickness $L$. In an
amorphous matter, constant $a$ equals
\begin{equation}\label{a-amorph}
a=\frac43\frac{L}{X_0},
\end{equation}
where the radiation length $X_0\approx (4Z^2\alpha r_e^2
n_{\text{a}} \ln\frac1{Z^{1/3}\alpha})^{-1}$ depends on the atomic
density $n_{\text{a}}$ and (net) nucleus charge $Z$
\cite{Rossi,PDG}. A similar estimate applies also for the incoherent
component of coherent bremsstrahlung in a crystal. For the case of
positron channeling, if their close collisions with atomic nuclei
are sufficiently rare, factor $Z^2n_{\text{a}}$ should be replaced
by the net interplanar electron density $n_e$. Finally, in an
undulator, $a$ is proportional to the density of the residual air,
though in high vacuum that can be negligible.

As was argued in the end of Sec.~\ref{subsec:Stat-indep-emission},
the integral of (\ref{dw1inc}), which is logarithmically divergent
in the UV, must be cut off at some photon energy commensurable with
the electron energy: $E=\kappa E_e$, with $\kappa\sim0.5$. We
implement this cutoff directly for approximation (\ref{dw1inc}),
adopting a simplified model:
\begin{equation}\label{dw1incoh-cutoff}
\frac{dw_{1\text{i}}}{d\omega_1}=\frac{a}{\omega_1}\theta(E-\omega_1),%\qquad E=\kappa E_e,
\end{equation}
where $\theta(z)$ is the Heaviside function.

\begin{figure}
\includegraphics{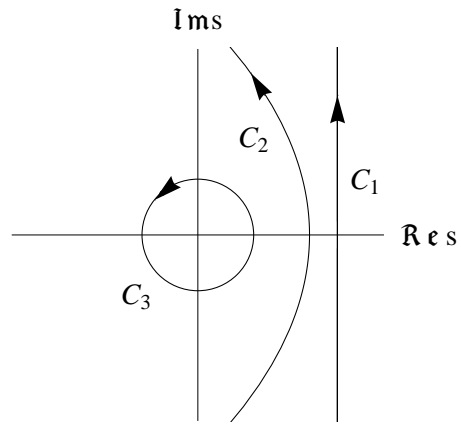}
\caption{\label{fig:contours} Integration contours for resummed
spectra. Line $\mathcal{C}_1$, used in Eq.~(\ref{generic-eq}),
(\ref{generic-eq-upperlimitomega}). Deformed contour
$\mathcal{C}_2$, used in Eqs.~(\ref{EulerGamma-def}),
(\ref{dw-first-interv}). Closed loop contour $\mathcal{C}_3$,
applicable for pure coherent radiation (see
Sec.~\ref{subsubsec:coh-in-fund-interv}).}
\end{figure}

A convenient starting point for evaluation of the resummed spectrum
corresponding to single-photon spectrum (\ref{dw1incoh-cutoff}) is
the energetically ordered contour integral representation,
Eq.~(\ref{generic-eq-upperlimitomega}). Therein, the $\delta$-term
is suppressed, since $W_0\to 0$, and for the present case it will be
neglected. Once (\ref{dw1incoh-cutoff}) is inserted to
Eq.~(\ref{generic-eq-upperlimitomega}),\footnote{Regarding the
random nature of incoherent bremsstrahlung, it may be unobvious
whether it is legitimate to resum it after the averaging, or only
prior to it (see Footnote \ref{exp-before-aver}). That depends on
how large are relative fluctuations of parameter $a$ per electron
passage. If the radiation formation length
$l_{\text{f}}=\frac{2E_e^2}{m^2\omega_1}$ is $\ll L$, which is
plausible under conditions of Footnote \ref{foot:no-medium-effects},
many electron scatterings in the target contribute to the radiation
independently, so fluctuations of $a$ are relatively small,
justifying the use of resummation for the averaged spectrum.
Averaging procedure for multiphoton radiation in a thin target will
be considered elsewhere.}  all $\omega$- and $E$-dependencies factor
out after a simple rescaling of the integration variables, giving
\begin{equation}\label{13}
\frac{dw_{\text{i}}}{d\omega}=\frac{a}\omega\left(\frac{\omega}{E}\right)^a
\Phi(a)\equiv\frac{dw_{1\text{i}}}{d\omega}\left(\frac{\omega}{E}\right)^a
\Phi(a)\qquad (\omega< E),
\end{equation}
with
\begin{equation}\label{Phi-def}
\Phi(a)=\frac1a\frac1{2\pi i}\int_{c-i\infty}^{c+i\infty}d\zeta
e^{\zeta+a\int_0^{1} \frac{d\xi}{\xi} \left(e^{-\zeta\xi}-1\right)
}.
\end{equation}
To evaluate $\Phi(a)$, one may regularize
$\int\frac{d\xi}{\xi}\left(e^{-\zeta\xi}-1\right)$ on the lower
limit, then split it in two integrals, and by Lemma~\ref{lemma1},
extend the upper limit of the integral involving $e^{-\zeta\xi}$ to
infinity. Thereafter, the IR regulator may again be sent to zero,
and the sum of $\xi$-integrals evaluates:
%\footnote{Structure
%(\ref{Phi-eval}) might be obtained more straightforwardly based on
%contour integral representation
%(\ref{generic-eq-upperlimitomega-byparts}), where integration by
%parts has already been made, but then applicability of
%Lemma~\ref{lemma1} may be less obvious.}
\begin{equation}\label{Phi-eval}
\Phi(a)=e^{a\int_0^\infty dz e^{-z}\ln z}\frac1a\frac1{2\pi
i}\int_{c-i\infty}^{c+i\infty}\frac{d\zeta}{\zeta^a} e^{\zeta}.
\end{equation}
The integrand of the latter contour integral is vanishing as
$\mathfrak{Re}\zeta\to-\infty$, so at complex infinity the contour
can be turned to the left from the imaginary axis (contour
$\mathcal{C}_2$ in Fig.~\ref{fig:contours}). Using representation
\begin{equation}\label{EulerGamma-def}
\frac1{\Gamma(a)}=\frac1{2\pi
i}\int_{\mathcal{C}_2}\frac{d\zeta}{\zeta^a}e^{\zeta}
\end{equation}
for the Euler gamma function, and definition
\[
-\int_0^\infty dz e^{-z}\ln
z\equiv-\Gamma'(1)=\gamma_{\text{E}}=0.577\ldots
\]
for Euler's constant, we get
\begin{equation}\label{Fa}
\Phi(a)=\frac{e^{-\gamma_\text{E} a}}{\Gamma(1+a)}.
\end{equation}
Function (\ref{Fa}) is manifestly positive and monotonously
decreasing (see Fig.~\ref{fig:Phi,nu}), its derivative in the origin
being zero:
\[
\Phi(a)=1-\frac{\pi^2}{12} a^2+\mathcal{O}(a^3).
\]
Other models for the single-photon spectrum suppression at large
$\omega_1$ yield equivalent results\footnote{For example, if instead
of remote but sharp cutoff (\ref{dw1incoh-cutoff}) one employs a
slow exponential factor,
\[
\frac{dw_{1\text{i}}}{d\omega_1}=\frac{a}{\omega_1}e^{-\lambda\omega_1}
\]
with $\lambda\sim E_e^{-1}$, the insertion of this form to
representation (\ref{generic-eq}), with the upper limit of
$\omega_1$-integral replaced by infinity, gives
\begin{equation}\label{gamma-distr-a}
\frac{dw_{\text{i}}}{d\omega}=\frac{a}{\omega}\frac{(\lambda\omega)^a}{\Gamma(1+a)}e^{-\lambda\omega}.
\end{equation}
Apart from factor $e^{-\lambda\omega}$ (which is close to unity when
$\omega\ll E_e$), this structure coincides with Eq.~(\ref{13}), once
one identifies $\lambda=e^{-\gamma_{\text{E}}}E^{-1}$. There, the
single-photon spectrum is expressed by a gamma-distribution
\cite{Feller} of index 0, while the resummed spectrum is again a
gamma-distribution, but of index $a$. That is a well-known example
of functional stability more general than L\'{e}vy stability (see,
e.g., \cite{random-walk}). One can as well arrive at result
(\ref{13}) from the side of L\'{e}vy-stable densities
\cite{Feller,Gnedenko-Kolmogorov}, for which the single-photon
distribution is strictly scale-invariant:
$\frac{dw_{1\text{i}}}{d\omega_1}\propto
\frac1{\omega_1^{1+\delta}}$; however, L\'{e}vy distributions at
finite $\delta$, in contrast to (\ref{13}) or (\ref{gamma-distr-a}),
do not reduce to a product of functions of a single variable.}.

Regarding the shape of the resulting distribution, two remarks are
in order. First, spectrum (\ref{13}) is suppressed compared to
single-photon spectrum (\ref{dw1inc}) everywhere in the region
$\omega<E$. At $\omega>E$, there formally develops an enhancement of
the spectrum, but it is not accurately described  by Eq.~(\ref{13}),
insofar as at $\omega\sim E$ quantum effects enhance, while at
$\omega=E_e$, the spectrum must strictly terminate at all. Hence,
there ansatz (\ref{dw1incoh-cutoff}) for the single-photon spectrum
breaks down, so our model description, though self-consistent
formally, is inadequate in domain $\omega\sim E$. With this
reservation, the uniform suppression of the resummed spectrum at
$\omega<E$ does not contradict, e.g., the conservation of the mean
photon energy at resummation [Eq.~(\ref{mean-omega})].
%It should also be noted that moments discussed in Sec.~\ref{subsec:moments} virtually have no sense for distribution (\ref{13}), because due to its slow decrease with $\omega$, they are dominated by the vicinity of the spectrum cutoff, where Eq.~(\ref{dw1inc}) is not expected to be accurate.

\begin{figure}
\includegraphics{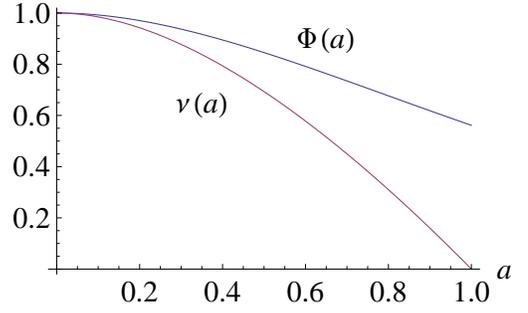}
\caption{\label{fig:Phi,nu} Graphs of functions $\Phi(a)$
[Eq.~(\ref{Fa})] and $\nu(a)$ [Eq.~(\ref{nu-through-digamma})]. Our
physical assumptions require that $a\ll1$ (see text).}
\end{figure}

Secondly, the dependence of the resummed spectrum on $\kappa$ enters
Eq.~(\ref{13}) through parameter $E=\kappa E_e$, resulting in an
overall factor $\kappa^{-a}$. Thus, in order for result (\ref{13})
not to be strongly sensitive to the actual value of $\kappa$, one
needs fulfillment of condition
\begin{equation}\label{all1}
a\ll1
\end{equation}
(otherwise, full kinetic equations for the electromagnetic cascade
will need to be solved). Since the accuracy of Eq.~(\ref{13}) is
$[\mathcal{O}(1)]^a=1+\mathcal{O}(a)$, function
$\Phi(a)=\Phi(0)+\mathcal{O}(a^2)$ may be safely approximated by
$\Phi(0)=1$. Then, one readily checks that
\begin{equation}\label{incoh-norm}
\int_0^{E}d\omega
\frac{dw_{\text{i}}}{d\omega}=\Phi(a)\underset{a\ll 1}\approx 1,
\end{equation}
in accord with Eq.~(\ref{total-probab-1}) (where, again, $W_0\to0$,
owing to IR divergence of $w_{1\text{i}}$). Property
(\ref{incoh-norm}) implies that despite the smallness of $a$,
function (\ref{13}) is not uniformly small, but rather is similar to
a $\delta$-function, peaking at $\omega\to0$ where function
(\ref{13}) blows up:
\begin{equation}\label{delta-lim-ato0}
\underset{a\to0}\lim \frac{dw_{\text{i}}}{d\omega}=\delta(\omega).
\end{equation}
To prove Eq.~(\ref{delta-lim-ato0}), and estimate the narrowness of
$\frac{dw_{\text{i}}}{d\omega}$ as a function of parameter $a$, it
may be expedient to evaluate a median energy
$\omega_{\text{i}\frac12}$ at which there accumulates half
probability of the resummed incoherent radiation, i.e.
\[
\int_0^{\omega_{\text{i}\frac12}}d\omega
\frac{dw_{\text{i}}}{d\omega}:=\frac12.
\]
That yields
\begin{equation}\label{omegaincoh12}
\omega_{\text{i}\frac12}=\left[2\Phi(a)\right]^{-1/a}E\lll E_e,
\end{equation}
which at small $a$ is exponentially small, thereby validating
relation (\ref{delta-lim-ato0}).

Finally,  the (regularized) photon multiplicity spectrum can be
evaluated for the present case via
Eq.~(\ref{mean-n-omega-convol-regul}). Inserting (\ref{dw1inc}) and
(\ref{13}) to (\ref{mean-n-omega-convol-regul}), and evaluating the
integral, we get:
\begin{equation}\label{mean-n-omega-incoh}
\bar n_{\text{i}}(\omega)=a\ln\frac{\omega}{\epsilon}+\nu(a),
\end{equation}
with
\begin{subequations}
\begin{eqnarray}\label{nu-def}
\nu(a)&=&a\int_0^1\frac{d\xi}{\xi}\left[\frac{1}{(1-\xi)^{1-a}}-1\right]\\
&=&1+a\left[\psi(1)-\psi(1+a)\right]\label{nu-through-digamma}\\
&=&1-\frac{\pi^2}6 a^2+\mathcal{O}(a^3),
\end{eqnarray}
\end{subequations}
involving digamma-function $\psi(z)=\Gamma'(z)/\Gamma(z)$. Hence,
photon multiplicity spectrum grows with $\omega$ strictly
logarithmically, and proves to be independent of the ultraviolet
(UV) cutoff $E$. The graph of function $\nu(a)$ is shown in
Fig.~\ref{fig:Phi,nu}. At low $a$, it is close to unity, which is
natural, because no more than one photon is typically emitted per
passing electron.

\subsection{Pure coherent radiation}\label{subsec:pure-coh}

Next we consider the case of pure coherent radiation. It is
principally different from the incoherent radiation case considered
above, for as it involves an integrable (IR- and UV-safe), although
discontinuous single-photon spectrum. Physically, that may
correspond to (gamma-ray) undulator radiation \cite{Moortgat-Pick},
or channeling radiation \cite{books-on-coh-sources}. As for coherent
bremsstrahlung, usually containing an admixture of incoherent
radiation component, too, it is more pertinent to the combined
radiation case discussed in the next subsection.

In what concerns application to channeling radiation, though, it
should be stressed that the spectrum of the latter strongly depends
on impact parameters of the charged particles, even when
non-channeled passage events are rejected. Hence, at passing to
multiphoton spectra, at first we have to perform the resummation for
a definite impact parameter value, and only at the final step
average the result over impact parameters\footnote{Another possible
issue is the radiation cooling
\cite{multiphot-CERN-Kirsebom,rad-cooling}, which potentially can
lead to inequivalence of photon emission at early and at late
channeling stages. This effect, however, should be negligible under
the condition of small radiative losses we presume.}. We can not
indulge into such specialized procedures here, leaving them for
future studies. So, in application to channeling radiation our
results in this paper will only be preliminary.

What we wish to reflect in our present study is the transverse
oscillatory motion of the radiating particle with respect to the
direction of high longitudinal momentum, taking place in all the
abovementioned cases. Thereat, the radiation spectrum shape depends
on the particle oscillation harmonicity\footnote{In case of
channeling, the harmonicity of the interplanar motion holds well
only for positrons, which repel from singularities of the continuous
potential created by atomic nuclei.}. In the simplest case of purely
harmonic and small-amplitude oscillatory motion, the electromagnetic
radiation spectrum has the structure (see
Appendix~\ref{app:coh-rad-sources})
\begin{equation}\label{dw1coh-real}
\frac{dw_{1\text{c}}}{d\omega_1}=bP\left(\frac{\omega_1}{\omega_0}\right)\theta(\omega_0-\omega_1)\qquad
(\omega_0\ll E),
\end{equation}
where coefficient $b$ in a straight crystal/undulator is
proportional to the radiator length $L$, and the profile function
reads \cite{books-on-coh-sources,Ter-Mik}
\begin{subequations}\label{f-def}
\begin{eqnarray}
P(z)&=& 1-2z+2z^2\\
&\equiv&z^2+(1-z)^2.\label{f-def-b}
\end{eqnarray}
\end{subequations}
[The latter form demonstrates the function positivity and symmetry
with respect to midpoint $z=\frac12$]. In case if aperture
collimation is imposed on the photon beam, in view of the
unambiguous relation between the photon energy $\omega_1$ and its
emission angle [see Eq.~(\ref{omega-theta-corresp})], function $P$
in Eq.~(\ref{dw1coh-real}) will change, but still remain finite
everywhere.

If the transverse oscillatory motion of the electron/positron
happens to be highly anharmonic, or is already relativistic, higher
harmonics in the single-photon spectrum develop, which can generally
be incorporated as
\begin{equation}\label{dw1coh-higher-harm}
\frac{dw_{1\text{c}}}{d\omega_1}=b\sum_{m}^\infty
P_m\left(\frac{\omega_1}{\omega_{0m}}\right)\theta(\omega_{0m}-\omega_1).
\end{equation}
%Typically, the magnitudes of functions $P_m$ decrease rapidly enough for existence of the moments defined in Sec.~\ref{subsec:moments}.
In particular, with the increase of the positron energy beyond
$E_e>100$ GeV, higher harmonics of channeling radiation in a crystal
proliferate, and ultimately render the spectrum a synchrotron-like
appearance. This case may also be regarded as universal; it was
dealt with in papers
\cite{Khokonov,BKS-simul,Kononets-Ryabov-simul}.

Henceforth, we will restrict ourselves to the simplest case of a
single-harmonic dipole radiation described by
Eq.~(\ref{dw1coh-real}). While in our equations the shape of $P(z)$
will be handled as generic (to allow for the possibility of aperture
collimation), in graphical illustrations specific form (\ref{f-def})
will be used throughout.

\begin{figure}
\includegraphics{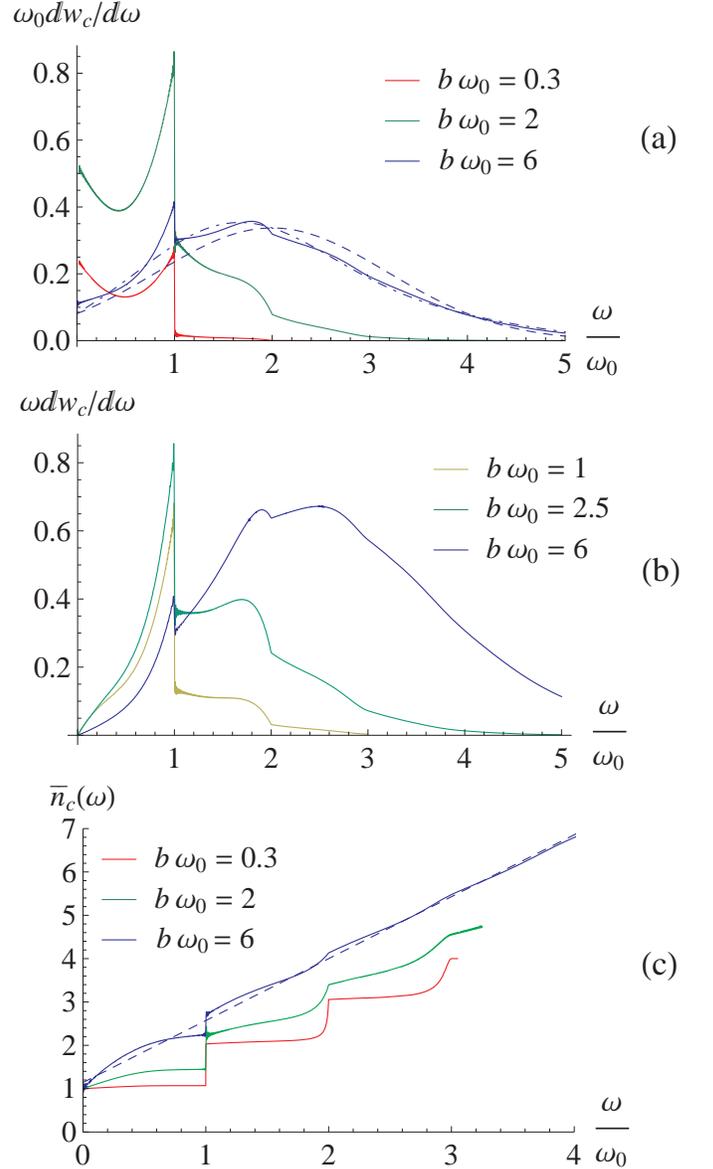}
\caption{\label{fig:pure-coh-spectrum} (a) Multiphoton probability
spectra computed for dipole harmonic coherent radiation defined
through Eqs.~(\ref{dw1coh-real}--\ref{f-def}) (solid curves). Red
curve, corresponding to intensity parameter $b\omega_0=0.3$, $\bar
n=\frac23b\omega_0=0.2$ [small deviations from the single-photon
spectrum (\ref{f-def})]. Green curve, the same for $b\omega_0=2$
(the highest summit reached by the fundamental maximum). Blue
curves, $b\omega_0=6$ (onset of high-intensity regime). Dashed blue
curve, Gaussian approximation (\ref{Gauss}). Dot-dashed blue curve,
corrected Gaussian approximation (\ref{Chebyshev}). (b) Multiphoton
energy spectrum $\omega\frac{dw_{\text{c}}}{d\omega}$ for the same
single-photon spectrum as in (a), and intensity parameter values
$b\omega_0=1$, 2.5, 6. The development of secondary maxima for the
energy spectrum is more spectacular than for the probability
spectrum $\frac{dw_{\text{c}}}{d\omega}$. (c) Photon multiplicity
spectrum described by Eqs.~(\ref{mean-n-omega-contour}),
(\ref{dw1coh-real}--\ref{f-def}). Solid curves correspond to same
parameters as in (a). Dashed blue line is high-intensity
approximation (\ref{mean-n-at-Gauss}).}
\end{figure}

Inserting entries (\ref{dw1coh-real}--\ref{f-def}) to Mellin
integrals (\ref{generic-eq}) and
(\ref{mean-n-omega-contour-non-casual}), and computing them
numerically, one can trace the evolution of the resummed spectrum
and the multiplicity spectrum with the increase of the intensity.
This is illustrated in Fig.~\ref{fig:pure-coh-spectrum} for a few
values of $b\omega_0$. The trends observed in those figures are:
smoothening of the spectral shape with the increase of the
intensity, the existence of an upper bound for the probability
distribution (Fig.~\ref{fig:pure-coh-spectrum}a), the appearance of
a second maximum in the energy spectrum at moderate intensity
(Fig.~\ref{fig:pure-coh-spectrum}b, green curve), and monotonic
increase of the photon multiplicity spectrum with $\omega$
(Fig.~\ref{fig:pure-coh-spectrum}c). Explaining the origin of those
features will occupy us next.

\subsubsection{Multiphoton coherent radiation spectrum in the fundamental energy interval}\label{subsubsec:coh-in-fund-interv}

When the single-photon spectrum is described by
Eq.~(\ref{dw1coh-real}), at sufficiently low radiation intensity,
the multiphoton spectrum will as well be concentrated within the
interval $0\leq\omega<\omega_0$ (termed fundamental) -- see
Fig.~\ref{fig:pure-coh-spectrum}a, red and green curves. So, it
seems natural to begin with evaluation of the multiphoton spectrum
in the fundamental interval.

The calculation is straightforward based on representation
(\ref{generic-eq}), written for finite $w_{1\text{c}}$ as
\begin{eqnarray}\label{wcoh-1stinterv}
\frac{dw_{\text{c}}}{d\omega}=e^{-w_{1\text{c}}}\frac1{2\pi
i}\int_{c-i\infty}^{c+i\infty}\!ds e^{s\omega+b\!\int_0^\infty
d\omega_1
P(\omega_1/\omega_0)e^{-s\omega_1}}\quad\\
(0<\omega<\omega_0).\qquad\qquad\qquad\qquad\nonumber
\end{eqnarray}
Here we dropped the $\delta$-term for $\omega>0$, and by virtue of
Lemma~\ref{lemma1}, the upper limit in the integral over $\omega_1$
was replaced by infinity, without the necessity to employ the
Heaviside step function, which therefore was omitted. If $P$ is a
polynomial, like (\ref{f-def}), the integral in the exponent of
(\ref{wcoh-1stinterv}) evaluates as a finite-order polynomial in
$s^{-1}$:
\begin{equation}\label{polynom-s}
\int_0^\infty d\omega_1
P\left(\frac{\omega_1}{\omega_0}\right)e^{-s\omega_1}=\sum_{k=0}^2\frac{P^{(k)}(0)}{s^{1+k}\omega_0^k},
\end{equation}
where $P^{(k)}$ stands for the $k$-th derivative of $P(z)$ w.r.t.
its argument. Inserting (\ref{polynom-s}) to (\ref{wcoh-1stinterv}),
and rescaling the integration variable to $s=\zeta/\omega_0$, yields
\begin{eqnarray}\label{dwc-oint}
\frac{dw_{\text{c}}}{d\omega}=\frac{e^{-w_{1\text{c}}}}{\omega_0}\frac1{2\pi
i}\oint_{\mathcal{C}_3} d\zeta
e^{\zeta\frac{\omega}{\omega_0}+b\omega_0\sum_{k=0}^2\frac{P^{(k)}(0)}{\zeta^{1+k}}}\\
(0<\omega<\omega_0).\qquad\qquad\nonumber
\end{eqnarray}
Benefiting from the exponential decrease of the integrand at
$\mathfrak{Re}\zeta\to-\infty$, along with the absence of cuts in
the complex $\zeta$-plane, we replaced the infinite integration
contour by a closed loop encircling the origin, where the only
singularity of the integrand resides (contour $\mathcal{C}_3$ in
Fig.~\ref{fig:contours}).

\paragraph{Infrared limit.}

Despite the closed integration contour, and representation of the
exponent in the integrand by only a few power terms,
integral~(\ref{dwc-oint}) generally does not permit classification
in terms of basic special functions.\footnote{The integral would
reduce to a modified Bessel function if there was only the 0-th term
in the sum over $k$, i.e., function $P(z)$ was constant. Such a
model may be suitable for quick estimates, but generally is not
acceptable numerically.} But it is instructive at least to obtain
its expansion about endpoint $\omega=+0$. There, one of the
exponentials tends to unity, $e^{\zeta\frac{\omega}{\omega_0}}\to1$,
and if the second exponential is expanded into Laurent series
\begin{equation}\label{exp-to-Laurent}
e^{b\omega_0\sum_{k=0}^2\frac{P^{(k)}(0)}{\zeta^{1+k}}}=\sum_{m=0}^\infty\frac{c_m}{\zeta^{m}},
\end{equation}
the only term surviving after the loop integration is
$\frac{c_1}\zeta=b\omega_0P(0)\frac{1}{\zeta}$, leaving the result
\begin{equation}\label{coh-omegato0}
\frac{dw_{\text{c}}}{d\omega}\bigg|_{\omega=+0}=bP(0)
e^{-w_{1\text{c}}}\equiv
e^{-w_{1\text{c}}}\frac{dw_{1\text{c}}}{d\omega_1}\bigg|_{\omega_1=+0}.
\end{equation}
This equation shows that in the limit $\omega\to0$, the
single-photon probability dominates, albeit in conjunction with
photon non-emission probability $e^{-w_{1\text{c}}}$. For the photon
multiplicity spectrum, Eq.~(\ref{coh-omegato0}), when inserted to
Eq.~(\ref{mean-n-omega-convol}) leads to Eq.~(\ref{mean-n-to1}).

To derive an $\mathcal{O}(\omega)$ correction to
Eq.~(\ref{coh-omegato0}), one needs to retain also a linear term in
the expansion of $e^{\zeta\frac{\omega}{\omega_0}}$:
\begin{equation}\label{coh-omegato0-corr}
\frac{dw_{\text{c}}}{d\omega}=e^{-w_{1\text{c}}}b\left\{\!P(0)\!+\!\frac{\omega}{\omega_0}\!\left[P'(0)\!+\!\frac{b\omega_0}2P^2(0)\right]\!\!+\!\mathcal{O}(\omega^2)\!\right\}\!.
\end{equation}
For our specific example (\ref{f-def}), $P'(0)=-2<0$. But for a
sufficiently high intensity, specifically
\begin{equation}\label{monot-after4}
b\omega_0>-\frac{2P'(0)}{P^2(0)}=4,
\end{equation}
the coefficient at the linear term in (\ref{coh-omegato0-corr})
turns positive, wherewith the resummed spectrum in the whole
fundamental interval becomes monotonously increasing. That may be
associated with the inception of high-intensity regime, which will
be examined in the next section.

From Eqs.~(\ref{coh-omegato0-corr}) and (\ref{mean-n-omega-convol}),
one similarly infers an expression for the photon multiplicity
spectrum
\begin{equation}\label{n0=1+}
\bar
n_{\text{c}}(\omega)=1+\frac{bP(0)}2\omega+\mathcal{O}(\omega^2).
\end{equation}
This shows that the slope in the origin does not depend on $P'(0)$,
and is always positive, in accord with Eq.~(\ref{mean-n-gtr1}). The
increase of the slope in the origin with the increase of parameter
$b$ agrees with Fig.~\ref{fig:pure-coh-spectrum}b.

\paragraph{Fundamental maximum.}\label{par:fund-max}

\begin{figure}
\includegraphics{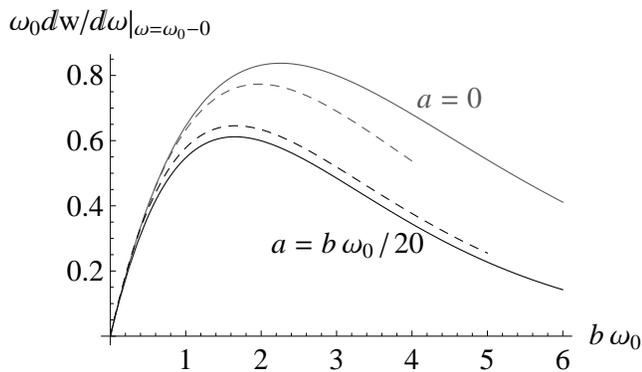}
\caption{\label{fig:intensity-in-max} The spectral intensity at the
fundamental maximum ($\omega=\omega_0-0$) vs. the radiator length
(in units of $b\omega_0$), for a single-photon spectrum described by
Eq.~(\ref{f-def}). Solid gray curve, pure coherent radiation
[Eq.~(\ref{max-int-vs-w1c})]. Dashed gray curve,
approximation~(\ref{40}). Solid black curve, the combination of
coherent and incoherent radiation [Eq.~(\ref{max-dw})] in proportion
$a=b\omega_0/20$, for $E=10\,\omega_0$ ($E_e\approx20\,\omega_0$).
Dashed black curve, approximation~(\ref{corr-1fund-max}).}
\end{figure}

The opposite endpoint of the fundamental interval,
$\omega=\omega_0-0$, is of prime practical significance, as long as
it represents the global spectral maximum at moderate radiation
intensity. Substituting $\omega=\omega_0$ in Eq.~(\ref{dwc-oint})
leaves
\begin{equation}\label{max-int-vs-w1c}
\omega_0\frac{dw_{\text{c}}}{d\omega}\bigg|_{\omega=\omega_0-0}=e^{-w_{1\text{c}}}\frac1{2\pi
i}\oint_{\mathcal{C}_3} d\zeta e^{\zeta+
b\omega_0\sum_{k=0}^2\frac{P^{(k)}(0)}{\zeta^{1+k}}}.
\end{equation}
Effectively, this is a function of a single dimensionless parameter
$b\omega_0 $, inasmuch as all $P^{(k)}(0)\sim1$. In
Fig.~\ref{fig:intensity-in-max}, the dependence of
(\ref{max-int-vs-w1c}) on $b\omega_0$ [with $P(z)$ defined by
Eq.~(\ref{f-def})] is displayed by the solid gray curve. Naturally,
with the increase of $b$, at first it rises proportionally, but
eventually the rise halts and ends up with a decrease, formally
owing to factor $e^{-w_{1\text{c}}}$. The ultimate suppression of
the resumed spectrum at any fixed energy $\omega$ is not surprising,
given the saturation of total probability (\ref{total-probab-1}) on
one hand, and the spread of the spectrum to higher $\omega$ on the
other hand.

To estimate the maximal summit of function (\ref{max-int-vs-w1c}),
one can adopt the following approach. To capture the first bend of
the nonlinear $b$-dependence, expand the contour integral in
(\ref{max-int-vs-w1c}) to second order in $b$ (which is equivalent
to keeping in Eq.~(\ref{generic-sum-int}) the first two terms):
%\begin{subequations}
\begin{eqnarray}
\omega_0\frac{dw_{\text{c}}}{d\omega}\bigg|_{\omega=\omega_0-0}&\simeq& e^{-w_{1\text{c}}}\bigg[b\omega_0\sum_{k=0}^2\frac{P^{(k)}(0)}{k!}\qquad\nonumber\\
&\,&+\frac{b^2\omega_0^2}2\sum_{k,l=0}^2\frac{P^{(k)}(0)P^{(l)}(0)}{(k+l+1)!}\bigg]\nonumber\\
%\qquad\equiv e^{-\omega_0b\int_0^1dzP(z)}\left[bP(1)+\omega_0b^2\frac12\int_0^1dzP(z)P(1-z)\right]\\
&=&P(1)e^{-b\omega_0\xi}\left(b\omega_0+\frac{b^2\omega_0^2}2\eta\right),\,\,\label{40}
\end{eqnarray}
%\end{subequations}
where
\begin{equation}\label{alpha-coh}
\xi=\int_0^1dzP(z),
\end{equation}
\begin{equation}\label{beta-coh}
\eta=\frac1{P(1)}\int_0^1dzP(z)P(1-z).
\end{equation}
The behavior of approximation (\ref{40}) is illustrated in
Fig.~\ref{fig:intensity-in-max} by the dashed gray curve. The
maximum of (\ref{40}) can be explicitly found by differentiation by
$b\omega_0$: it is situated at
\begin{equation}\label{loc-max}
b_\star\omega_0=\frac1{\xi}-\frac1\eta+\sqrt{\frac1{\xi^2}+\frac1{\eta^2}}.
\end{equation}
For $P(z)$ described by Eq.~(\ref{f-def}) (wherewith $\xi=\frac23$,
$\eta=\frac7{15}$), Eq.~(\ref{loc-max}) yields
$b_\star\omega_0\approx2.0$, at which
$\omega_0\frac{dw_{\text{c}}}{d\omega}\big|_{\omega=\omega_0-0}=0.77$.
For comparison, the maximum of exact expression
(\ref{max-int-vs-w1c}) is achieved at $b\omega_0\approx2.26$, and
amounts 0.84 (see Fig.~\ref{fig:intensity-in-max}, solid gray
curve). Hence, at any radiation intensity, the probability spectrum
$\frac{dw}{d\omega}$ of multiphoton coherent radiation at practice
never exceeds unity\footnote{It may exceed unity in principle,  if
profile of single-photon spectrum $P(z)$ is such that $\xi$ is
exceptionally small, while $P(1)$ is sizable. Those are typical
conditions for strong aperture collimation. But that case is rather
trivial, because multiphoton effects are then suppressed as a
whole.} -- in marked contrast with the unlimited growth of the
single-photon spectrum. [At that, the energy spectrum
$\omega\frac{dw}{d\omega}$ can indefinitely grow, $\sim\sqrt{b}$ --
see Sec.~\ref{subsubsec:centr}, Eq.~(\ref{Gauss})].

%The spectrum shape at $\omega_0 P(0)=2.5$ is shown in %Fig.~\ref{fig:high-intensity}.

\subsubsection{Discontinuity at $\omega=\omega_0$}\label{subsubsec:disc-dwc}

Besides peaking at $\omega=\omega_0$, the coherent radiation
spectrum encounters at this point a discontinuity, which is actually
easier to evaluate. For our simplified case (\ref{f-def}), the
discontinuity of the single-photon spectrum (defined to be a
positive quantity) just equals the height of its maximum:
\begin{equation}\label{disc1}
\Delta_{1\text{c}}|_{\omega_0}=\frac{dw_{1\text{c}}}{d\omega_1}\bigg|_{\omega_1=\omega_0-0}-\frac{dw_{1\text{c}}}{d\omega_1}\bigg|_{\omega_1=\omega_0+0}=bP(1).
\end{equation}
With the account of  emission of several photons, the spectrum
beyond $\omega_0$ deviates from zero, so higher terms of series
(\ref{generic-sum-int}) demand consideration. But all terms with
$n\geq2$ prove to be continuous\footnote{That can be supported by
the following argument. Albeit in $n$-th term of
(\ref{generic-sum-int}) the integrand function $\prod_{k=1}^n
P(\omega_k/\omega_0)\theta(\omega_k-\omega_0)$ is discontinuous on
every face of an $n$-cubic domain, but the entering
$\delta$-function restricts the integration domain to a slicing
plane $\omega=\sum_k\omega_k$, which is oblique and nowhere parallel
to any face of the cube for $n\geq2$ (see
Fig.~\ref{fig:integr-domain}b). Thereby, for $n\geq 2$ no
discontinuities in the $\omega$-dependence can arise.}. Hence, the
discontinuity of the total spectrum just equals that of the first
term. This term yet involves the total photon non-emission
probability factor
\begin{equation}\label{disc}
\Delta_\text{c}|_{\omega_0}=\Delta_{1\text{c}}
|_{\omega_0}e^{-w_{1\text{c}}}=bP(1)e^{-b\omega_0\int_0^1 dzP(z)},
\end{equation}
which embodies all the multiphoton effects on the discontinuity.

At multiples of the fundamental energy, $\omega=n\omega_0$ with
integer $n\geq2$, the resummed spectrum remains continuous, but its
$(n-1)$-th derivative encounters a discontinuity (as can be noticed
already from Figs.~\ref{fig:pure-coh-spectrum}). That also follows
from the analysis of the phase space available for integration for
terms of series (\ref{generic-sum-int}a,b). Furthermore, even in
case if function $P(z)$ is vanishing at $z\to1$, in representation
(\ref{generic-sum-int}) the function discontinuity at
$\omega=\omega_0$ cancels, but discontinuities of its derivatives
persist. We leave the proof of those statements to the reader.

As regards the discontinuities of $\bar n_{\text{c}}(\omega)$,
Fig.~\ref{fig:pure-coh-spectrum}c shows that they are similar to
those of $\frac{dw_{\text{c}}}{d\omega}$, but have opposite sign
[because $\bar n_{\text{c}}(\omega)$ contains $dw/d\omega$ in the
denominator]. At low radiation intensity, the multiplicity spectrum
exhibits a step-like behavior, approximately amounting to the
smallest integer greater than $\omega/\omega_0$. That is traced to
the fact that in an interval $n-1<\omega/\omega_0<n$
$(n\in\mathbb{N})$, the $n$-photon component dominates. As the
intensity increases, function $\bar n_{\text{c}}(\omega)$ smoothens
out, given that a $n$-photon component becomes competitive with
fewer-photon components even in intervals where those components are
not yet extinct.

%\newpage

\subsection{Combination of coherent and incoherent radiation}\label{subsec:coh+incoh}

Having explored separately the shapes of resummed coherent and
incoherent radiation spectra, we are now in a position to examine
their nonlinear interplay. The physical example when those
components are both significant is coherent bremsstrahlung,
occurring when an electron or positron crosses a family of
crystalline planes at an above-critical (although small) angle.
Thereat, the continuous interplanar potential acts on the electron
periodically, evoking coherent radiation
\cite{books-on-coh-sources,Ter-Mik,Diambrini-Palazzi}. At the same
time, the incoherent scattering on atomic nuclei in the planes
causes incoherent bremsstrahlung, similar to that in amorphous
matter. In a satisfactory approximation, the single-photon radiation
spectrum may be expressed as a sum of two
non-interfering\footnote{In principle, an interference term between
coherent and incoherent radiation components exists (see, e.g.,
\cite{Ter-Mik,Diambrini-Palazzi}), but it remains minor compared
either to coherent, or to incoherent component almost everywhere.
More rigorously, the interference term can be included to
$\frac{dw_{1\text{c}}}{d\omega_1}$, because it is IR- and UV-safe,
but that would complicate the structure of $P(z)$. On the other
hand, if there is a coherent contribution to the electron r.m.s.
deflection angle, it ought to be included to
$\frac{dw_{1\text{i}}}{d\omega_1}$.} parts:
\begin{equation}\label{coh+inc}
\frac{dw_1}{d\omega_1}=\frac{dw_{1\text{i}}}{d\omega_1}+\frac{dw_{1\text{c}}}{d\omega_1},
\end{equation}
where $\frac{dw_{1\text{i}}}{d\omega_1}$ is given by
Eq.~(\ref{dw1inc}). In graphical illustrations, we will continue using Eqs.~(\ref{dw1coh-real}--\ref{f-def}) for $\frac{dw_{1\text{c}}}{d\omega_1}$, i.e., neglect higher harmonics. In some cases, those can be small indeed even for coherent bremsstrahlung (`one-point' spectra, see \cite{Diambrini-Palazzi}). At the same time, we saw in the previous section that the second maximum in the multiphoton spectrum can be generated even in the absence of secondary harmonics in the single-photon spectrum, so it will be expedient to check whether their effect is anyhow affected by incoherent radiation. %In principle, even if such a
%separation is impossible for the whole range of $\omega_1$, one can
%always conduct it asymptotically.

For what concerns the ratio of magnitudes of coherent and incoherent
radiation components, for coherent bremsstrahlung it obeys a
relation
\begin{equation}\label{a/bomega0}
\frac{a}{b\omega_0}\simeq C \chi \qquad (\text{coh. bremsstr.}),
\end{equation}
with $\chi$ the misalignment angle between the electron momentum and
the co-oriented family of atomic planes, and $C\sim10$ [cf.
Eqs.~(\ref{a-amorph}) and (\ref{dw-CB-Fmax})]. At typical
$\chi\sim10^{-2}\div10^{-4}$, ratio (\ref{a/bomega0}) is pretty
small. Nevertheless, effects of incoherent radiation in the spectrum
may be quite noticeable. That is illustrated by
Fig.~\ref{fig:high-intensity}, where in spite of rather large
initial ratio $b\omega_0/a=20$, the incoherent bremsstrahlung
plateau in the multiphoton energy spectrum
$\omega\frac{dw}{d\omega}$ is only a few times lower than the
spectrum height in the maximum (Fig.~\ref{fig:high-intensity}b).
Handling the effects of incoherent radiation proves to be most
convenient with the aid of a convolution representation derived
below.

%, as we are going to demonstrate.

\subsubsection{Convolution representations}

Once decomposition (\ref{coh+inc}) is inserted to
Eq.~(\ref{generic-eq}), the exponential in the integrand splits into
a product of two factors, one depending on
$\frac{dw_{1\text{i}}}{d\omega}$, and the other on
$\frac{dw_{1\text{c}}}{d\omega}$. Expressing then each of those
factors via Eq.~(\ref{generic-exponentiation}) through Laplace
transforms of the corresponding resummed radiation spectra, and
doing the contour integral, we arrive at a convolution
relation\footnote{This may be viewed as an analog of the
Chapman-Kolmogorov identity \cite{Feller,random-walk}, with the
proviso that instead of separating the contributions by their time
order, we sorted them according to their shape (coherent or
incoherent). Albeit we abstracted from the process development in
time, the Chapman-Kolmogorov equation is an indicator of Markovian,
or random-walk character of the process. The correspondence with
random walks will be detalized further in
Sec.~\ref{subsec:random-walk}.}:
\begin{eqnarray}\label{wcoh-times-winc}
\frac{dw}{d\omega}=\int_0^{\omega}d\omega'\frac{dw_\text{i}}{d\omega'}\frac{dw_\text{c}}{d\omega''}\bigg|_{\omega''=\omega-\omega'}+e^{-w_{1\text{c}}}\frac{dw_\text{i}}{d\omega}\\
+\underset{\hookrightarrow0}{e^{-w_{1\text{i}}}}\frac{dw_\text{c}}{d\omega}.\nonumber
\end{eqnarray}
The term in the second line vanishes if $w_{1\text{i}}$
diverges\footnote{\label{foot:winc-to-inf}At practice, when the
incoherent component receives a physical IR cutoff, factor
$e^{-w_{1\text{i}}}$ may be not really small, so the last term in
(\ref{wcoh-times-winc}) may remain sizable. To be more quantitative,
$w_{1\text{i}}\sim a\ln\frac{E_e}{\epsilon}$ with $\epsilon$ the IR
cutoff value. But for cutoffs of different nature, and practical
energies $E_e$, ratio ${E_e}/{\epsilon}$ can hardly exceed $10^5$,
wherewith $w_{1\text{i}}\lesssim a\ln10^5\sim10\,a$. Hence,
$e^{-w_{1\text{i}}}$ can really be vanishing only provided
$a\gtrsim0.1$. We will restrict our consideration to the idealized
case when this condition is met. The extension to finite
$w_{1\text{i}}$ is straightforward, and basically resembles the
situation described in the previous subsection.}, whereupon
expression (\ref{wcoh-times-winc}) becomes linear in $dw_\text{i}$.
Still, the resulting convolution is non-vanishing at $a\to0$,
because, as we saw in Sec.~\ref{subsec:incoh-bremsstr}, the
multiphoton incoherent radiation spectrum in this limit tends to a
$\delta$-function, not to zero [see Eq.~(\ref{delta-lim-ato0})].
More precisely, if median $\omega_{\text{i}\frac12}$ defined by
Eq.~(\ref{omegaincoh12}) proves to be much smaller than $\omega_0$,
convolution (\ref{wcoh-times-winc}) will be close to
$\frac{dw_{\text{c}}}{d\omega}$.

Observing further that in Eq.~(\ref{wcoh-times-winc})
$\frac{dw_\text{i}}{d\omega'}$ is a smooth function, whereas the
coherent radiation spectrum
$\frac{dw_\text{c}}{d\omega''}\big|_{\omega''=\omega-\omega'}$ is
not, it may be beneficial to integrate in (\ref{wcoh-times-winc}) by
parts:
\begin{eqnarray}\label{}
\frac{dw}{d\omega}=e^{-w_{1\text{c}}}\frac{dw_\text{i}}{d\omega}+\frac{dw_\text{c}}{d\omega''}\bigg|_{\omega''=0}\int_0^{\omega}d\omega'''\frac{dw_\text{i}}{d\omega'''}\qquad\qquad\nonumber\\
-\int_0^{\omega}d\omega'
\left(\int_0^{\omega'}d\omega'''\frac{dw_\text{i}}{d\omega'''}\right)\!\frac{d}{d\omega'}\!\left(\frac{dw_\text{c}}{d\omega''}\bigg|_{\omega''=\omega-\omega'}\right)\!.\quad
\end{eqnarray}
Invoking here Eq.~(\ref{coh-omegato0}) along with identity
$\int_0^{\omega}d\omega'''\frac{dw_\text{i}}{d\omega'''}=\Phi(a)\left(\frac{\omega}{E}\right)^a$,
the convolution relation recasts
\begin{eqnarray}\label{convol-byparts}
\frac{dw}{d\omega}=e^{-w_{1\text{c}}}\Phi(a)\left(\frac{\omega}{E}\right)^a\left[\frac{a}{\omega}+bP(0)\right]\qquad\qquad\quad\nonumber\\
-\Phi(a)\int_0^{\omega}d\omega'
\left(\frac{\omega'}{E}\right)^a\frac{d}{d\omega'}\left(\frac{dw_\text{c}}{d\omega''}\bigg|_{\omega''=\omega-\omega'}\right).
\end{eqnarray}
The merit of the latter representation is that it picks up discrete
contributions from points at which
$\frac{dw_\text{c}}{d\omega''}\big|_{\omega''=\omega-\omega'}$
encounters discontinuities, transforming to $\delta$-function-like
terms for its derivative.

\begin{figure}
\vspace{5mm}
\includegraphics{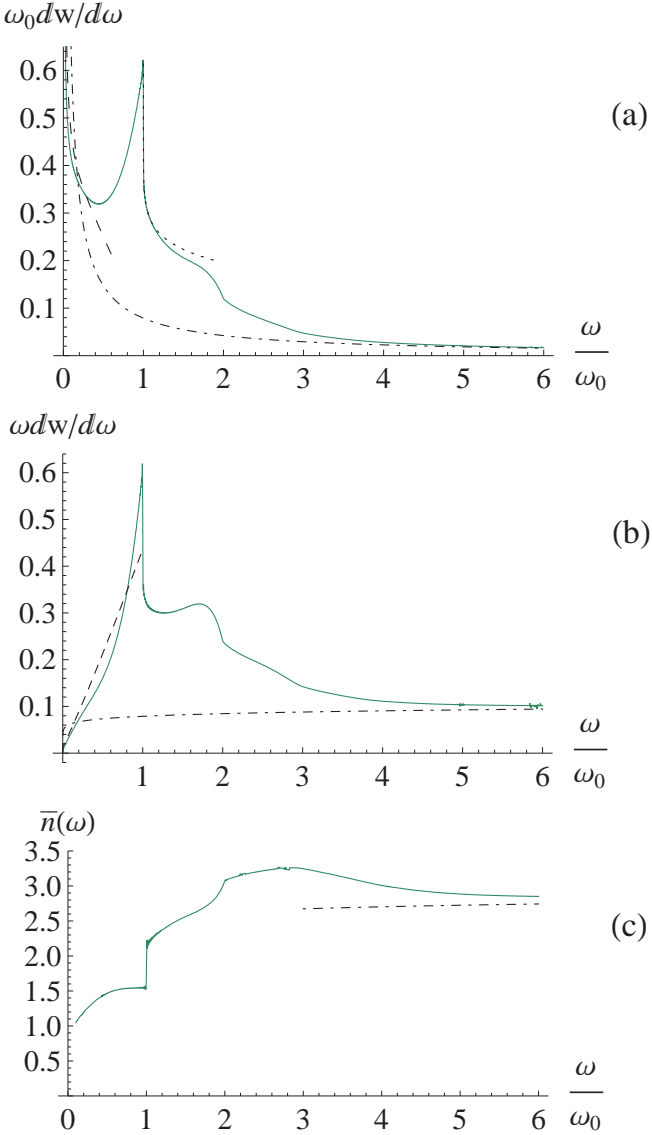}
\caption{\label{fig:high-intensity} (a) Multiphoton radiation
spectrum involving coherent component with magnitude $b\omega_0=2$,
and incoherent component with $a=0.1$. The UV cutoff parameter is
chosen at $E=10\,\omega_0$. Solid green curve, the exact result
obtained by numerical evaluation of the contour integral.
Long-dashed black curve, corrected low-$\omega$ approximation
(\ref{dw-small-omega-NLO}). Dotted black curve, approximation next
to the coherent emission edge, Eq.~(\ref{smearing-of-disc}).
Dot-dashed black curve, large-$\omega$ asymptotics
(\ref{simple-large-omega}). (b) The energy spectrum for same
conditions as in (a). Short-dashed black curve, low-$\omega$
approximation (\ref{dw-small-omega}). Dot-dashed black curve,
large-$\omega$ asymptotics (\ref{simple-large-omega}). (c) The
photon multiplicity spectrum for same parameters as in (a,b), and IR
cutoff $\epsilon=0.1\,\omega_0$. Solid green curve, exact result.
Dot-dashed black curve, large-$\omega$ approximation
(\ref{n-omega-asympt-large}). } \vspace{5mm}
\end{figure}

In the limit $\omega\to0$, the integral term in
Eq.~(\ref{convol-byparts}), vanishes due to the integration interval
shrinkage, leaving
\begin{equation}\label{dw-small-omega}
\frac{dw}{d\omega}=e^{-w_{1\text{c}}}\Phi(a)\left(\frac{\omega}{E}\right)^a\left[\frac{a}{\omega}+bP(0)+\mathcal{O}(\omega)\right].
\end{equation}
Clearly, here the leading term $\frac{a}\omega$ represents the
contribution due to incoherent bremsstrahlung, but it receives an
extra suppression by factor $e^{-w_{1\text{c}}}$ depending on the
coherent radiation. Hence, the incoherent component can not be
asserted to dominate in this limit alone, in contrast to the case of
large $\omega$ (see Sec.~\ref{subsubsec:large-omega-asympt} below).
The behavior of asymptotic approximation (\ref{dw-small-omega}) is
illustrated in Fig.~\ref{fig:high-intensity}b by the short-dashed
black curve.

In a similar fashion, one can derive a convolution relation for the
multiplicity spectrum, which reads
\begin{eqnarray}\label{convol-for-n-omega}
\bar n(\omega)\frac{dw}{d\omega}=\int_0^{\omega}d\omega'\frac{dw_\text{c}}{d\omega'}\frac{dw_\text{i}}{d\omega''}\bigg|_{\omega''=\omega-\omega'}\qquad\qquad\nonumber\\
\times\left[\bar n_\text{c}(\omega')+\bar n_\text{i}(\omega-\omega')\right]\\
+e^{-w_{1\text{c}}}\bar
n_\text{i}(\omega)\frac{dw_\text{i}}{d\omega}
+\underset{\hookrightarrow0}{e^{-w_{1\text{i}}}}\bar
n_\text{c}(\omega)\frac{dw_\text{c}}{d\omega},\nonumber
\end{eqnarray}
with $\bar n_\text{c}(\omega)$ and $\bar n_\text{i}(\omega)$ the
corresponding multiplicity spectra for pure coherent and incoherent
components. In particular, from Eqs.~(\ref{convol-for-n-omega}) and
(\ref{wcoh-times-winc}) one observes that both $\frac{dw}{d\omega}$,
and $\frac{dw_\text{i}}{d\omega}$ involve $E$-dependent factors
$E^{-a}$, which cancel between l.h.s. and r.h.s. of
Eq.~(\ref{convol-for-n-omega}). Thus, $\bar n(\omega)$ is
independent of $E$, as anticipated.

\subsubsection{Spectrum of combined radiation in the fundamental interval}\label{subsubsec:comb-in-fund-int}

For an efficient coherent radiator, the coherent spectral component
must prevail in the fundamental interval $0<\omega\leq\omega_0$,
anyway. It is thus instructive to repeat the analysis of
Sec.~\ref{subsubsec:coh-in-fund-interv} for a mixture of coherent
and incoherent radiation components in the fundamental interval. We
will barely sketch it here, emphasizing the distinctions from the
pure coherent radiation case.

Once expressions (\ref{coh+inc}), (\ref{dw1inc}),
(\ref{dw1coh-real}) are inserted to
Eq.~(\ref{generic-eq-upperlimitomega}), manipulations similar to
those used at derivation of Eq.~(\ref{dwc-oint}) yield
\begin{eqnarray}\label{dw-first-interv}
\frac{dw}{d\omega}\!=\!\frac{e^{-\gamma_{\text{E}}a-w_{1\text{c}}}}{\omega_0}\!\left(\frac{\omega_0}{E}\right)^{\!a}\!\frac1{2\pi
i}\!\int_{\mathcal{C}_2}\! \frac{d\zeta}{\zeta^a}
e^{\zeta\!\frac{\omega}{\omega_0}+b\omega_0\!\sum_{k=0}^2\!\!\frac{P^{(k)}\!(0)}{\zeta^{1+k}}}\quad\\
(\omega<\omega_0).\qquad\qquad\qquad\qquad\nonumber
\end{eqnarray}
Here exponential decrease of the integrand at
$\mathfrak{Re}\zeta\to-\infty$ justifies deformation of the
integration contour to shape $\mathcal{C}_2$ (see
Fig.~\ref{fig:contours}), but its complete enclosure is impossible
because of the presence of branching factor $\zeta^{-a}$.
%At $P\equiv 0$
%representation (\ref{dw-first-interv}) turns to Eq.~(\ref{13}),
%while at $a=0$, to Eq.~(\ref{dwc-oint}).

%Let us again investigate behavior of the multiphoton spectrum at the endpoints of the fundamental interval. Investigation of limit $\omega\to0$ promises to be even more important than for the case of pure coherent radiation, since it receives an enhancement due to incoherent component.

\paragraph{Infrared limit.}

When $\omega\to+0$, contour integral (\ref{dw-first-interv}) is
dominated by large $\zeta$. Then, the leading-order term in
asymptotic expansion (\ref{exp-to-Laurent}) reproduces result
(\ref{dw-small-omega}). Retaining next-to-leading order terms yields
an $\mathcal{O}(\omega)$ correction:
\begin{eqnarray}\label{dw-small-omega-NLO}
\frac{dw}{d\omega}=e^{-w_{1\text{c}}}\Phi(a)\left(\frac{\omega}{E}\right)^a\Bigg\{\frac{a}{\omega}+bP(0)\qquad\qquad\quad\nonumber\\
+\frac{b\omega}{(1+a)\omega_0}\left[P'(0)+\frac{b\omega_0}2P^2(0)\right]+\mathcal{O}(\omega^2)\Bigg\}.
\end{eqnarray}
Compared to (\ref{coh-omegato0-corr}), apart from the overall factor $\left(\frac{\omega}{E}\right)^a$, there emerges a term $\frac{a}{\omega}$, whereas other terms remain essentially the same. The term in the second line of Eq.~(\ref{dw-small-omega-NLO}) changes its sign under the same condition (\ref{monot-after4}), so inception of high-intensity regime is essentially independent of $a$. The behavior of asymptotic approximation (\ref{dw-small-omega-NLO}) is illustrated in Fig.~\ref{fig:high-intensity}a by the gray long-dashed curve. %Approximation (\ref{dw-small-omega}) has a minimum at
%\begin{equation}\label{min-dw-first-interv}
%\omega=\frac{1-a}{bP(0)},
%\end{equation}
%while (\ref{dw-small-omega-NLO}) features none, unless condition
%(\ref{monot-after4}) is satisfied. Although exact probability
%distribution always has a minimum on the fundamental interval,  it
%can only be described by Eq.~(\ref{min-dw-first-interv}) under
%condition (\ref{monot-after4}).

%For $\bar n(\omega)$ the small-$\omega$ asymptotics can be inferred from ..., but rather obvious:
%\begin{equation}\label{n-omega-aggr-small-omega}
%\bar n(\omega)=1+a\ln\frac{\omega}{\epsilon}+\frac{bP(0)}2
%\omega+\mathcal{O}(...),
%\end{equation}
%where the correction is $\int^{\omega}_{\epsilon}d\omega_1\frac{dw_1}{d\omega_1}$ (?).

\paragraph{Reduction of the fundamental maximum by hard incoherent bremsstrahlung.}

In the point of the spectral fundamental maximum,
$\omega\to\omega_0-0$, integral (\ref{dw-first-interv}) reduces to
\begin{eqnarray}\label{max-dw}
\frac{dw}{d\omega}\bigg|_{\omega=\omega_0-0}&=&\frac{e^{-\gamma_{\text{E}}a-w_{1\text{c}}}}{\omega_0}\left(\frac{\omega_0}{E}\right)^a\nonumber\\
&\,&\times\frac1{2\pi i}\int_{\mathcal{C}_2} \frac{d\zeta}{\zeta^a}
e^{\zeta+b\omega_0\sum_{k=0}^2\frac{P^{(k)}(0)}{\zeta^{1+k}}}.\qquad
\end{eqnarray}
Since $a$ and $b$ are both proportional to the target thickness, the
increase of the latter at fixed $\chi$ [cf. Eq.~(\ref{a/bomega0})]
corresponds to a simultaneous increase of $b$ and $a$, at ratio
$a/b$ held fixed. For an exemplary ratio $a/b=\omega_0/20$, the
spectral intensity in the fundamental maximum is illustrated in
Fig.~\ref{fig:intensity-in-max} by solid gray curve. Compared to the
pure coherent radiation spectrum, its maximum is achieved at a
slightly lower value of $b$, and its height is somewhat lower, too.
This owes primarily to factor $\left(\frac{\omega_0}{E}\right)^a$,
where $\frac{\omega_0}{E}$ is a small number, but there can also be
other appreciable effects of $a$.

To assess changes in the height and location of the fundamental
maximum due to incoherent radiation, one can, as in
Sec.~\ref{subsubsec:coh-in-fund-interv}, expand the contour integral
in (\ref{max-dw}) to second order in $b$ and $a$:
\begin{eqnarray*}\label{}
\frac{dw}{d\omega}\bigg|_{\omega=\omega_0-0}\approx
e^{-w_{1\text{c}}}\left(1-\gamma_{\text{E}}a-a\ln\frac{E}{\omega_0}\right)\Bigg[\frac1{\omega_0\Gamma(a)}\quad\qquad\nonumber\\
+b\sum_{k=0}^2\frac{P^{(k)}(0)}{\Gamma(k+1+a)}+\frac{b^2\omega_0}2\sum_{k,l=0}^2\frac{P^{(k)}(0)P^{(l)}(0)}{\Gamma(k+l+2+a)}
\Bigg].
\end{eqnarray*}
It may suffice to account for corrections in $a$ within the leading
logarithmic accuracy, which gives
\begin{eqnarray}\label{corr-1fund-max}
\frac{dw}{d\omega}\bigg|_{\omega=\omega_0-0}\approx e^{-w_{1\text{c}}} \Bigg\{\!\!\left(1-a\ln\frac{E_e}{\omega_0}\right)\!\left[\frac{a}{\omega_0}+bP(1)\right]\,\,\,\nonumber\\
+\frac{b^2\omega_0}2\int_0^1dzP(z)P(1-z) \Bigg\}\quad
\end{eqnarray}
(see Fig.~\ref{fig:intensity-in-max}, dashed black curve). Here term
$-a\ln\frac{E_e}{\omega_0}$ can be sizable due to the large
logarithm, despite the condition $a\ll1$. Owing to its negative
sign, that correction reduces the height of the fundamental coherent
radiation maximum, in addition to factor $e^{-w_{1\text{c}}}$.

In the spirit of Sec.~\ref{subsubsec:coh-in-fund-interv}, for
approximation (\ref{corr-1fund-max}) one can determine the target
thickness at which the spectral radiation intensity in the
fundamental maximum reaches its highest summit. It is still given by
formula (\ref{loc-max}), with parameter $\xi$ unaltered, but
parameter $\eta$ modifies to
\begin{equation}\label{eta-coh+incoh}
\eta=\frac{\int_0^1 dz
P(z)P(1-z)-\frac{2a}{b\omega_0}\ln\frac{E_e}{\omega_0}}{P(1)+\frac{a}{b\omega_0}}.
\end{equation}
Corrections in the numerator and denominator of
(\ref{eta-coh+incoh}) both result in lowering of $\eta$, and
therethrough in an earlier turnover of the fundamental maximum, as
confirmed by Fig.~\ref{fig:intensity-in-max}. The largest effect, of
course, stems from the logarithmic term in (\ref{eta-coh+incoh}).

%\newpage

\subsubsection{Regulation of spectrum discontinuities by soft incoherent bremsstrahlung}\label{subsubsec:regul-of-disc}

A principally interesting effect of incoherent radiation concerns
discontinuities of the coherent radiation spectrum. As we saw in
Sec.~\ref{subsubsec:disc-dwc}, the discontinuity at
$\omega=\omega_0$ is damped by factor $e^{-w_{\text{c}}}$. With the
addition of the incoherent bremsstrahlung component, this factor
actually turns to zero, wherefore the discontinuity must be
nullified\footnote{See, however, Footnote \ref{foot:winc-to-inf}.}.
Our analysis of the spectrum behavior then needs extension to a
non-vanishing vicinity of point $\omega=\omega_0$.

In vicinity of a singularity of the coherent part of the spectrum,
it is convenient to use representation (\ref{convol-byparts}). When
$\omega$ slightly exceeds $\omega_0$, the integrand of
(\ref{convol-byparts}) receives an extra contribution proportional
to a $\delta$-function
\[
\frac{d}{d\omega'}\left(\frac{dw_\text{c}}{d\omega''}\bigg|_{\omega''=\omega-\omega'}\right)\underset{\omega'\approx\omega-\omega_0}\simeq
-\Delta_{\text{c}}|_{\omega_0}\delta(\omega-\omega_0-\omega'),
\]
where $\Delta_{\text{c}}|_{\omega_0}$ is given by Eq.~(\ref{disc}).
Integration of the $\delta$-singularity yields a sizable
contribution even at an infinitesimal extension of $\omega$ beyond
$\omega_0$:
\begin{equation}\label{smearing-of-disc}
\frac{dw}{d\omega}\simeq
\frac{dw}{d\omega'}\bigg|_{\omega'=\omega_0}-\Delta_\text{c}|_{\omega_0}\!\left(\frac{\omega-\omega_0}{E}\right)^{\!a}\theta(\omega-\omega_0)+\mathcal{O}(\omega-\omega_0).
\end{equation}
Eq.~(\ref{smearing-of-disc}) shows that for $a>0$, factor
$\left(\frac{\omega-\omega_0}{E}\right)^a$ vanishes as
$\omega\to\omega_0+0$, indeed nullifying the discontinuity brought
by factor $\theta(\omega-\omega_0)$, and making the whole spectrum
everywhere continuous. But the derivative of factor
$\left(\frac{\omega-\omega_0}{E}\right)^a$ in point
$\omega=\omega_0$ diverges, wherewith the resulting spectrum
features a sharp spike. The behavior of approximation
(\ref{smearing-of-disc}) is shown in Fig.~\ref{fig:high-intensity}a
by the dotted black curve.

It should be realized, though, that the described subtle effect
should be obscured in experiments with limited energy resolution and
finite angular divergence of the initial beam, which smear the
fundamental peak. Besides that, the filling of the dip adjacent to
the coherent radiation maximum is also partially provided by the
pure coherent multiphoton spectrum, whose derivative beyond
$\omega_0$ is negative (see Fig.~\ref{fig:pure-coh-spectrum}a).
Thus, verification of threshold behavior (\ref{smearing-of-disc})
may be feasible only with a rather perfect beam, and in a
sufficiently thick target, when $a$ becomes sizable.

%At practice, the latter effect may actually be even more important.
%In fact, filling in of the first coherent spectrum minimum was observed in experiments [ ]. At practice, of course, this effect is obscured by the finite beam divergence and the multiple scattering in the target.

%Discontinuities due to higher harmonics (\ref{dw1coh-higher-harm}) can be treated similarly.

%What if $\bar n(\omega)$ is IR regularized?

\subsubsection{Large-$\omega$ asymptotics}\label{subsubsec:large-omega-asympt}

To accomplish our analysis in different regions, let us consider the
large-$\omega$ asymptotics $\overline{\omega_1}_{\text{c}},
\omega_0\ll\omega$ (while still obeying condition $\omega\ll E$). In
this limit, the incoherent bremsstrahlung spectrum decreases by a
power law, whereas the multiphoton coherent radiation spectrum, in
general, decreases exponentially (see Appendix~\ref{app:periph}).
Therefore, if Eq.~(\ref{wcoh-times-winc}) is rewritten as
\begin{equation*}\label{wcoh-times-winc-rewritten}
\frac{dw}{d\omega}=\int_0^{\omega}d\omega'\frac{dw_\text{c}}{d\omega'}\frac{dw_\text{i}}{d\omega''}\bigg|_{\omega''=\omega-\omega'}+e^{-w_{1\text{c}}}\frac{dw_\text{i}}{d\omega},
\end{equation*}
the main contribution to the integral over $\omega'$ comes from
vicinity of the lower integration limit. Thus, we can expand
$\frac{dw_{\text{i}}}{d\omega''}$ to Taylor series about point
$\omega''=\omega-\overline{\omega_{1\!}}_{\text{c}}$, and replace
the upper integration limit by infinity:
\begin{eqnarray*}\label{}
\frac{dw}{d\omega}&=& \frac{dw_\text{i}}{d\omega''}\bigg|_{\omega''=\omega-\overline{\omega_{1\!}}_{\text{c}}}\int_0^{\infty}d\omega'\frac{dw_\text{c}}{d\omega'}%\qquad\qquad\qquad\qquad
\nonumber\\
&\,&+\frac{d}{d\omega''}\frac{dw_\text{i}}{d\omega''}\bigg|_{\omega''=\omega-\overline{\omega_{1\!}}_{\text{c}}}\int_0^{\infty}d\omega'\frac{dw_\text{c}}{d\omega'}\left(\overline{\omega_{1\!}}_{\text{c}}-\omega'\right)+\ldots\nonumber\\
&\,& \qquad\qquad\qquad\qquad\qquad
+e^{-w_{1\text{c}}}\frac{dw_\text{i}}{d\omega}.%\qquad\qquad
\end{eqnarray*}
The integral in the second line vanishes due to
Eq.~(\ref{mean-omega}), leaving
\begin{equation}\label{large-omega-corr}
\frac{dw}{d\omega}=
\frac{a\Phi(a)}{E^a}\Bigg[\frac{1-e^{-w_{1\text{c}}}}{(\omega-\overline{\omega_{1\!}}_{\text{c}})^{1-a}}+\frac{e^{-w_{1\text{c}}}}{\omega^{1-a}}%\qquad \nonumber\\
+\mathcal{O}\!\left(\!\frac{\overline{\omega^2_{1\!}}_{\text{c}}}{(\omega-\overline{\omega_{1\!}}_{\text{c}})^{3-a}}\!\right)\!\!\Bigg].
\end{equation}
Furthermore, in the limit $\omega\gg
\overline{\omega_{1\!}}_{\text{c}}$, Eq.~(\ref{large-omega-corr})
boils down to
\begin{equation}\label{simple-large-omega}
\frac{dw}{d\omega}\simeq
\frac{dw_\text{i}}{d\omega}\left[1+\mathcal{O}\left(\frac{\overline{\omega_{1\!}}_{\text{c}}}{\omega}\right)\right],
\end{equation}
which is shown in Figs.~\ref{fig:high-intensity}a,b by dot-dashed
black curves. Eq. (\ref{simple-large-omega}) implies that the
combined spectrum ultimately decreases by the same power law as its
(resummed) incoherent bremsstrahlung component. That property is in
contrast with the infrared limit (\ref{dw-small-omega}), which yet
involves factor $e^{-w_{1\text{c}}}$ of coherent photon non-emission
probability [note the difference between green and black long-dashed
curves in Fig.~\ref{fig:high-intensity}a at small $\omega$].
%Dependence (\ref{large-omega-corr}) is illustrated in Fig.~\ref{fig:high-intensity}a by
%blue dot-dashed curve, while (\ref{simple-large-omega}) -- . The agreement further improves with
%the increase of $b$.

Finally, it is straightforward to derive large-$\omega$ asymptotics
for the photon multiplicity spectrum, with the aid of
Eq.~(\ref{convol-for-n-omega}). Due to factor
$\frac{dw_{\text{c}}}{d\omega'}$ in the integrand of
(\ref{convol-for-n-omega}), again, the dominant contribution comes
from the lower integration limit, where it is acceptable to replace
$\frac{dw_\text{i}}{d\omega''}\big|_{\omega''=\omega-\omega'}=
\frac{dw_\text{i}}{d\omega}\left[1+\mathcal{O}\left(\frac{\omega'}{\omega}\right)\right]$,
$\bar n_\text{i}(\omega-\omega')=\bar
n_\text{i}(\omega)\left[1+\mathcal{O}\left(\frac{a\omega'}{\omega}\right)\right]$,
and replace the upper integration limit by infinity. Therewith,
Eq.~(\ref{convol-for-n-omega}) yields
\begin{eqnarray}\label{83}
\bar
n(\omega)\frac{dw}{d\omega}\approx\frac{dw_\text{i}}{d\omega}\left[\bar
n_\text{i}(\omega)+\int_0^{\infty}d\omega'
\frac{dw_\text{c}}{d\omega'}\bar n_\text{c}(\omega')\right],
\end{eqnarray}
where $\bar n_{\text{i}}(\omega)$ is given by
Eq.~(\ref{mean-n-omega-incoh}). (Terms involving
$e^{-w_{1\text{c}}}$ canceled mutually.) The integral entering
(\ref{83}) equals $w_{1\text{c}}=\bar n_{\text{c}}$, as can be
checked with the aid of Eq.~(\ref{mean-n-omega-contour-non-casual}).
Canceling in (\ref{83}) the overall factors according to
Eq.~(\ref{simple-large-omega}), we are left with
\begin{equation}\label{n-omega-asympt-large}
\bar n(\omega)=\left[\bar n_{\text{c}}+\bar
n_{\text{i}}(\omega)\right]\left[1+\mathcal{O}\left(\frac{\overline{\omega_{1\!}}_{\text{c}}}{\omega}\right)\right].
\end{equation}

The behavior of the photon multiplicity spectrum is shown in
Fig.~\ref{fig:high-intensity}c by the solid green curve. At large
$\omega$, it flattens out, and ultimately enters the regime of
logarithmic growth as predicted by Eq.~(\ref{n-omega-asympt-large})
(dot-dashed black curve). Yet before entering the asymptotic regime,
as Fig.~\ref{fig:high-intensity}b shows, the multiplicity spectrum
first achieves a maximum, then passes through a shallow minimum. The
theory for such a behavior will be provided in
Sec.~\ref{subsec:conv-inc}.

\subsection{Identification of multiphoton effects against other effects in measured spectra}\label{subsec:confus-effects}

The preceding analysis revealed that multiphoton emission effects
redistribute the coherent radiation spectrum, elevating it beyond
the fundamental energy $\omega_0$, at the expense of reduction below
$\omega_0$. At practice, though, one should yet be aware of
existence of other physical processes which can produce
superficially similar effects. In conclusion of this section, we
will briefly discuss such effects, too.

First of all, next to the coherent emission edge $\omega_0$, the
spectrum may receive an enhancement not only from multiphoton
effects, but also from higher harmonics (\ref{dw1coh-higher-harm}).
At that, higher harmonics produce also 3rd, 4th maxima, etc.,
whereas multiphoton effects can only give rise to a relatively
narrow second maximum between $\omega_0$ and $2\omega_0$. So, if no
maxima are observed beyond $2\omega_0$, the simplest way to judge
about the origin of the secondary peak would be to refer to the
photon multiplicity spectrum: the greater its discontinuity at
$\omega\approx\omega_0$, the greater the significance of multiphoton
effects. If measurement of $\bar n(\omega)$ is unfeasible, an
indirect criterion may be used: the relative contribution of higher
harmonics does not vary with the extent of the radiator, whereas
that of multiphoton effects does. Therefore, comparing the spectrum
shapes at two different radiator lengths\footnote{Instead of
actually increasing the radiator length, it may suffice to
self-convolve the spectrum from the same radiator two or more times
(the procedure adopted in \cite{relevance-of-multiphot,Bavizhev}).},
and assessing the length-dependence nonlinearity (cf.
Sec.~\ref{subsec:Ln}), one can judge about the origin of the
spectrum enhancement beyond $\omega_0$.

%the coherent radiation peak smearing, which can
%stem either from multiphoton effects, or from the angular divergence
%of the initial electron beam (and its additional increase in the
%target due to multiple scattering). The direct discrimination
%criterion is that multiphoton effects will produce a step in $\bar
%n(\omega)$ at $\omega=\omega_0$, whereas the beam divergence will
%not. Besides that, the beam divergence would also smear out the cusp
%itself (if the experiment statistics allows this effect to be
%discerned).
%Thirdly, the competing effects have different dependence on the
%crystal orientation: according to Eq.~(\ref{a/bomega0}), the
%misalignment angle determines the strength of the incoherent
%radiation relative to coherent one, and therethrough, the strength of
%discontinuity smearing due to multiphoton effects. In contrast, the
%relative smearing $\Delta\omega_0/\omega_0\approx\Delta\chi/\chi$
%[cf.~Eq.~(\ref{dw-CB-Fmax})] due to the beam divergence (the
%line-width effect) is independent of the crystal orientation.

Another effect is  the difference between asymptotics of the
resummed energy spectrum $\omega\frac{dw}{d\omega}$ at $\omega\to0$
and at $\omega\to\infty$ [see Fig.~\ref{fig:high-intensity}b, and
Eqs.~(\ref{simple-large-omega}) and (\ref{dw-small-omega})]. This
may either owe to multiphoton effects, or to incomplete
LPM-suppression of radiation in a finite-thickness target [existing
as well in an amorphous matter (TSF-effect), see \cite{TSF} and
refs. therein]. Again, unambiguous discrimination between those
effects may rely on the photon multiplicity spectrum behavior: if
within the suppression region $\bar n(\omega)$ rises significantly
above unity (cf., e.g., Fig.~\ref{fig:high-intensity}c), the
suppression origin ought to be attributed to multiphoton effects,
otherwise, it is more likely to be due to LPM-like effects.
Discrimination criteria based on the target thickness dependence
appear to be nonlinear both for multiphoton and incomplete LPM
suppressions, and therefore are to be used cautiously.

%In any event, at interpretation of measured radiation spectra it it
%necessary to account for effects of the beam spread, along with
%studies of target thickness and orientation dependencies. These
%aspects are left beyond the scope of the present paper. In the next
%section we will specialize to conditions of high radiation
%intensity, when the significance of multiphoton effects is
%doubtless.

%\newpage

\section{High photon multiplicity limit}\label{sec:high-intensity}

The measure of significance of multiphoton effects in a resummed
radiation spectrum is given by the mean photon multiplicity, defined
by Eqs.~(\ref{mean-n-def}), (\ref{w1-def}). For the dominant
coherent radiation component (\ref{dw1coh-real}--\ref{f-def}), it
estimates as
\begin{equation}\label{mult-param}
w_{1\text{c}}\sim b\omega_0.
\end{equation}
From the practical viewpoint, it is important further to estimate
how large this parameter can be for 3 basic coherent gamma-radiation
source types: coherent bremsstrahlung, channeling radiation, and
undulator radiation. Generic expressions for product $b\omega_0$ for
mentioned cases are quoted in Appendix \ref{app:coh-rad-sources}, so
it is now left to assess the entering parameters for conditions of
former and future experiments.

For coherent bremsstrahlung, parameter $b\omega_0$ is described by
Eq.~(\ref{dw-CB-Fmax}). Early coherent bremsstrahlung spectrum
measurements operated with relatively thin crystals
($L\sim10^{-1}\div1$ mm) and relatively large misalignment angles
$\chi\sim 10^{-2}\div10^{-3}$ rad, at which $w_{1\text{c}}$ was
comfortably small. But with the advent of more practical radiation
sources having $L\sim1$ cm and $\chi\sim10^{-4}$ rad
\cite{Medenwaldt-Gauss,w1c>1}, formidable values $w_{1\text{c}}>10$
were reached (corresponding to an extensive electromagnetic shower).

For channeling radiation, the reference equation is
(\ref{dw-chann-fin}). It tells that at multi-GeV positron energies,
and $L\gtrsim L_{\text{d}}/10$, the photon multiplicity must achieve
values $w_{1\text{c}}\gtrsim1$. Multiphoton effects in channeling
radiation experiments were found appreciable already when dealing
with moderately high energies $E_e\sim10$ GeV, and
moderate-thickness targets, $L\sim0.1\,\text{mm}\sim
L_{\text{d}}/50$, corresponding to $w_{1\text{c}}\gtrsim0.1$
\cite{relevance-of-multiphot,relevance-of-multiphot2}. In more
recent CERN channeling experiments \cite{multiphot-CERN-Kirsebom},
with $E_e\sim10^2$ GeV and $L\sim1\,\text{mm}$, photon
multiplicities reached the order of 5, and the measured spectra were
apparently loosing the features characteristic of coherent radiation
(see also \cite{Bavizhev}). It should be realized, though, that at
$E_e>100$ GeV, channeling radiation becomes non-dipole, wherewith
simple form (\ref{dw1coh-real}--\ref{f-def}) for the single-photon
spectrum does not apply [although contour integral representations
(\ref{generic-eq}), (\ref{generic-eq-upperlimitomega}) hold
generally].

For undulator radiation, we must refer to Eq.~(\ref{dw-undul}). For
the newest and forthcoming undulators characterized by
$N\gtrsim10^2$ and $K\sim0.5$ \cite{Moortgat-Pick}, parameter
$b\omega_0$ must be $\gtrsim1$. The only question is whether
$\omega_0$ belongs to the gamma-range, for which the calorimetric
method of spectrum measurement is pertinent. In the SLAC experiment
E-166 \cite{E-166}, $\omega_0$ reached $7.5$ MeV, though only under
moderate photon multiplicity $w_{1\text{c}}=0.35$. For the TESLA
design ($E_e=250$ GeV, $L=135$ m, $N=10^4$, $K=1$), $\omega_0$ must
rise to 25 MeV, and correspond to $w_{1\text{c}}\sim10^2$
\cite{Moortgat-Pick}. Such photon energies are admittedly well into
the gamma domain.
%Note that for optical undulator radiation, the multiphoton regime is a
%common situation, permitting Hunbury-Brown--Twiss interferometry,
%which also calls for extension to X-ray and gamma-domain
%\cite{HBT-gamma}.

The present estimates show that virtually for all practical coherent
radiation sources it appears both feasible and beneficial to enter
the regime $w_{1\text{c}}\gtrsim1$. In some cases, it may be optimal
to confine to values $w_{1\text{c}}\sim1\div2$, when the spectrum is
not strongly multiphoton yet (see Fig.~\ref{fig:intensity-in-max}).
In other cases, such as positron sources
\cite{Moortgat-Pick,Artru-Chehab}, $w_{1\text{c}}$ is demanded to be
large indeed. In either case, exploring the asymptotic limit of high
$w_{1\text{c}}$ is of fundamental interest. Its study will occupy us
in the remainder of this section.

\subsection{Random walk interpretation for the multiphoton spectrum. Anomalous diffusion}\label{subsec:random-walk}

To get feel of the trends for the resummed radiation spectrum
behavior at high photon multiplicity, let us first take a look at
spectral moments introduced in Sec.~\ref{subsec:moments}. Relations
(\ref{mean-omega}), (\ref{mean-omega2-alt}) suggest that with the
increase of the radiator length, multiphoton spectrum moments grow
proportionally:
\[
\overline{\omega}=\overline{\omega_{1\!}} \propto b\omega_0^2 ,
\]
\[
\overline{(\omega-\overline{\omega})^2 }=\overline{\omega_1^2}
\propto b\omega_0^3
\]
(and normally $b\propto L$). Therewith, the ratio of the width to
mean value decreases:
\begin{equation}\label{g2}
\frac{\sqrt{\overline{(\omega-\overline{\omega})^2 }}
}{\overline{\omega}}\propto \frac1{\sqrt{b\omega_0
 }}.
\end{equation}
That decrease implies that the multiphoton spectrum becomes more
sharply peaked. Furthermore, from Eq.~(\ref{mean-omega4}) one infers
that the so-called skewness \cite{random-walk}
\begin{equation}\label{skewness}
\gamma_3=\frac{\overline{(\omega-\overline{\omega})^3}}{\overline{
(\omega-\overline{\omega})^2 }^{\,3/2}}=\frac{\overline{
\omega_1^3}}{\overline{ \omega_1^2}^{\,3/2}}\propto
\frac1{\sqrt{b\omega_0 }}
\end{equation}
decreases as well, while the kurtosis
\begin{equation}\label{kurtosis}
\gamma_4=\frac{\overline{(\omega-\overline{\omega})^4}}{\overline{
(\omega-\overline{\omega})^2
}^{\,2}}=3+\frac{\overline{\omega_1^4}}{\overline{\omega_1^2}^{\,2}}=3+\mathcal{O}\left[\left(b\omega_0
\right)^{-1}\right]
\end{equation}
tends to a constant value 3 (characteristic of a
Gaussian).\footnote{One can notice that according to
Eq.~(\ref{kurtosis}), ratio $\gamma_4$ appears to be always greater
than 3, and diverges at small $b$, even though the underlying single
photon spectrum may well be leptokurtic [like that defined by
Eqs.~(\ref{dw1coh-real}--\ref{f-def})]. That owes to our definition
of the moments including term $W_0\delta(\omega)$ in the weighting
distribution, which makes any distribution platikurtic. In the
high-intensity limit, where $W_0\to 0$ exponentially, this
definition suits us, anyway.
%This owes merely to the fact that the weighting is carried
%out with an unnormalized probability distribution. The normalized
%kurtosis equals $\left(1-e^{-w_1}\right)\gamma_4$, and for the
%leptokurtic primary spectrum (\ref{dw1coh-real}--\ref{f-def}), in
%the low-intensity limit it equals $9/5<3$. But in the high intensity
%limit, the normalization tends to unity exponentially, whereas
%${\left\langle \omega_1^4\right\rangle}{\left\langle\omega_1^2\right\rangle^{-2}}$
%decreases only by a power law, so, ultimately, the normalized
%kurtosis exceeds 3, subsequently
%approaching 3 from above, anyway.
}

The above observations look natural from the point that the process
of statistically independent photon emission is a kind of a random
walk \cite{Feller,Poisson-rad,random-walk,anom-diff} in the energy
space. For our case, the walk is one-sided, with continuously
distributed step size, yet the total probability of each step is
less than unity. To substantiate the analogy further,
$dw_1/d\omega_1$ in Eq.~(\ref{generic-eq}) may be thought of as
depending on the target thickness $L$. Then, differentiating both
sides of the equation by $L$ yields a kinetic equation
\begin{subequations}\label{kinetic-eq}
\begin{eqnarray}\label{kinetic-eq1}
\frac{\partial}{\partial L}\frac{dw}{d\omega}=\int_0^E
d\omega_1\left(\frac{dw}{d\omega'}\bigg|_{\omega'=\omega-\omega_1}-\frac{dw}{d\omega}\right)\frac{\partial}{\partial
L}\frac{dw_1}{d\omega_1}\nonumber\\
+W_0(L)\frac{\partial}{\partial L}\frac{dw_1}{d\omega},\qquad
\end{eqnarray}
where kernel $\frac{\partial}{\partial L}\frac{dw_1}{d\omega_1}$
stands for the radiative cross-section. If we take into account that
in the first term of (\ref{kinetic-eq1})
$\frac{dw}{d\omega'}\big|_{\omega'=\omega-\omega_1}$ equals zero for
$\omega_1>\omega$, we may also rewrite the equation as
\begin{eqnarray}\label{kinetic-eq2}
\frac{\partial}{\partial L}\frac{dw}{d\omega}
=\int_0^\omega d\omega_1 \frac{dw}{d\omega'}\bigg|_{\omega'=\omega-\omega_1}\frac{\partial}{\partial L}\frac{dw_1}{d\omega_1}\qquad\quad\nonumber\\
-\frac{dw}{d\omega}\int_0^{E}d\omega_1\frac{\partial}{\partial
L}\frac{dw_1}{d\omega_1}+W_0(L)\frac{\partial}{\partial
L}\frac{dw_1}{d\omega}.
\end{eqnarray}
\end{subequations}
The initial condition for Eq.~(\ref{kinetic-eq1}) or
(\ref{kinetic-eq2}) is
\begin{equation}\label{kinetic-eq-init-cond-0}
\frac{dw}{d\omega}\bigg|_{L=0}\equiv0.
\end{equation}

It is noteworthy that the last term in Eqs.~(\ref{kinetic-eq}) makes
them inhomogeneous with respect to $\frac{dw}{d\omega}$. That term
is necessary to describe the change of the normalization with $L$
according to Eq.~(\ref{total-probab-1}). At small $L$, when
$W_0(L)\approx W_0(0)=1$, the inhomogeneous term dominates, but at
large $L$, it vanishes exponentially. For the electron distribution
function
\begin{equation}\label{def-Pie}
\Pi(E_e-\omega)=\frac{dw}{d\omega}+W_0\delta(\omega)
\end{equation}
(see Sec.~\ref{subsec:cont-int-repr}), the kinetic equation will be
strictly homogeneous:
\begin{subequations}\label{kinetic-eq-for-Pie}
\begin{eqnarray}\label{kinetic-eq-for-Pie1}
\frac{\partial}{\partial L}\Pi(E_e-\omega)=\int_0^E d\omega_1 \frac{\partial}{\partial L}\frac{dw_1}{d\omega_1}\qquad\qquad\qquad\quad\, \nonumber\\
\times\left[\Pi(E_e-\omega+\omega_1)-\Pi(E_e-\omega)\right] \\
\equiv\int_0^\omega d\omega_1 \Pi(E_e-\omega+\omega_1)\frac{\partial}{\partial L}\frac{dw_1}{d\omega_1}\quad \nonumber\\
-\Pi(E_e-\omega)\int_0^E d\omega_1 \frac{\partial}{\partial
L}\frac{dw_1}{d\omega_1}.\qquad\label{kinetic-eq-for-Pie2}
\end{eqnarray}
\end{subequations}
with initial condition
\begin{equation}\label{kinetic-eq-init-cond-for-Pie}
\Pi(E_e-\omega)\big|_{L=0}\equiv\delta(\omega).
\end{equation}
In the literature, the kinetic equation is usually quoted in the
form of electromagnetic cascade equation
(\ref{kinetic-eq-for-Pie}b), but the function entering thereto is
sometimes also called the multiphoton spectrum (which may be
misleading). In the presence of significant incoherent radiation,
though, the inhomogeneous term should vanish, anyway, insofar as
$W_0\to0$.

Note, incidentally, that integro-differential equation
(\ref{kinetic-eq-for-Pie}) is well-suited for numerical solution by
Monte-Carlo method, allowing one to simulate a multiphoton radiation
spectrum without actually computing separate $n$-photon components.
Alternative forms (\ref{kinetic-eq}a,b) are suitable for that
purpose, too, but granted that their initial condition
(\ref{kinetic-eq-init-cond-0}) is non-singular, they must be
solvable as well by non-random finite-difference methods. In fact,
knowledge of $\frac{\partial}{\partial L}\frac{dw}{d\omega}$ may be
sufficient to extract also the photon multiplicity spectrum:
Assuming that the single-photon spectrum is proportional to $L$,
i.e. $\frac{dw_1}{d\omega_1}=L\frac{\partial}{\partial
L}\frac{dw_1}{d\omega_1}$, Eq.~(\ref{kinetic-eq2}) rewrites
\begin{eqnarray}\label{rewrit}
L\frac{\partial}{\partial L}\frac{dw}{d\omega}
&=&W_0(L)\frac{dw_1}{d\omega}+\int_0^\omega d\omega_1 \frac{dw}{d\omega'}\bigg|_{\omega'=\omega-\omega_1}\frac{dw_1}{d\omega_1}\nonumber\\
&\,&-\frac{dw}{d\omega}\int_0^{E}d\omega_1\frac{dw_1}{d\omega_1}.
\end{eqnarray}
Then, noticing that the first two terms on the r.h.s. of
(\ref{rewrit}) coincide with the r.h.s. of
Eq.~(\ref{mean-n-omega-convol}), one arrives at a relation
\begin{equation}\label{n-omega-through-LdLlndwdomega}
\bar n(\omega)=w_{1}+L\frac{\partial}{\partial
L}\ln\frac{dw}{d\omega}
\end{equation}
(where for IR regularization, $w_{1}$ may need to be replaced by
$\int_{\epsilon}^E d\omega_1 \frac{dw_1}{d\omega_1}$).

The linear homogeneous 1st-order integro-differential form of
equation (\ref{kinetic-eq-for-Pie}) provides the correspondence
between our multiphoton emission process and continuous random walks
(a rather general category of Markov processes)\footnote{The
Markovian (memoryless) character of the radiation process may seem
to be at odds with the physical dependence of the radiation spectrum
on the entire electron trajectory. There is no contradiction here,
since the memoryless character is understood in the sense that the
radiating electron does not `remember' the negligible photon
recoils. So, kernel $\frac{\partial}{\partial
L}\frac{dw_1}{d\omega_1}$ in principle may depend on $L$, i.e., on
the electron history, in an arbitrary way, but its further promotion
to multiphoton spectrum $\frac{dw}{d\omega}(L)$ is sought as
solution of a Markovian differential equation with respect to
increase of $\frac{dw_1}{d\omega_1}(L)$, or simply $L$.}.
Eqs.~(\ref{kinetic-eq}) actually describe a \emph{driven} random
walk, although it is equivalent to a free one, and belongs to the
Markovian process category, anyway.

Now, let us turn to the case of large $L$. Then, $W_0$ becomes
small, and the radiation process definitely enters the free random
walk regime. Intuitively, for long random walks, detail of the
single-step distribution should fade away after a large number of
steps. The Central Limit Theorem \cite{Feller,Gnedenko-Kolmogorov}
asserts that for a sufficiently long random walk, the particle
probability distribution tends to a Gaussian, involving only two
parameters: the mean value and the variance of the single-step
distribution (single-photon spectrum). It must be realized, however,
that under the presence of incoherent bremsstrahlung component, the
first moments defining the Gaussian diverge. That is a familiar
situation when the gaussianity breaks down, and the diffusion
becomes anomalous \cite{anom-diff}. For such a case, there exists a
Generalized Central Limit Theorem \cite{Feller,Gnedenko-Kolmogorov},
but it is not really relevant for our conditions, as long as the
incoherent component intensity, quantified by parameter $a$, is not
supposed to become large ($L\ll X_0$).

A reasonable way out of the present situation may be to treat the
incoherent radiation component as a single-jump process, rather than
a multistep random walk. To this end, one may first cope with the
pure coherent radiation in the high multiplicity regime
independently, where the problem of divergent moments is not
encountered (that may also be of intrinsic interest in application
to undulator radiation, where $a$ is often negligible). At the final
step of the calculation, the contribution from the incoherent
radiation component can be restored via convolution relation
(\ref{wcoh-times-winc}). Following this approach, in the next
subsection we deal with the pure coherent radiation case.

\subsection{Normal diffusion for pure coherent radiation component}\label{subsec:Normal-diffusion}

Since typical $\omega$ in the coherent radiation spectrum at  high
intensity are $>\omega_0$, energetically ordered representations
offer in this case no advantage. It is thus simpler to issue from
contour integral representation (\ref{generic-eq}),
\begin{equation}\label{cont-int-for-wc}
\frac{dw_{\text{c}}}{d\omega}=\frac1{2\pi
i}\int_{c-i\infty}^{c+i\infty}ds e^{s\omega+\int_0^{\omega_0}
d\omega_1
\frac{dw_{1\text{c}}}{d\omega_1}\left(e^{-s\omega_1}-1\right)},
\end{equation}
where we explicitly introduced the coherent emission edge $\omega_0$
as the finite upper limit for $\omega_1$-integration. Obviously, as
$\frac{dw_{1\text{c}}}{d\omega_1}$ grows large, the exponential
integrand of (\ref{cont-int-for-wc}) steepens as a function of $s$,
peaking somewhere between the integration contour endpoints. Such a
situation suggests application of the steepest descent method, which
is a common tool in the random walk theory. For our case, though,
the corresponding procedure yet involves certain subtleties which
will be highlighted below.

\subsubsection{The steepest descent approximation}\label{subsubsec:steepest-descent}

The steepest descent method (see, e.g., \cite{Olver}) determines the
asymptotics of a contour integral by deforming its integration
contour in the complex plane so that it passes through a point, to
both sides from which the integrand decreases by its absolute value,
without significant oscillations. For an analytic integrand, this
must be a regular extremum (saddle point in the complex plane),
where the derivative vanishes. Hence the equation for saddle point
$s_0$ of the integrand of (\ref{cont-int-for-wc}) emerges
as\footnote{Since the integrand of (\ref{cont-int-for-wc}) involves
no preexponential, the equation for the saddle point is basically
unequivocal.}
\begin{eqnarray}\label{saddle-point-eq}
\frac{\partial}{\partial s}\left[s\omega+\int_0^{\omega_0} d\omega_1 \frac{dw_{1\text{c}}}{d\omega_1}\left(e^{-s\omega_1}-1\right) \right]\Bigg|_{s=s_0}\nonumber\\
=\omega-\int_0^{\omega_0} d\omega_1
\omega_1\frac{dw_{1\text{c}}}{d\omega_1}e^{-s_0\omega_1}=0.
\end{eqnarray}

\begin{figure}
\includegraphics[width=\columnwidth]{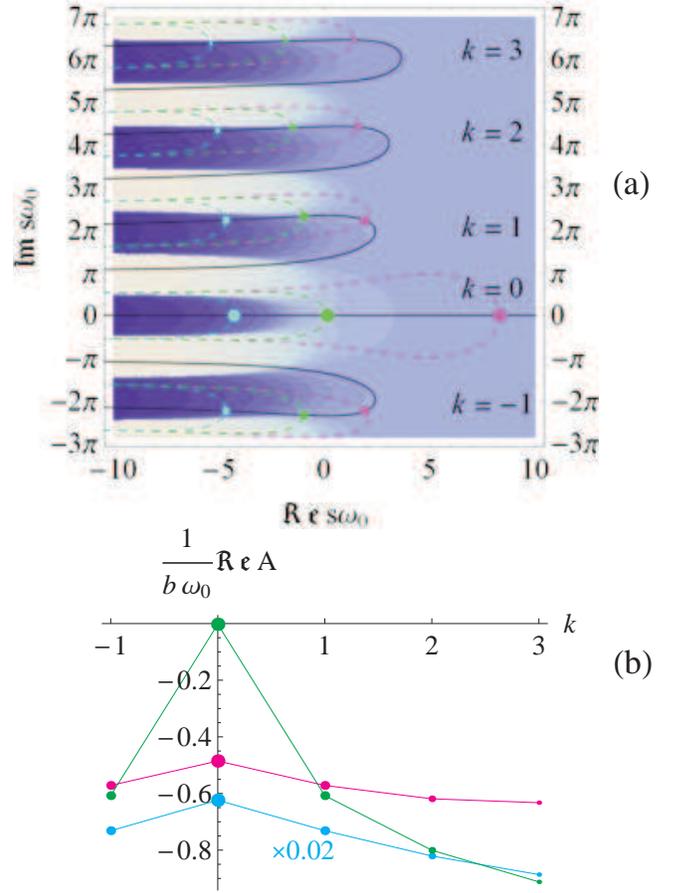}%{Multiple-saddle-points8}
%[width=\columnwidth]
%\includegraphics[width=70mm]{Re-Plot3D}
%\includegraphics{Re-at-k}
\caption{\label{fig:Re-Plot3D} (a) The distribution of saddle points
in the complex plane of $s$. Solid black curves are solutions of
equation $\mathfrak{Im}\int_0^{\omega_0} d\omega_1
\omega_1\frac{dw_{1\text{c}}}{d\omega_1}e^{-s_0\omega_1}=0$
(saddle-point trajectories). Dashed curves are solutions of equation
$\mathfrak{Re}\int_0^{\omega_0} d\omega_1
\omega_1\frac{dw_{1\text{c}}}{d\omega_1}e^{-s_0\omega_1}=\omega$
(magenta, $\omega=0.01{b\omega^2_0}$; green,
$\omega=0.3b\omega^2_0$; cyan, $\omega=10b\omega^2_0$).
Intersections of solid curves with dashed ones give saddle points at
a given $\omega/b$. The density plot, function
$\frac1{b\omega_0}\mathfrak{Re}A(s)$, for $A(s)$ given by
Eq.~(\ref{H}) (brighter regions correspond to greater values of
$\mathfrak{Re}A$). (b) Values of function
$\frac1{b\omega_0}\mathfrak{Re}A(s)$ in the saddle points. [For
$\omega=10b\omega^2_0$ (cyan), the plotted function is multiplied by
0.02.]}
\end{figure}

Once solution $s_0=s_0(\omega)$ to Eq.~(\ref{saddle-point-eq}) is
found, the exponent in Eq.~(\ref{cont-int-for-wc}) is further
expanded around this point to second order:
\begin{eqnarray}\label{HD}
s\omega+\int_0^{\omega_0} d\omega_1 \frac{dw_{1\text{c}}}{d\omega_1}\left(e^{-s\omega_1}-1\right)\quad\qquad\qquad\qquad\qquad\nonumber\\
=A[s_0(\omega)]+\frac12 B[s_0(\omega)](s-s_0)^2%\nonumber\\
+\mathcal{O}\left[(s-s_0)^3\right]\!.\quad
\end{eqnarray}
To express coefficients $A$ and $B$, we write in the exponent of
(\ref{cont-int-for-wc})
\begin{eqnarray*}\label{}
e^{-s\omega_1}&=&e^{-s_0\omega_1}e^{(s_0-s)\omega_1}\nonumber\\
&\simeq&
e^{-s_0\omega_1}\left[1+(s_0-s)\omega_1+\frac12(s_0-s)^2\omega_1^2\right],
\end{eqnarray*}
and notice that term $-s\omega_1$, when convolved with
$\frac{dw_{1\text{c}}}{d\omega_1}$, cancels with term $s\omega$ by
virtue of Eq.~(\ref{saddle-point-eq}). That produces the expected
structure (\ref{HD}), wherefrom we read off
\begin{equation}\label{H}
A[s_0(\omega)]=\int_0^{\omega_0} d\omega_1
\frac{dw_{1\text{c}}}{d\omega_1}\left[e^{-s_0\omega_1}(1+s_0\omega_1)-1\right],
\end{equation}
and
\begin{equation}\label{D}
B[s_0(\omega)]=\int_0^{\omega_0} d\omega_1
\omega_1^2\frac{dw_{1\text{c}}}{d\omega_1}e^{-s_0\omega_1}=-\frac1{s_0}\frac{\partial
A}{\partial s_0}.
\end{equation}
The neglect in the exponent of (\ref{cont-int-for-wc}) of Taylor
terms higher than quadratic reduces the integrand to a Gaussian, and
the contour integral evaluates
\begin{subequations}
\begin{equation}\label{dwdomega-saddle}
\frac{dw_\text{c}}{d\omega}\approx\frac{e^{A}}{\sqrt{2\pi B}}.
\end{equation}

%By experience of deriving asymptotics of simplest special functions
%expressed by contour integrals, the saddle-point approximation often
%gives plausible results even when the exponent is not particularly
%large. But in our case, we deal with edgy integrands, so application
%of the saddle-point approximation here needs more caution.

One should yet be aware that the saddle-point equation in the
complex plane may have several solutions. For our edgy single-photon
spectrum (\ref{dw1coh-real}), the number of solutions is actually
infinite. To demonstrate that, the distribution of saddle points in
the complex $s$-plane, and their motion with the variation of
$\omega$, is portrayed in Fig.~\ref{fig:Re-Plot3D} for specific
single-photon spectrum profile (\ref{dw1coh-real}--\ref{f-def}).
There, black solid curves correspond to the $\omega$-independent
imaginary part of Eq.~(\ref{saddle-point-eq}), giving saddle points
trajectories. Dashed color curves correspond to the real part of
Eq.~(\ref{saddle-point-eq}); their intersections with the solid
curves give saddle point locations, which depend on $\omega$. The
figure suggests that for any $\omega$, the number of saddle points
must be infinite, so instead of Eq.~(\ref{dwdomega-saddle}) one
ougth to write
\begin{equation}\label{dwdomega-saddle-sum}
\frac{dw_\text{c}}{d\omega}\approx\sum_{k=-\infty}^{\infty}\frac{e^{A_k}}{\sqrt{2\pi
B_k}},
\end{equation}
\end{subequations}
where the summation runs over all the saddle points, and
$A_{-k}=A_k^*$, $B_{-k}=B_k^*$.

Nevertheless, in the limit $A\propto b\to\infty$, series
(\ref{dwdomega-saddle-sum}) must be dominated by a term with the
greatest $\mathfrak{Re}A$. The diagram in Fig.~\ref{fig:Re-Plot3D}b
shows the values of $\frac1{b\omega_0}\mathfrak{Re}A_k$ in saddle
points. For any $\omega$, the saddle point at $k=0$ appears to be
the highest, and therefore asymptotically dominant. Thus, in the
limit $b\to \infty$, Eq.~(\ref{dwdomega-saddle}) is recovered, with
$s_0(\omega)$ being the unique \emph{real} solution of
Eq.~(\ref{saddle-point-eq}).
%At moderate $b$,
%though, we expect the accuracy of single-term approximation
%(\ref{dwdomega-saddle}) to diminish, which is in line with the
%existence of sharp features at moderate $b$. In the present paper,
%we will restrict ourselves solely to the contribution from the leading
%saddle point.

Another issue is the robustness of the saddle point approximation
itself. That approximation breaks down in vicinity of the point
where coefficient $B$ vanishes, i.e. at small $\omega$ corresponding
to large $s_0$. To determine the condition of applicability of the
saddle-point approximation, one needs to assess higher terms in
$(s-s_0)$ in Eq.~(\ref{HD}). That leads to the inequality
\begin{equation}\label{saddle-point-cond}
w_{1\text{c}}\frac{\omega}{\omega_0}\gg1\qquad\quad (\text{saddle
point approx.}).
\end{equation}
The l.h.s. thereof admits interpretation as an order of magnitude of
the number of emitted photons with energies below $\omega$. Thus,
Ineq.~ (\ref{saddle-point-cond}) implies that for the diffusion mode
to develop, the same `destination' (total energy) $\omega$ must be
reachable by many radiative scenarios with commensurable and sizable
probability.

If all the premises listed above are fulfilled,
Eq.~(\ref{dwdomega-saddle}) offers an explicit solution for the
resummed spectrum, provided the real solution to equation
(\ref{saddle-point-eq}) is found and substituted to
Eqs.~(\ref{H}--\ref{D}). An impediment is that even with simplest
shapes of $\frac{dw_{1\text{c}}}{d\omega_1}$, like (\ref{f-def}) or
a uniform distribution, equation (\ref{saddle-point-eq}) is
transcendental. On the other hand, in the crudest approximation, it
may suffice to know its behavior around single point corresponding
to the maximum of expression (\ref{dwdomega-saddle}). The latter
region, called central, will be scrutinized in the next paragraph.

%\newpage

\subsubsection{Central region}\label{subsubsec:centr}

The problem of finding the maximum of distribution
(\ref{dwdomega-saddle}) is essentially equivalent to tracing that
for exponent $A$, as long as the exponential should be an
asymptotically steeper function than its preexponential factor.
Inspection of integral (\ref{H}) reveals that for any positive
kernel $\frac{dw_{1\text{c}}}{d\omega_1}$, the integral achieves a
maximum in point $s_0=0$, since there the $s_0$-dependent factor
$\left[e^{-s_0\omega_1}(1+s_0\omega_1)-1\right]$ turns to zero,
while being negative elsewhere. From Eq.~(\ref{saddle-point-eq}),
one finds that value $s_0=0$ corresponds to
$\omega=\int_0^{\omega_0} d\omega_1 \omega_1
\frac{dw_{1\text{c}}}{d\omega_1}=\overline{\omega_{1\!}}_{\text{c}}$.
Expansion the of r.h.s. of Eq.~(\ref{H}) by deviations of $s_0$ from
zero (beginning from a quadratic term) yields
\begin{equation}\label{Hcoh-quadr}
A[s_0(\omega)]= -\frac12 \overline{\omega^2_{1\!}}_{\text{c}}
s_0^2+\mathcal{O}\left(\overline{\omega^3_{1\!}}_{\text{c}}s_0^3\right).
\end{equation}
If Eq.~(\ref{saddle-point-eq}) in turn is linearized in $s_0$,
\begin{equation}\label{saddle-point-eq-lin}
\omega=
\overline{\omega_{1\!}}_{\text{c}}-\overline{\omega^2_{1\!}}_{\text{c}}
s_0+\mathcal{O}\left(\overline{\omega^3_{1\!}}_{\text{c}}s_0^2\right),
\end{equation}
it solves as
\begin{equation}\label{saddle-point-eq-lin-soln}
s_0(\omega)\simeq
\frac{\overline{\omega_{1\!}}_{\text{c}}-\omega}{\overline{\omega^2_{1\!}}_{\text{c}}}.
\end{equation}
Combining Eqs.~(\ref{Hcoh-quadr}) and
(\ref{saddle-point-eq-lin-soln}), we obtain the explicit asymptotic
expression for the probability spectrum in vicinity of the maximum
(central region):
\begin{equation}\label{Gauss}
\frac{dw_\text{c}}{d\omega}\simeq \frac{1}{\sqrt{2\pi
\overline{\omega^2_{1\!}}_{\text{c}}}}e^{-\rho^2/2},
\end{equation}
with
\begin{equation}\label{scal-var}
\rho=\frac{\omega-\overline{\omega_{1\!}}_{\text{c}}}{\sqrt{\overline{\omega^2_{1\!}}_{\text{c}}}}.
\end{equation}
As anticipated, this limiting distribution is merely a Gaussian,
with the mean value and variance complying with
Eqs.~(\ref{mean-omega}--\ref{mean-omega2-alt}), for an arbitrary
shape of $\frac{dw_{1\text{c}}}{d\omega}$ (see, e.g.,
\cite{Akh-Shul-shower-ZhETF,Kolch-Uch}). Since dependencies
$\frac{dw_\text{c}}{d\omega}(\omega)$ for different shapes and
magnitudes of $\frac{dw_{1\text{c}}}{d\omega_1}$ and intensities are
related by a linear change of the variables (i.e., the multiphoton
spectrum shape is invariant under changes of
$\frac{dw_{1\text{c}}}{d\omega_1}$), the central region is also
called scaling one \cite{anom-diff}. So far, the approach to
Gaussian asymptotics remains largely unexplored experimentally,
although spectra resembling Gaussians were obtained, e.g., in
\cite{multiphot-CERN-Kirsebom,Medenwaldt-Gauss,w1c>1}.

Approximation (\ref{Gauss}) holds as long as in
Eq.~(\ref{saddle-point-eq-lin}) nonlinear terms in $s_0$ keep
relatively small, i.e.
\begin{equation}\label{central-region}
|\rho|\ll\frac{\overline{\omega^2_{1\!}}_{\text{c}}^{3/2}}{\overline{\omega^3_{1\!}}_{\text{c}}}\sim\sqrt{b\omega_0}.
\end{equation}
Since here the r.h.s. is $\gg1$, the Gaussian approximation must
work in a sufficiently broad interval about its maximum.

The corresponding photon multiplicity spectrum $\bar
n_{\text{c}}(\omega)$ is further evaluated by referring to
representation (\ref{mean-n-omega-convol}). Its second term vanishes
due to practically vanishing $W_0$, or due to vanishing
$\frac{dw_{1\text{c}}}{d\omega_1}$ for $\omega>\omega_0$, while in
the first term, the integration is effectively limited to interval
$0<\omega_1\leq\omega_0$, where factor
$\frac{dw_{1\text{c}}}{d\omega_1}$ differs from zero. Noting that
within such an interval, the second factor
$\frac{dw_{\text{c}}}{d\omega'}\big|_{\omega'=\omega-\omega_1}$
varies slowly, one can Taylor-expand it to first order, say, about
the origin:
\begin{equation*}\label{}
\frac{dw_\text{c}}{d\omega}\bar
n_{\text{c}}(\omega)\approx\int_0^{\omega_0}d\omega_1\frac{dw_{1\text{c}}}{d\omega_1}\left(\frac{dw_\text{c}}{d\omega}-\omega_1\frac{d^2w_\text{c}}{d\omega^2}\right),
\end{equation*}
wherewith
\begin{equation}\label{}
\bar n_{\text{c}}(\omega)\approx
w_{1\text{c}}-\overline{\omega_{1\!}}_{\text{c}}\frac{d}{d\omega}\ln\frac{dw_\text{c}}{d\omega}.
\end{equation}
Inserting here expression (\ref{Gauss}), we obtain an utterly simple
result:
\begin{equation}\label{mean-n-at-Gauss}
\bar n_{\text{c}}(\omega)\approx
w_{1\text{c}}+\frac{\overline{\omega_{1\!}}_{\text{c}}}{\overline{\omega^2_{1\!}}_{\text{c}}}\left(\omega-\overline{\omega_{1\!}}_{\text{c}}\right).
\end{equation}
The linear function in the r.h.s. of (\ref{mean-n-at-Gauss}) has
positive intercept $\bar n_{\text{c}}(0)\approx
w_{1\text{c}}-\frac{\overline{\omega_{1\!}}_{\text{c}}^2}{\overline{\omega^2_1}_{\text{c}}}$,
and slope\footnote{In the central region, the photon multiplicity
spectrum slope appears to be independent of $b$, but at small
$\omega$, its slope is $\propto b$ [see Eq.~(\ref{n0=1+})].}
$\frac{\partial}{\partial\omega}\bar
n_{\text{c}}(\omega)\approx\frac{\overline{\omega_{1\!}}_{\text{c}}}{\overline{\omega^2_1}_{\text{c}}}>\frac1{\omega_0}$;
the latter implies that $\bar n_{\text{c}}(\omega)$ increases by
more than unity per each interval $\omega_0 (n-1)<\omega<\omega_0
n$. It is also evident that in vicinity of the Gaussian maximum,
\begin{equation}\label{n-to-wi1c}
\bar n_{\text{c}}(\overline{\omega_{1\!}}_{\text{c}})\approx
w_{1\text{c}},
\end{equation}
which, according to (\ref{mean-n-def}), amounts the mean photon
multiplicity for the whole coherent radiation process. An
approximately linear increase of $\bar n_{\text{c}}(\omega)$ in the
range of typical $\omega$ was observed in experiment
\cite{multiphot-CERN-Kirsebom}. There was also observed a turnover
of $\bar n_{\text{c}}(\omega)$, one of the possible reasons for
which will be explained later.

In Fig.~\ref{fig:pure-coh-spectrum}, the exact resummed spectrum
shape  for radiation intensity parameter $b\omega_0=6$ (blue solid
curve) is compared with the behavior of Gaussian approximation
(\ref{Gauss}), (\ref{mean-n-at-Gauss}) (blue dashed curve). At the
chosen value of $b\omega_0$, a perfect convergence for the resummed
spectrum (Fig.~\ref{fig:pure-coh-spectrum}a) is not reached yet. At
the same time, for $\bar n_{\text{c}}(\omega)$
(Fig.~\ref{fig:pure-coh-spectrum}c), the agreement is already
excellent. Remarkably, the linearity of the latter function holds
well even far away from the central region, where the spectrum is no
longer Gaussian [$s_0(\omega)$ dependence gets highly nonlinear, and
variation of the preexponent $B$ becomes significant, as well].
%We can not explain this phenomenon here.

\begin{figure}
\includegraphics{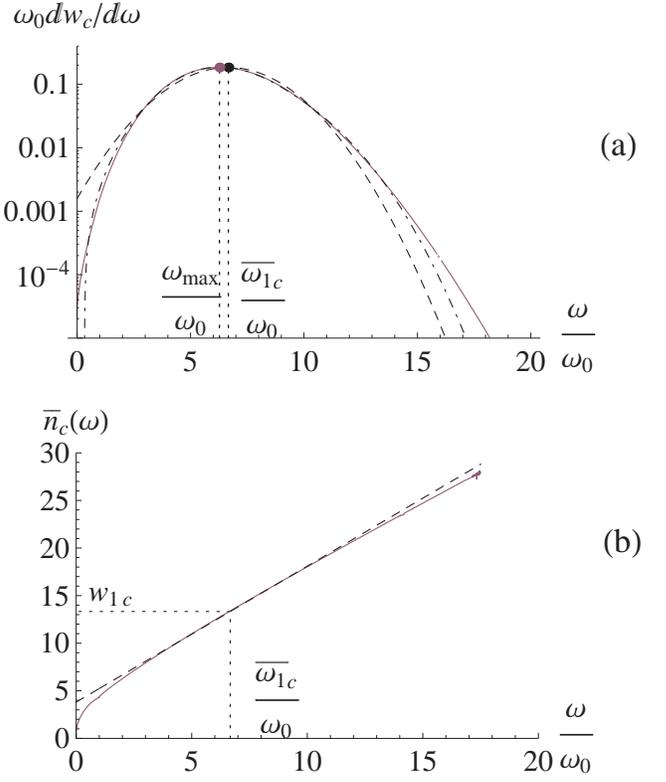}
\caption{\label{fig:LogPlot} (a) Log plot of the resummed pure
coherent radiation spectrum, at intensity parameter $b\omega_0=20$.
Solid purple curve, exact distribution evaluated by
Eqs.~(\ref{cont-int-for-wc}), (\ref{dw1coh-real}--\ref{f-def}).
Dashed black curve, Gaussian approximation (\ref{Gauss}). Dot-dashed
black curve, Gaussian
approximation with Chebyshev correction (\ref{Chebyshev}). %Blue dashed curve, NLL large-$\omega$ asymptotics  based on (\ref{s0c-log-1}). Blue dot-dashed curve, NNLL based on Eq.~(\ref{s0c-log-2}).
(b) Photon multiplicity spectrum under same conditions. Dashed black
line, approximation (\ref{mean-n-at-Gauss}).}
\end{figure}

To improve the accuracy of the diffusive approximation, one may be
prompted to derive a correction to Eq.~(\ref{Gauss}). To this end,
note that Eq.~(\ref{Gauss}) corresponds to expanding the integral in
the exponent in Eq.~(\ref{cont-int-for-wc}) to second order in $s$.
A natural improvement thus results if in the expansion of the
exponentiated integral a cubic term in $s$ is retained:
\begin{equation}\label{}
\frac{dw_\text{c}}{d\omega}\simeq\frac1{2\pi
i}\int^{c+i\infty}_{c-i\infty} ds
e^{s(\omega-\overline{\omega_{1\!}}_{\text{c}})+\frac12
s^2\overline{\omega^2_1}_{\text{c}}-\frac16
s^3\overline{\omega^3_1}_{\text{c}}}.
\end{equation}
The latter contour integral is reducible to an Airy function, but
exponentiation of a small correction may be an excess of precision.
If we simply linearize the exponential,
\[
e^{-\frac16 s^3\overline{\omega^3_1}_{\text{c}}}\simeq 1-\frac16
s^3\overline{\omega^3_1}_{\text{c}},
\]
and evaluate the contour integral, it leads to an additive
correction dating back to Chebyshev (see, e.g.,
\cite{Gnedenko-Kolmogorov})\footnote{Note that in
\cite{Gnedenko-Kolmogorov}, the definition of Hermite polynomials is
non-standard. We adhere to standard definition \cite{Abr-Stig}.}:
\begin{equation}\label{Chebyshev}
\frac{dw_\text{c}}{d\omega}\simeq \frac{e^{-\rho^2/2}}{\sqrt{2\pi
\overline{\omega^2_{1\!}}_{\text{c}}}}\left[1+\frac{\gamma_3}{12\sqrt2}H_3\left(\frac{\rho}{\sqrt{2}}\right)\right],
\end{equation}
with $\gamma_3$ defined by Eq.~(\ref{skewness}), and
\begin{equation}\label{H3}
H_3(z)=-e^{z^2}\frac{d^3}{dz^3}e^{-z^2}=8z^3-12z
\end{equation}
the Hermite polynomial of order 3 \cite{Abr-Stig,Olver}.
%\begin{eqnarray}\label{Airy}
%\frac{dw_\text{c}}{d\omega}&\approx&
%\left(\frac{2}{\left\langle \omega_1^3 %\right\rangle_{\text{c}}}\right)^{1/3}
%e^{\frac{\left\langle \omega_1^2 \right\rangle_{\text{c}}}{\left\langle %\omega_1^3 \right\rangle_{\text{c}}}\left(\omega-\left\langle %\omega_1\right\rangle_{\text{c}}+\frac13\frac{\left\langle %\omega_1^2\right\rangle_{\text{c}}^2}{\left\langle %\omega_1^3\right\rangle_{\text{c}}}\right)}\nonumber\\
%&\,&\times\text{Ai}\left[\left(\frac{2}{\left\langle \omega_1^3 %\right\rangle_{\text{c}}}\right)^{1/3}\left(\omega-\left\langle %\omega_1\right\rangle_{\text{c}}+\frac12\frac{\left\langle %\omega_1^2\right\rangle_{\text{c}}^2}{\left\langle %\omega_1^3\right\rangle_{\text{c}}}\right)\right].
%\end{eqnarray}
The behavior of approximation (\ref{Chebyshev}) is illustrated in
Fig.~\ref{fig:pure-coh-spectrum} and Fig.~\ref{fig:LogPlot} (for a
greater $b\omega_0$) by dot-dashed curves, and is appreciably closer
to the exact distribution than a simple Gaussian (dashed curves).

The evaluated skewness correction mildly breaks the scaling, and
serves to describe the residual asymmetry of the spectrum. One of
its consequences is the redshift of the spectral maximum location to
the value
\begin{equation}\label{omega-max-skew}
\omega_{\max}=\overline{\omega_{1\!}}_{\text{c}}-\frac{\overline{\omega^3_1}_{\text{c}}}{2\overline{\omega^2_{1\!}}_{\text{c}}}+\mathcal{O}(b^{-1}),
\end{equation}
where the correction term is independent of the radiation intensity
[for the spectral shape (\ref{f-def}),
$\frac{\overline{\omega^3_{1\!}}_{\text{c}}}{2\overline{\omega^2_{1\!}}_{\text{c}}}=\frac{11}{28}\omega_0$].
However, the correction to the height of the maximum in this
approximation is zero:
\begin{equation}\label{max-Gauss-skew}
\max \frac{dw_\text{c}}{d\omega}=\frac1{\sqrt{2\pi
\overline{\omega^2_1}_{\text{c}}}}\left[1+\mathcal{O}(b^{-2})\right].
\end{equation}

Fig.~\ref{fig:LogPlot}a also indicates that even though the Gaussian
approximation, especially with skewness correction
(\ref{Chebyshev}), works fairly well, away from the central region
the falloff law differs from the Gaussian, and the scaling property
gets strongly violated. In principle, this deviation can still be
calculated based on the steepest descent method of
Sec.~\ref{subsubsec:steepest-descent}, provided one copes with the
saddle point equation at $s\neq0$. Approximations derived along
these lines for sub- and trans-central regions are described in
Appendix~\ref{app:periph}.

%\newpage

\subsection{Convolution with incoherent component}\label{subsec:conv-inc}

Let us finally inspect how does the multiphoton spectrum shape
modify upon incorporation of an incoherent component. As in
Sec.~\ref{subsec:coh+incoh}, this is accomplished by convolving the
evaluated pure coherent radiation contribution with pure incoherent
component (\ref{13}) via Eq.~(\ref{wcoh-times-winc}). Given the
concentration of the coherent radiation probability in the central
region, it may be reasonable to employ in (\ref{wcoh-times-winc})
the simplest Gaussian form (\ref{Gauss}). That leaves us with the
expression
\begin{equation}\label{1st-conv}
\frac{dw}{d\omega}\approx \frac{a\Phi(a)}{\sqrt{2\pi
\overline{\omega^2_1}_{\text{c}}}}\int_0^\omega
\frac{d\omega_1}{\omega_1}\left(\frac{\omega_1}{E}\right)^a
e^{-\frac{\left(\omega-\overline{\omega_{1\!}}_{\text{c}}-\omega_1\right)^2}{2\overline{\omega^2_1}_{\text{c}}}}.
%\nonumber\\
%&\,&+e^{-w_{1\text{c}}}\frac{a\Phi(a)}{\omega}\left(\frac{\omega}{E}\right)^a.
\end{equation}
(Term $e^{-w_{1\text{c}}}\frac{dw_{\text{i}}}{d\omega}$ has been
neglected as exponentially small at large $w_{1\text{c}}$.)

The encountered integral still depends on rather many parameters,
but under the high-intensity condition
$\overline{\omega_1\!}_{\text{c}}\gg\sqrt{\overline{\omega^2_1}_{\text{c}}}$,
their number can be effectively reduced. The upper limit of the
integral is always on the right of the maximum of the Gaussian
entering the integrand, hence at the upper integration limit, the
Gaussian is decreasing. Granted that this decrease is rapid, one can
replace this integration limit by infinity, unless the integration
interval is too short. More precisely, the Gaussian varies across
the integration interval significantly if
%First of all, the Gaussian
%in the integrand of Eq.~(\ref{1st-conv}) is much smaller on the
%upper integration limit $\omega_1=\omega$ than in its maximum
%$\omega_1=\omega-\overline{\omega_{1\!}}_{\text{c}}$. So, if the
%maximum of the Gaussian falls within the integration domain, i.e.,
%$\omega-\overline{\omega_{1\!}}_{\text{c}}>0$, it certainly gives
%the main contribution to the integral. Then, without the loss of
%accuracy, the upper limit may be replaced by infinity. In the
%opposite case $\omega-\overline{\omega_{1\!}}_{\text{c}}<0$, the
%Gaussian is decreasing throughout the integration domain, so the
%dominant contribution to the integral stems from vicinity of the
%lower integration limit. Given the high rate of the decrease, again,
%the upper integration limit may be replaced by infinity, provided
\begin{equation}\label{exp-exp}
e^{-\frac{\left(\omega-\overline{\omega_{1\!}}_{\text{c}}\right)^2}{2\overline{\omega^2_{1\!}}_{\text{c}}}}\gg
e^{-\frac{\overline{\omega_{1\!}}_{\text{c}}^2}{2\overline{\omega^2_{1\!}}_{\text{c}}}},
\end{equation}
i.e.,
\begin{equation}\label{small-omega-range}
\omega\gg\frac{\overline{\omega^2_{1\!}}_{\text{c}}}{\overline{\omega_{1\!}}_{\text{c}}}.
\end{equation}
Condition (\ref{small-omega-range}) is not very restrictive as long
as its r.h.s. is independent of the radiation intensity, while its
l.h.s. is typically $\sim\overline{\omega_{1\!}}_{\text{c}}\propto
b$.

Thus, in the high photon multiplicity limit, it appears legitimate
virtually at all $\omega$ to replace the upper integration limit in
Eq.~(\ref{1st-conv}) by infinity.  Therewith, utilizing the value of
the integral
\begin{equation}\label{D-def}
\int_0^\infty
\frac{ds}{s^{1-a}}e^{-(s-\rho)^2/2}=\Gamma(a)e^{-{\rho^2}/4}D_{-a}(-\rho)
\end{equation}
defining parabolic cylinder function $D_{-a}$ \cite{Abr-Stig,Olver},
we arrive at the closed-form representation
\begin{equation}\label{parab-cyl}
\frac{dw}{d\omega}\approx \frac{\Gamma(1+a)\Phi(a)}{\sqrt{2\pi
}E^a\overline{\omega^2_{1\!}}_{\text{c}}^{\frac{1-a}2}}e^{-\rho^2/4}D_{-a}\left(-\rho\right).
%+e^{-w_{1\text{c}}}\frac{a\Phi(a)}{\omega}\left(\frac{\omega}{E}\right)^a.
\end{equation}
Since function (\ref{parab-cyl}), similarly to (\ref{Gauss}),
depends on $\omega$ only through a linearly related variable $\rho$,
this may be regarded as extension of the scaling property beyond the
vicinity of the maximum. But it should be remembered that the
spectrum shape also depends on parameter $a$, so here we have a
2-parameter scaling. The corresponding functions, though, differ
from L\'{e}vy distributions, rather being intermediate between
L\'{e}vy distributions and Gaussian distributions.

\begin{figure}
\includegraphics[width=\columnwidth]{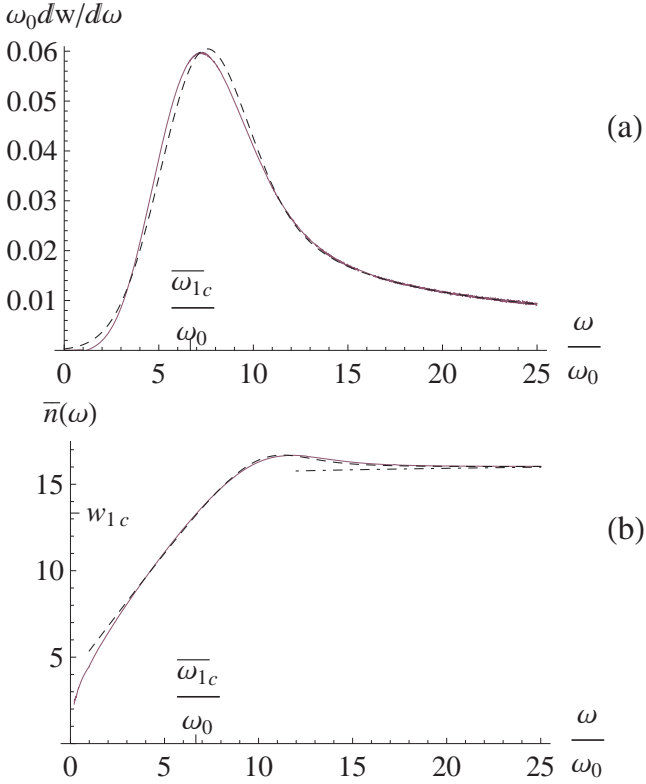}
\caption{\label{fig:LogLogPlot-a} (a) Resummed radiation spectrum
for coherent component described by
Eqs.~(\ref{dw1coh-real}--\ref{f-def}) with $b\omega_0=20$, and
incoherent radiation component with $a=0.3$, $E=100\,\omega_0$.
Solid purple curve, exact distribution; dashed black curve,
parabolic cylinder approximation (\ref{parab-cyl}). (b) The photon
multiplicity spectrum for same conditions as (a), and
$\epsilon=0.1\omega_0$. Dashed black curve, parabolic cylinder
approximation (\ref{convol-for-n-omega2}). Dot-dashed black curve,
high-$\omega$ asymptotics (\ref{n-omega-asympt-large}).}
\end{figure}

The accuracy of approximation (\ref{parab-cyl}) may be assessed from
Fig.~\ref{fig:LogLogPlot-a}a. There one observes that the parabolic
cylinder approximation (dashed black curve) holds fairly well almost
everywhere, except the small-$\omega$ limit.
%On the other hand, the incomplete gamma-function
%approximation (red dashed curve) is more tenable, though not quite perfect as well, because of breakdown of the
%gaussianity in this region, which might be improved, say, by
%utilizing Chebyshev correction (\ref{Chebyshev}). In the latter
%case, the final result is again expressible in terms of incomplete
%gamma-functions, but inevitably becomes more cumbersome, and will
%not be quoted here.

Based on the obtained compact formula (\ref{parab-cyl}), we may now
assess the deviation from the Gaussian behavior in the central
region. In our equations, parameter $a$ must always remain small,
and it is straightforward to check from definition (\ref{D-def})
that $\underset{a\to+0}\lim D_{-a}(-\rho)=e^{-\rho^2/4}$, wherefore,
at $a\to0$ distribution (\ref{parab-cyl}) returns to Gaussian form
(\ref{Gauss}). In the next approximation, the maximum of function
(\ref{D-def}) satisfies the equation
\begin{eqnarray}\label{eq-for-zmax}
\frac{\partial}{\partial \rho}\int_0^\infty
\frac{ds}{s^{1-a}}e^{-(s-\rho)^2/2}\qquad\qquad\qquad\qquad\qquad\qquad\,\,\nonumber\\
=\int_0^\infty \!ds s^a e^{-(s-\rho)^2/2} -\rho\int_0^\infty\!
\frac{ds}{s^{1-a}}e^{-(s-\rho)^2/2}%\nonumber\\
=0.\quad
\end{eqnarray}
At small $a$, the solution of Eq.~(\ref{eq-for-zmax}) is small, too,
and in the linear approximation it equals
\begin{equation}\label{}
\rho_{\max}=\frac{\int_0^\infty ds e^{-s^2/2}}{\int_0^\infty
dss^{a-1}e^{-s^2/2}}+\mathcal{O}(a^2)=a\sqrt{\frac\pi2}+\mathcal{O}(a^2).
\end{equation}
In terms of $\omega$, it casts as
\begin{equation}\label{max-arg-1}
\omega_{\max}\simeq\overline{\omega_{1\!}}_{\text{c}}+a\sqrt{\frac{\pi
}2\overline{\omega^2_{1\!}}_{\text{c}}}.
\end{equation}
One may as well add here the skewness correction
(\ref{omega-max-skew}) (provided both corrections are small),
getting:
\begin{equation}\label{max-arg-tot}
\omega_{\max}\simeq\overline{\omega_{1\!}}_{\text{c}}-\frac{\overline{\omega^3_{1\!}}_{\text{c}}}{2\overline{\omega^2_{1\!}}_{\text{c}}}+a\sqrt{\frac{\pi
}2\overline{\omega^2_{1\!}}_{\text{c}}}.
\end{equation}
Note that those corrections have opposite signs, i.e., the
incoherent radiation component gives a blueshift, while skewness a
redshift, and they partially compensate each other. Evaluating the
second derivative in the maximum, one can also show that the effect
of $a$ is to broaden the spectrum.

At large $\omega$, the spectrum deviates from a Gaussian strongly,
developing a power-law `tail'. The parabolic cylinder function at
large values of its argument has asymptotics
\[
e^{-\rho^2\!/4}D_{-a}(-\rho)\!\underset{\rho\to+\infty}=\!\frac{\sqrt{2\pi}}{\Gamma(a)\rho^{1-a}}\!\left[1\!+\!\mathcal{O}\!\left(\rho^{-2}\right)\!\right]+\mathcal{O}\!\left(\!e^{-\rho^2\!/2}\right)\!,
\]
entailing for the spectrum
\begin{equation}\label{parab-cyl-large-omega}
\frac{dw}{d\omega}\underset{\omega\to+\infty}\simeq
\frac{a\Phi(a)}{E^a(\omega-\overline{\omega_{1\!}}_{\text{c}})^{1-a}}.
\end{equation}
That agrees with Eq.~(\ref{large-omega-corr}) modulo exponentially small terms. %In the log-log plot of Fig.~\ref{fig:LogLogPlot-a}a, this power-law asymptotics corresponds to the straight-line asymptote at large $\omega$.

The calculation of $\bar n(\omega)$ proceeds along the same lines
when based on convolution relation (\ref{convol-for-n-omega}).
Employing there Gaussian approximations (\ref{Gauss}),
(\ref{mean-n-at-Gauss}) along with power-law and logarithmic
expressions (\ref{13}), (\ref{mean-n-omega-incoh}) for the pure
incoherent component, we are led to the integral representation
\begin{eqnarray}
\bar n(\omega)\approx\frac1{dw/d\omega}\frac{a\Phi(a)}{\sqrt{2\pi
\overline{\omega^2_{1\!}}_{\text{c}}}}\qquad\qquad\qquad\qquad\qquad\quad\nonumber\\
\times\!\int_0^\omega
\frac{d\omega'}{\omega'}\left(\frac{\omega'}{E}\right)^a
e^{-\frac{\left(\omega-\overline{\omega_{1\!}}_{\text{c}}-\omega'\right)^2}{2\overline{\omega^2_{1\!}}_{\text{c}}}}\Bigg[\nu(a)+a\ln\frac{\omega'}{\epsilon}\qquad\nonumber\\
+w_{1\text{c}}+\frac{\overline{\omega_{1\!}}_{\text{c}}}{\overline{\omega^2_{1\!}}_{\text{c}}}\left(\omega-\overline{\omega_{1\!}}_{\text{c}}-\omega'\right)\Bigg].\quad
\end{eqnarray}
By the same reasoning as for Eq.~(\ref{1st-conv}), the upper
integration limit here may be extended to infinity. Evaluation of
the integral gives the result
\begin{equation}
\bar n(\omega)\approx
w_{1\text{c}}+\frac{\overline{\omega_{1\!}}_{\text{c}}}{\sqrt{\overline{\omega^2_{1\!}}_{\text{c}}}}R_a(\rho)
+S_a(\rho)+a\ln\frac{\sqrt{\overline{\omega^2_{1\!}}_{\text{c}}}}{\epsilon},\label{convol-for-n-omega2}
\end{equation}
where
\begin{equation}\label{R-def}
R_a(\rho)=\rho-a\frac{D_{-1-a}(-\rho)}{D_{-a}(-\rho)},
\end{equation}
and
\begin{equation}\label{S-def}
S_a(\rho)=a\frac{\partial}{\partial a}\ln
D_{-a}(-\rho)-a\gamma_{\text{E}}.
\end{equation}

\begin{figure}
\includegraphics{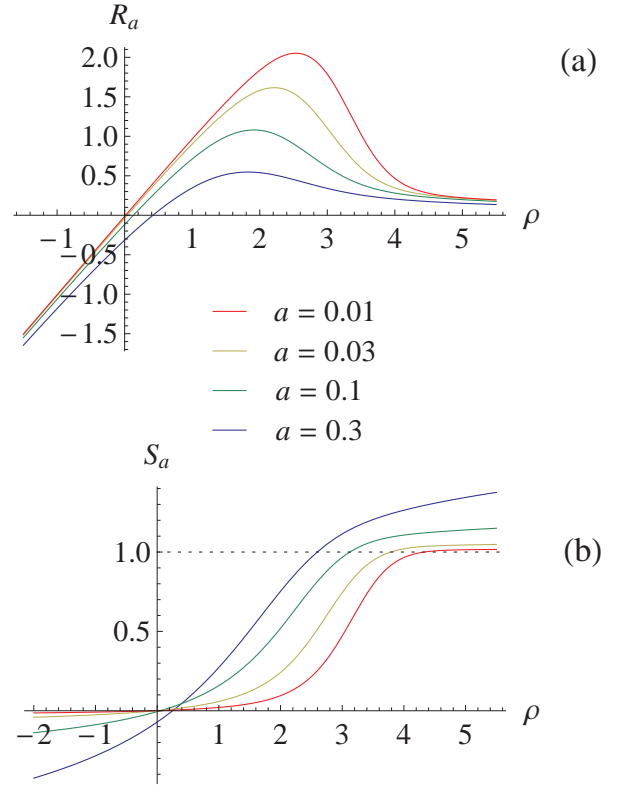}
\caption{\label{fig:Ra,Sa} (a) Plots of function $R_{a}(\rho)$, for
parameter $a=0.01$, 0.03, 0.1, 0.3 (top to bottom). (b) Plots of
$S_{a}(\rho)$ for same values of $a$. }
\end{figure}

Shapes of functions (\ref{R-def}), (\ref{S-def}) for several values
of parameter $a$ are illustrated in Figs.~\ref{fig:Ra,Sa}a,b.
Function $S_a(\rho)$ appears to be monotonous, though step-like,
whereas $R_a(\rho)$ features a hump. The maximum of $R_a(\rho)$
rises with the decrease of parameter $a$, though only
logarithmically. Note that function $R_a$ enters
Eq.~(\ref{convol-for-n-omega2}) with a large coefficient, thence the
shape of the maximum of $\bar n(\omega)$ is primarily described by
$R_a(\rho)$. An additive correction comes from $S_a(\rho)$, whose
unit jump in the turnover region of $R_a(\rho)$ partially
compensates the drop of the photon multiplicity spectrum beyond its
maximum.

The analysis of behavior of functions (\ref{R-def}), (\ref{S-def})
is alleviated by the knowledge of their values in the origin,
\begin{equation}\label{R-0}
R_a(0)=-\frac{a}{\sqrt2}\frac{\Gamma\left(\frac{1+a}2\right)}{\Gamma\left(1+\frac{a}2\right)}
%=-a\sqrt{\frac{\pi}2}+\mathcal{O}\left(a^2\right),
\end{equation}
\begin{equation}\label{S-0}
S_a(0)=-a\left[\frac12\psi\left(\frac{1+a}2\right)+\frac12\ln2+\gamma_{\text{E}}\right],
\end{equation}
and asymptotics at large $|\rho|$:
\begin{equation}\label{R-large-neg-z}
R_a(\rho)=\rho-\mathcal{O}(a/\rho)\qquad (\rho\to-\infty),
\end{equation}
\begin{equation}\label{R-large-z}
R_a(\rho)=\frac{2(1-a)}{\rho}+\mathcal{O}\left(\rho^{-3},
\frac{\rho}{a}e^{-\rho^2/2}\right)\qquad (\rho\to+\infty),
\end{equation}
\begin{equation}\label{S-large-neg-z}
S_a(\rho)\simeq -a\ln|\rho|\qquad (\rho\to-\infty),
\end{equation}
\begin{equation}\label{S-large-z}
S_a(\rho)\simeq a\ln \rho+\nu(a)\qquad (\rho\to+\infty),
\end{equation}
with $\nu(a)$ defined by Eq.~(\ref{nu-def}). At $\rho\to+\infty$,
i.e. asymptotic trans-central $\omega$, function $S_a(\rho)$
dominates, and along with terms
$w_{1\text{c}}+a\ln\frac{\sqrt{\overline{\omega^2_{1\!}}_{\text{c}}}}{\epsilon}$
reproduces result (\ref{n-omega-asympt-large}).

Finally, we recall that the humpy structure of $\bar n(\omega)$
manifested itself already at moderate radiation intensity (cf.
Fig.~\ref{fig:high-intensity}c). That may be non-accidental, since
according to Ineq.~(\ref{saddle-point-cond}), the accuracy of the
saddle-point approximation improves at large $\omega$, where the
maximum of $\bar n(\omega)$ is achieved.

%At small $\omega$, the behavior of the spectrum is also power-like, or, rather, it is a sum of two power functions: ...

%In conclusion of this section, note that with the account of incoherent radiation component, one can develop unified treatment ... based on any of the contour integral representations ( ). We will not develop these alternative views here.

%\newpage

\section{Reconstruction of single-photon spectrum from the multiphoton one}\label{sec:reconstr}

Although measurements of radiation spectra by the calorimetric
method are ubiquitous, for further applications of the extracted
photon beam it is often preferable to describe it in terms of photon
number spectrum (\ref{class}), anyway. Needless to say, the rate of
any reaction (such as $e^+e^-$ creation or nuclear
reactions)\footnote{The latter reactions also serve for measurement
of the gamma-quantum beam polarization, which is not discussed in
this paper.} caused by a photon beam in a target is proportional to
the photon number per energy interval, irrespective of temporal
correlations
between photons. %\footnote{Unless the multiphoton flow is so intense
%that it constitutes a strong electromagnetic wave at the target.
%However, such conditions in the gamma-range are not within
%experimental reach nowadays.}
Besides that, if the radiation spectrum is used for diagnostics of
the radiator, the single-photon spectrum is more directly related to
the classical radiation spectrum and the charged particle
trajectory, and hence to the radiator parameters, than the
multiphoton one.

Thus, there arises a need for reconstructing single-photon spectra
from calorimetrically measured spectra. Such a problem had been
addressed some time ago in \cite{Kolchuzhkin}, where a procedure for
reconstructing first moments (the mean and the variance) of the
single-photon spectra from those of a multiphoton spectra was
proposed. Here we will develop a reconstruction procedure for the
complete spectrum. The prerequisite, however, is fair independence
of the spectrum on initial state parameters to be averaged over,
wherewith the outcoming multiphoton spectrum is representable purely
by contour integral (\ref{generic-eq}). As was argued in
Sec.~\ref{subsec:pure-coh}, such conditions can be met for coherent
bremsstrahlung or undulator radiation, but are unlikely for
channeling radiation.

Let us begin with the observation that integral relation
(\ref{generic-exponentiation}), with the aid of
Eq.~(\ref{total-probab-1}) can be cast in form where the l.h.s.
depends only on $\frac{dw_1}{d\omega_1}$, while the r.h.s. only on
$\frac{dw}{d\omega}$:
\begin{equation}\label{90}
e^{\int_0^{E} d\omega_1
\frac{dw_1}{d\omega_1}\left(e^{-s\omega_1}-1\right)} =1+\int_0^{E_e}
d\omega \frac{dw}{d\omega}\left(e^{-s\omega}-1\right).
\end{equation}
(As long as $\frac{dw}{d\omega}$ is supposed to be inferred from
experimental data, the upper limit of integration over $\omega$ is
set to $E_e$, which is the physical end of the spectrum.) To further
obtain a linear integral equation for the single-photon spectrum, it
suffices to take the logarithm of both sides of Eq.~(\ref{90}):
\begin{subequations}
\begin{eqnarray}\label{simpler}
\int_0^{E} d\omega_1
\frac{dw_1}{d\omega_1}\left(e^{-s\omega_1}-1\right)
\qquad\qquad\qquad\qquad\nonumber\\
=\ln\left[1+\int_0^{E_e} d\omega
\frac{dw}{d\omega}\left(e^{-s\omega}-1\right)\right].
\end{eqnarray}
After integration in the l.h.s. by parts, this can also be written
\begin{eqnarray}\label{finite}
-s\int_0^{E} d\omega_1
e^{-s\omega_1}\int^{E}_{\omega_1}d\omega'_1\frac{dw_1}{d\omega'_1}\qquad\qquad\qquad\quad
\nonumber\\
=\ln\left[1+\int_0^{E_e} d\omega
\frac{dw}{d\omega}\left(e^{-s\omega}-1\right)\right].\,\,
\end{eqnarray}
\end{subequations}

If $w_1$ is IR-finite, one can deal with Eq.~(\ref{simpler})
arranged with the aid of relation (\ref{total-probab-1}) in a simple
form
\begin{equation}\label{108}
\int_0^{E} d\omega_1 \frac{dw_1}{d\omega_1}e^{-s\omega_1}
=\ln\frac{1+\int_0^{E_e} d\omega
\frac{dw}{d\omega}\left(e^{-s\omega}-1\right)}{1-\int_0^{E_e}
d\omega \frac{dw}{d\omega}}.
\end{equation}
Since here the l.h.s. is merely a Laplace transform of
$\frac{dw_1}{d\omega_1}$, it inverts in the standard way:
\begin{subequations}\label{110}
\begin{eqnarray}
\frac{dw_1}{d\omega_1}\!
&=&\!\frac1{2\pi i}\int^{c+i\infty}_{c-i\infty}\!ds e^{s\omega_1}\!\ln\!\left(\!1\!+\!\frac{\int_0^{E_e} d\omega \frac{dw}{d\omega}e^{-s\omega}}{1-\int_0^{E_e} d\omega \frac{dw}{d\omega}}\!\right)\,\,\,\label{log(1+...)}\nonumber\\
\\
&\equiv&\!\frac1{2\pi i}\int^{c+i\infty}_{c-i\infty}\!ds
e^{s\omega_1}\!\ln\!\left(\int_0^{E_e} d\omega
\frac{dw}{d\omega}e^{-s\omega}+W_0\!\right)\,\,\nonumber\\
&\,&-\ln W_0\delta(\omega_1) \label{reconstr}
\end{eqnarray}
\[
(\omega_1<E).
\]
Now, for $\omega_1>0$ (when the $\delta$-term does not contribute),
IR divergences do not appear in Eq.~(\ref{reconstr}), so the initial
assumption of IR finiteness of $w_1$ may be relaxed.
Eqs.~(\ref{110}) might as well be derived from Eq.~(\ref{finite}),
which is a bit more lengthy (solving a 1st-order differential
equation), but does not need presuming IR finiteness of $w_1$. If
$W_0$ is finite, integral (\ref{log(1+...)}) converges better, since
the logarithm in the integrand vanishes at infinite integration
limits. On the contrary, if $W_0=1-\int_0^{E_e} d\omega
\frac{dw}{d\omega}\to 0$, representation (\ref{log(1+...)}) is
untenable, and (\ref{reconstr}) (without the $\delta$-term) is the
representation of choice.

From our experience with the resummation procedure in
Sec.~\ref{sec:2}, one may expect ambiguity in the contour integral
representation for the reconstructed spectrum, as well. Indeed,
applying Lemma~\ref{lemma1} with $\mathcal{F}(z)=\ln(1+z/W_0)$, or
$\mathcal{F}(z)=\ln (W_0+z)$,\footnote{At $W_0=0$, the argument of
the logarithm, $z(s)=\int_0^{E_e}d\omega
\frac{dw}{d\omega}e^{-s\omega}\underset{s\to\infty}\sim s^{-a}$,
tends to zero as $s\to\infty$. Thereat,
$|\mathcal{F}'[z(s)]|=\frac1{|z(s)|}$ increases, but only by a power
law, $\sim s^a$, still satisfying the conditions of
Lemma~\ref{lemma1}.} one can see that the upper integration limit
for the Laplace transform of $\frac{dw}{d\omega}$ in (\ref{110}) may
be an arbitrary number greater than $\omega_1$, e.g. equal
$\omega_1+0$:
\begin{eqnarray}
\frac{dw_1}{d\omega_1} &=&\frac1{2\pi
i}\int^{c+i\infty}_{c-i\infty}ds
e^{s\omega_1}\ln\!\left(\int_0^{\omega_1+0} \!d\omega
\frac{dw}{d\omega}e^{-s\omega}+W_0\right)\label{reconstr1}\nonumber\\
&\,&-\ln W_0\delta(\omega_1). \label{reconstr-caus}
\end{eqnarray}
\end{subequations}
Relation $\omega_1>\omega$ in integral (\ref{reconstr-caus}) may
seem to be physically counterintuitive, because it is energetically
\emph{anti}-ordered. But as we emphasized in
Sec.~\ref{subsec:cont-int-repr}, the energetic ordering property
does not hold exactly for our distributions, anyway.

Another way to arrive at Eq.~(\ref{reconstr-caus}) is to expand the
logarithm in Eq.~(\ref{log(1+...)}) to power series, and evaluate
the contour integral termwise:
\begin{subequations}\label{reconstr-power-series-together}
\begin{eqnarray}\label{reconstr-power-series}
\frac{dw_1}{d\omega_1}
=\sum^\infty_{n=1}\frac{(-1)^{n-1}}{n\left(1-\int_0^{E_e} d\omega
\frac{dw}{d\omega}\right)^n}\int_0^{E_e} d\Omega_1\frac{dw}{d\Omega_1}\ldots\nonumber\\
\times\int_0^{E_e}
d\Omega_n\frac{dw}{d\Omega_n}\delta\left(\omega_1-\sum_{k=1}^n\Omega_k\right).\qquad
\end{eqnarray}
Here, as in passing from Eq.~(\ref{generic-sum-int}) to
(\ref{generic-sum-int-upper-omega}), one may replace upper limits
$E_e$ by $\omega_+0$:
\begin{eqnarray}\label{reconstr-power-series-caus}
\frac{dw_1}{d\omega_1}
=\sum^\infty_{n=1}\frac{(-1)^{n-1}}{n\left(1-\int_0^{E_e} d\omega
\frac{dw}{d\omega}\right)^n}\int_0^{\omega_1+0}d\Omega_1\frac{dw}{d\Omega_1}\ldots\nonumber\\
\times\int_0^{\omega_1+0}d\Omega_n\frac{dw}{d\Omega_n}\delta\left(\omega_1-\sum_{k=1}^n\Omega_k\right).\qquad
\end{eqnarray}
\end{subequations}
In contrast to Eq.~(\ref{generic-sum-int}), however, series
(\ref{reconstr-power-series-together}) involves no factorial in the
denominator, whence it must have a finite convergence radius, such
that $w_1\lesssim1$. The latter property is natural, as long as
resummed spectra are always bounded from above (see
Sec.~\ref{par:fund-max}, \ref{subsubsec:comb-in-fund-int}). Series
(\ref{reconstr-power-series-together}) at sufficiently small
intensities of $w_1$ may prove to be actually even more convenient
than contour integral representations (\ref{110}). In particular, in
the limit $\omega_1\to0$, only the first term of
(\ref{reconstr-power-series-caus}) survives, giving
\begin{equation}\label{}
\frac{dw_1}{d\omega_1}\bigg|_{\omega_1=0}
=\frac{\frac{dw}{d\omega}\big|_{\omega=0}}{1-\int_0^{E_e} d\omega
\frac{dw}{d\omega}},
\end{equation}
which is equivalent to Eq.~(\ref{coh-omegato0}). Keeping
$\mathcal{O}(\omega_1)$ terms, one can derive an analog of
Eq.~(\ref{coh-omegato0-corr}), etc.

For application of contour integral representations (\ref{110}), it
is also highly desirable to map singularities\footnote{Strictly
speaking, the singularities arise only in the limit
$E_e\to\infty$.}, as well as sectors of growth and decrease of the
integrand, to which the integration contour can be adjusted. As long
as the radiation spectrum obeys Eq.~(\ref{108}), the integrand
singularities in Eqs.~(\ref{110}) must coincide with those of
Laplace transform of the single-photon spectrum. But the latter
transform may not have singularities in the right half-plane, since
$\int^{\infty} d\omega_1 \frac{dw_1}{d\omega_1}e^{-s\omega_1}$
converges at any positive $\mathfrak{Re}s$, so it is an analytic
function of $s$ at $\mathfrak{Re}s>0$. In case if with the increase
of $\omega_1$, $\frac{dw_1}{d\omega_1}$ decreases faster than
exponentially [e.g., strictly vanishes beyond a coherent emission
edge $\omega_0$, like in Eq.~(\ref{dw1coh-real})], the l.h.s. of
(\ref{108}) exists at any $s$, so no singularities in the $s$ plane
can emerge at all. In case if $\frac{dw_1}{d\omega_1}$ decreases
with $\omega_1$ exponentially, its Laplace transform possesses
singularities at finite negative $\mathfrak{Re}s$. Finally, if
$\frac{dw_1}{d\omega_1}$ decreases by a power law, the l.h.s. of
(\ref{108}) diverges at $s<0$, so a singularity is situated at point
$s=0$.

As concerns the convergence sectors, the integrand of
(\ref{log(1+...)}) obviously diverges at $\mathfrak{Re}s\to+\infty$
because of factor $e^{s\omega_1}$. At $\mathfrak{Re}s\to-\infty$,
the integrand may decrease only provided the argument of the
logarithm, after analytic continuation to the negative
$\mathfrak{Re}s$ half-plane, does not increase faster than
exponentially. In the latter case, the integration contour may be
deformed to shape $\mathcal{C}_2$ (see Fig.~\ref{fig:contours}). But
that appears to be impossible in the important case when
$\frac{dw_1}{d\omega_1}$ is anticipated to decrease at
$\omega_1\to\infty$ faster than exponentially (e.g., pure
single-harmonic coherent radiation). Then, as we noted above, the
integrand of (\ref{110}) may not have singularities in the whole
complex plane, so if the integration contour might be deformed to
$\mathcal{C}_2$, the result of the integration would identically
equal zero, which is a clear contradiction. In this case, the only
option for the integration contour in Eqs.~(\ref{110}) is a line
parallel to the imaginary axis ($\mathcal{C}_1$).

At practice, when $\frac{dw}{d\omega}$ is measured as a number of
events per bin of deposited energy, in Eq.~(\ref{reconstr}) the
integral involving $\frac{dw}{d\omega}$ must be replaced by the
corresponding finite sum. That is harmless for evaluation of the
contour integrals, since the resulting function of $s$ remains
analytic in the $s$ plane. But it is mandatory that the spectrum
measurements are absolute, providing correct normalization of the
spectrum.

%The reconstruction efficiency depends on the sensitivity of the log entering Eq.~(\ref{reconstr}) to its argument. When $\frac{dw}{d\omega}$ is small, the logarithm behaves as $\ln\left[1+\int_0^\infty d\omega \frac{dw}{d\omega}\left(e^{-s\omega}-1\right)\right]\simeq \int_0^\infty d\omega \frac{dw}{d\omega}\left(e^{-s\omega}-1\right)$, which means perfect sensitivity and a linear relation; in this case, obviously, $\frac{dw_1}{d\omega_1}\approx \frac{dw}{d\omega_1}$.

At high intensity, when any resummed coherent spectrum tends to
Gaussian form, our reconstruction method will ultimately become
unsustainable. To appreciate the encountered difficulties, consider
a toy model for the single-photon spectrum:
\begin{equation}
\frac{dw_1}{d\omega_1}=be^{-\Lambda\omega_1}
\end{equation}
(with $E_e\to\infty$). The corresponding resummed spectrum evaluates
in terms of a modified Bessel function:
\begin{equation}
\frac{dw}{d\omega}=\sqrt{\frac{b}{\omega}}e^{-\Lambda\omega-\frac{b}{\Lambda}}I_1(2\sqrt{b\omega}).
\end{equation}
Laplace transform thereof equals
\[
\int_0^\infty d\omega
\frac{dw}{d\omega}e^{-s\omega}=e^{-\frac{b}\Lambda}\left(-1+e^{\frac{b}{\Lambda+s}}\right).
\]
Now suppose that a calorimetric spectrum measurement gave instead of
$\frac{dw}{d\omega}$ some slightly deviating distribution
$\frac{dw}{d\omega}+\delta \frac{dw}{d\omega}$. Inserting this to
Eq.~(\ref{log(1+...)}) leads to the reconstructed underlying
spectrum
\[
\frac{dw_1}{d\omega_1}\!=\!\frac1{2\pi
i}\!\int^{c+i\infty}_{c-i\infty}\!ds
e^{s\omega_1}\!\ln\!\frac{e^{-\frac{bs}{\Lambda(\Lambda+s)}}\!+\!\int_0^\infty
\!d\omega\!
\left(e^{-s\omega}\!-\!1\right)\!\delta\frac{dw}{d\omega}}{e^{-\frac{b}\Lambda}-\int_0^\infty
d\omega \delta\frac{dw}{d\omega}}.
\]
Clearly, as $b$ increases, signal term
$e^{-\frac{bs}{\Lambda(\Lambda+s)}}$ becomes exponentially small,
and ultimately inferior to the error term; hence, for sufficiently
large $b/\Lambda$ the method will loose efficiency. The best one can
do in this situation, probably, is to extract first spectral moments
as in Ref.~\cite{Kolchuzhkin}, which may appear insufficient for
complete characterization of a highly structured single-photon
spectrum.

In the latter straitened circumstances, it may also be worth
engaging information contained in the photon multiplicity spectrum
$\bar n(\omega)$. Examination of structure of integral
(\ref{mean-n-omega-contour-non-casual}) shows that it is more
complicated than (\ref{generic-eq}), and does not allow expressing
$dw_1/d\omega_1$ explicitly. But in the high-multiplicity limit,
when the reconstruction method based on $dw/d\omega$ is obstructed,
one can at least extract from $\bar n(\omega)$ an additional
parameter $w_{1\text{c}}$, say, through Eq.~(\ref{n-to-wi1c}). Its
knowledge, along with the knowledge of moments
$\overline{\omega_{1\!}}_{\text{c}}$ and
$\overline{\omega^2_{1\!}}_{\text{c}}$ (and possibly
$\overline{\omega^3_{1\!}}_{\text{c}}$) extracted from
$\frac{dw}{d\omega}$, may constrain the shape of $dw_1/d\omega_1$
more tightly.

%\newpage
%.

%\newpage

\section{Summary and outlook}\label{sec:summary}

The techniques developed in the present paper can be further
employed to calculate spectra of specific coherent radiation
sources, including channeling radiation. For undulator radiation,
and partially for coherent bremsstrahlung, the results presented in
this paper may already be applicable directly. Let us summarize the
main physical lessons learned at this stage:
\begin{itemize}
    \item
The fundamental maximum of the multiphoton spectrum does not rise
indefinitely with the increase of the radiator length $L$, but
eventually saturates, and subsequently decreases. Instead, there is
an elevation of the spectrum beyond the  fundamental maximum, and
its general spread (Sec.~\ref{sec:coh+incoh}). The high-$\omega$
spectral tail due to incoherent bremsstrahlung is the least affected
by multiphoton effects, and its ratio w.r.t. the fundamental maximum
thereby increases.
    \item
Multiphoton effects alone can give rise to a second maximum in the
coherent radiation spectrum, at $\omega\approx2\omega_0$, even when
secondary harmonics in the single-photon spectrum are absent.
    \item
      The low-$\omega$ part of the multiphoton spectrum, as well as discontinuities of the spectrum for coherent radiation, are suppressed by the factor of photon non-emission probability (Sec.~\ref{sec:coh+incoh}). Therein, manifestations of the incoherent radiation component in the spectrum are particularly pronounced.
    \item
 At high radiation intensity (in a long radiator),
the resummed spectrum of pure coheernt radiation tends to a Gaussian
form. But under presence of an incoherent component, it features a
power-law tail towards large $\omega$
(Sec.~\ref{sec:high-intensity}). Such a diffusion regime may be
regarded as \emph{weakly} anomalous (characterized by $0<a\ll1$).
The corresponding limiting multiphoton spectrum is described by a
parabolic cylinder function, being an intermediate asymptotic case
between Gaussian and L\'{e}vy distributions.
    \item
        The photon multiplicity spectrum $\bar n(\omega)$ is an informative observable complementary to $\frac{dw}{d\omega}$, and is measurable with the modern state of the technology. The growth of $\bar n(\omega)$ with $\omega$ at low radiation intensity is stepwise, whereas at high intensity it smoothens out and tends to a linear law. In the region where the incoherent contribution overtakes the coherent one, $\bar n(\omega)$ saturates, and slightly subsides, whereupon continues rising, but only logarithmically (Secs.~\ref{sec:coh+incoh},
\ref{sec:high-intensity}).
    \end{itemize}

Promising are also the formulas for reconstruction of the
single-photon spectrum from the calorimetrically measured one,
derived in Sec.~\ref{sec:reconstr}. The applications thereof will be
considered elsewhere.

It should also be remembered that although the present framework is
well suited for fast evaluation of multiphoton emission effects, it
is limited to conditions of soft radiation, which do not yet cover
all the variety of interesting physical problems (e.g., hard
coherent electromagnetic cascades, with deposited energies close to
$E_e$, investigated in experiments \cite{Belkacem}). Nonetheless,
for practical coherent radiation sources, as discussed in
Appendix~\ref{app:coh-rad-sources} and the beginning of
Sec.~\ref{sec:high-intensity}, conditions of soft radiation are
quite typical. At the same time, albeit many experiments in the soft
radiation domain found multiphoton effects to be appreciable, none
of those experiments addressed multiphoton effects systematically.
The need for such dedicated experimental studies thus remains acute.

%Equally well, there remain unodsolved theoretical
%problems, such as incorporation of the electron-positron kinetic
%component within the analytic approach, and extension of the
%resummation procedure beyond the soft radiation domain
%(non-Poissonian regime).

%More dedicated tests, including the nonlinear length dependence, maximal summit of the fundamental maximum, may appear in the future.

%At the same time in course of our study, we made some simplifications, which would be desirable to redress in the future.

%\subsection*{Acknowledgements}
%\ack

%.....

%I am indebted to V.A. Maisheev for arousing my interest in this
%problem and thank ... for

%\newpage

\appendix

\section{Evaluation of the effective UV cutoff parameter}\label{app:kappa}

In Sec.~\ref{subsec:incoh-bremsstr}, we introduced parameter
$E=\kappa E_e$ as an effective UV cutoff for the single-photon
scale-invariant incoherent bremsstrahlung spectrum. The value of
this parameter can be inferred from a theory accurately describing
acts of large radiative energy loss. For our purposes, we need only
the NLL soft asymptotics of this theory. As long as we deal solely
with spectrum $\frac{dw}{d\omega}$ (and are not concerned with the
photon multiplicity spectrum), the coherent radiation component at
large $\omega$ can be entirely neglected. Thereby, we return to the
well-known electromagnetic cascade theory in an amorphous medium,
but with the due account of the electron spin.

For the single-photon bremsstrahlung spectrum, instead of
(\ref{dw1incoh-cutoff}) we take the ultra-relativistic and
completely screened approximation of the Bethe-Heitler formula:
\begin{equation}\label{BH-spectrum}
\frac{dw_{1\text{i}}}{d\omega_1}=\frac{a}{\omega_1}W_{\text{BH}}\left(\frac{\omega_1}{E_e}\right),
\end{equation}
where \cite{Rossi}
\begin{equation}\label{WBH-def}
W_{\text{BH}}(z)=1-z+\frac34 z^2.
\end{equation}
For simplicity, the pair production process will be neglected, as
before. Then, the kinetic equation for electron energy distribution
$\Pi(\mathcal E)$ remains one-component:
\begin{eqnarray}\label{shower-eq}
\frac{\partial\Pi(\mathcal E,a)}{\partial a}=\int_{\mathcal E}^{E_e}\frac{d\mathcal E'}{\mathcal E'-\mathcal E}\Pi(\mathcal E',a)W_{\text{BH}}\left(\frac{\mathcal E'-\mathcal E}{\mathcal E'}\right)\nonumber\\
-\Pi(\mathcal E,a)\int_0^{\mathcal E} \frac{d\mathcal E'}{\mathcal
E-\mathcal E'}W_{\text{BH}}\left(\frac{\mathcal E-\mathcal
E'}{\mathcal E}\right),
\end{eqnarray}
where dimensionless parameter $a$ is proportional to the target
thickness [see Eq.~(\ref{a-amorph})]. Integro-differential equation
(\ref{shower-eq}) is endowed with the initial condition
(\ref{kinetic-eq-init-cond-for-Pie}):
\begin{equation}\label{shower-init-cond}
\Pi(\mathcal E,0)=\delta(\mathcal E-E_e).
\end{equation}

Once the solution of Eq.~(\ref{shower-eq}--\ref{shower-init-cond})
is attained, the multiphoton radiation spectrum, equal to the
radiative energy loss spectrum, is determined by the correspondence
rule (\ref{def-Pie}):
\begin{equation}\label{rel-with-multiphot-spectr}
\frac{dw_{\text{i}}}{d\omega}=\Pi(E_e-\omega,a)\qquad [W_0(a>0)=0],
\end{equation}
and for this function we are interested in the limit $\omega\ll
E_e$. Note that by a change of integration variables,
Eq.~(\ref{shower-eq}) may can be recast
\begin{eqnarray*}\label{shower-eq2}
\frac{\partial}{\partial a}\Pi(E_e-\omega,a)=\int^{\omega}_{0}\frac{d\omega_1}{\omega_1}\Pi(E_e-\omega+\omega_1,a)\qquad\qquad\nonumber\\
\times W_{\text{BH}}\left(\frac{\omega_1}{E_e-\omega+\omega_1}\right)\qquad\nonumber\\
-\Pi(E_e-\omega,a)\int_0^{E_e-\omega}\frac{d\omega_1}{\omega_1}W_{\text{BH}}\left(\frac{\omega_1}{E_e-\omega}\right),
\end{eqnarray*}
which turns to recoilless kinetic equation (\ref{kinetic-eq2}) in
the limit $\omega_1\ll E_e$, when $W_{\text{BH}}(z)\to
W_{\text{BH}}(0)=1$. However, form (\ref{shower-eq}) is more
convenient for solution.

Solution of Eq.~(\ref{shower-eq}) is obtained by applying Mellin
transform
\begin{equation}\label{Mell}
\Pi(s,a)=\int_0^{E_e} d\mathcal E{\mathcal E}^{s-1}\Pi(\mathcal
E,a),
\end{equation}
in terms of which the integro-differential equation becomes ordinary
differential:
\begin{equation}\label{shower-eq-Mell}
\frac{\partial}{\partial a}\Pi(s,a)=-\mu(s)\Pi(s,a),
\end{equation}
with $a$-independent coefficient
\begin{equation}\label{mu-def}
\mu(s)=\int_0^1 dz\frac{1-z^{s-1}}{1-z}W_{\text{BH}}(1-z)
\end{equation}
and initial condition
\begin{equation}\label{shower-init-cond-Mell}
\Pi(s,0)=E_e^{s-1}.
\end{equation}
Integrating Eq.~(\ref{shower-eq-Mell}), and applying the inverse
Mellin transform $\Pi(\mathcal E,a)=\frac1{2\pi
i}\int_{c-i\infty}^{c+\infty}ds {\mathcal E}^{-s}\Pi(s,a)$, leads to
the solution
\begin{equation}\label{Pi-solution}
\Pi(\mathcal E,a)=\frac1{E_e}\frac1{2\pi
i}\int_{c-i\infty}^{c+\infty}ds e^{-a\mu(s)+s\ln(E_e/\mathcal E)}.
\end{equation}
By correspondence rule (\ref{rel-with-multiphot-spectr}), the
multiphoton bremsstrahlung spectrum ensues
\begin{eqnarray}\label{Pi-solution}
\frac{dw_{\text{i}}}{d\omega}=\frac1{E_e}\frac1{2\pi i}\int_{c-i\infty}^{c+\infty}ds e^{-a\mu(s)+s\ln\frac1{1-\omega/E_e}}\qquad\quad\nonumber\\
\simeq \frac1{\omega}\frac1{2\pi i}\int_{c-i\infty}^{c+\infty}d\zeta
e^{\zeta-a\mu(\zeta E_e/\omega)} \qquad (\omega\ll E_e).\quad
\end{eqnarray}
In the last line, we linearized $\ln\frac1{1-\omega/E_e}\simeq
\frac{\omega}{E_e}$ for small ${\omega}/{E_e}$ we are concerned
with, and changed the integration variable to $\zeta=-s\omega/E_e$.

Function $\mu$ at large arguments needed for us increases
logarithmically:
\begin{eqnarray}\label{mu-NLL}
\mu(s)\equiv W_{\text{BH}}(0)\int_0^1 dz\frac{1-z^{s-1}}{1-z}\qquad\qquad\qquad\qquad\qquad\nonumber\\
+\int_0^1 dz\frac{1-z^{s-1}}{1-z}[W_{\text{BH}}(1-z)-W_{\text{BH}}(0)]\qquad\qquad\nonumber\\
=\ln s+\gamma_{\text{E}}+\!\int_0^1\!\frac{dz'}{z'}[W_{\text{BH}}(z')\!-\!W_{\text{BH}}(0)]+\mathcal{O}\!\left(s^{-1}\right)\!.\nonumber\\
\end{eqnarray}
Inserting this approximation to Eq.~(\ref{Pi-solution}), and doing
the elementary contour integral, from comparison with Eq.~(\ref{13},
\ref{Fa}) we ultimately infer the value for $\kappa$:
\begin{equation}\label{kappa58}
\kappa=e^{\int_0^1\frac{dz'}{z'}[W_{\text{BH}}(z')-W_{\text{BH}}(0)]}=e^{-5/8}\approx0.5.
\end{equation}
According to Eq.~(\ref{kappa58}), the effective UV cutoff must be
set at about half the initial electron energy. With this value for
$\kappa$, our Eqs.~(\ref{13}, \ref{Fa}) agree with Eqs.~(2.11),
(2.19) of \cite{BK}.

%\section{Proof of identity (\ref{lemma})}\label{app:proof}

\section{Estimates of photon multiplicities for coherent radiation sources}\label{app:coh-rad-sources}

In this Appendix, we will collect equations for spectral parameters
of main coherent radiation sources, needed for numerical estimates
of significance of multiphoton effects in
Sec.~\ref{sec:high-intensity}.

Any coherent radiation source we consider may be viewed as an
ultrarelativistic electron (or positron) radiating under the
influence of periodic force $F(t)$ (having period $T$), directed
predominantly transverse to the electron's large mean momentum
(which is taken parallel to $Oz$). If the radiation is of dipole
type, i.e. produced by a weakly accelerated electron remaining
non-relativistic in its average rest frame (e.r.f.), the emitted
electromagnetic radiation in this frame will be monochromatic. In
e.r.f., period $T$ is Lorentz-contracted to $\widetilde T=T/\gamma$
($\gamma$ stands for the electron Lorentz factor), corresponding to
frequency\footnote{In this section, we work in the system of units
$\hbar=c=1$, where frequency and mass have dimension of energy,
while the electron charge is dimensionless.}
$\widetilde\omega_0=\frac{2\pi}{\widetilde T}$. The radiation in the
e.r.f. is not strictly isotropic, as long as the driving force is
known to be transverse w.r.t. $Oz$. For our purposes, we need only
the angular distribution averaged over the azimuthal angles, which
will thus obey a $\propto1+\cos^2\tilde\theta$ law, where
$\tilde\theta$ is the angle between $Oz$ and the photon emission
direction in e.r.f. This information suffices to write the
angle-differential radiation intensity:
\begin{equation}\label{I-dipole}
d\tilde I=e^2\left|\frac1{\widetilde T}\int^{\widetilde
T/2}_{-\widetilde T/2}d\tilde t\frac{\widetilde F(\tilde t)}m
e^{-{2\pi i\tilde t}/{\widetilde
T}}\right|^2\frac{1+\cos^2\tilde\theta}{2}\sin\tilde\theta
d\tilde\theta,
\end{equation}
where $m$ and $e$ are the electron mass and charge. [To check the
numerical coefficient, note that integration of (\ref{I-dipole})
over photon emission angles with $\int_0^{\pi}
\frac{1+\cos^2\tilde\theta}{2}\sin\tilde\theta
d\tilde\theta=\frac43$, and identification of $\frac{\widetilde
F(\tilde t)}m$ with the electron acceleration, leads to the famous
Larmor formula.]

Eq.~(\ref{I-dipole}) gives an average (over the oscillation period)
energy emitted by the electron in its rest frame per unit time.
Dividing this by the photon energy $\widetilde\omega_0$ gives the
probability density, and the total probability is obtained by
multiplying by the period value and the number of periods $N$:
\begin{equation*}\label{}
dw_{1\text{c}}=\frac{\widetilde T N}{\widetilde\omega_0}d\tilde
I\equiv \frac{N}{2\pi}\widetilde T^2d\tilde I.
\end{equation*}
The latter quantity must be Lorentz invariant, which is opportune
for expressing the desired radiation spectrum in the lab.

The advantage of referring to e.r.f. variables is that
$\cos\tilde\theta$ is related to the photon momentum projection on
the light front in $z$-direction, and therethrough to the photon
energy in the lab frame:
\begin{equation}\label{omega-theta-corresp}
\frac{\omega_1}{\omega_0}=\frac{1+\cos\tilde\theta}{2},
\end{equation}
where $\omega_0$ is the maximal photon energy in the lab,
corresponding to $\tilde\theta=0$. Expressing $\cos\tilde\theta$
from (\ref{omega-theta-corresp}), inserting it to (\ref{I-dipole}),
and differentiating, we convert the angular distribution in e.r.f.
to the spectral distribution in the lab:
\begin{equation}\label{152}
dw_{1\text{c}}=\frac{e^2N}{\pi}\left|\int^{\widetilde
T/2}_{-\widetilde T/2}d\tilde t\frac{\widetilde F(\tilde t)}m
e^{-{2\pi i\tilde t}/{\widetilde T}}\right|^2
P\left(\frac{\omega_1}{\omega_0}\right)\frac{d\omega_1}{\omega_0},
\end{equation}
with $P(z)$ being function (\ref{f-def}), thus explaining its
origin. Product $d\tilde t\widetilde F(\tilde t)=dp_\perp$ entering
the time integral is Lorentz invariant, too, so it can be written as
$dt F(t)$, where $F$ is the force, and $t$ the time in the lab.
Comparing (\ref{152}) with Eq.~(\ref{dw1coh-real}), we infer
\begin{equation}\label{dw-lab}
b\omega_0 =\frac{e^2}{\pi}\left|\int^{T/2}_{-T/2}d t\frac{F(t)}m
e^{-{2\pi i t}/{ T}}\right|^2 N,
\end{equation}
where $T$ is the motion period in the lab, and obviously, $N=L/T$.

Although product $b\omega_0$ is all we need for estimates of the
photon multiplicity, it is also expedient to evaluate the photon
energy scale $\omega_0$ independently. The latter is obtained from
e.r.f. value $\widetilde\omega_0=\frac{2\pi\gamma}{T}$ by boosting
along $Oz$ with the Lorentz factor $\gamma$. That results in
multiplication by $\sqrt{\frac{1+v}{1-v}}\approx2\gamma$, so the
maximal photon energy in the lab equals
\begin{equation}\label{2gamma2}
\omega_0 \approx2\gamma^2\frac{2\pi}{T},
\end{equation}
exceeding by factor of $2\gamma^2$ the driving force frequency in
this frame.

The analysis of photon multiplicity for specific radiation sources
further reduces to estimating in Eq.~(\ref{dw-lab}) the driving
force magnitude and period. Note that generally, the square of the
integral in (\ref{dw-lab}) is $\sim T^2$, while $N\sim T^{-1}$.
Thus, in total, the radiation source brightness is proportional to
its period. On the other hand, according to Eq.~(\ref{2gamma2}), the
photon energy scales as $T^{-1}$, though that is generally
compensable by an appropriate increase in $\gamma$.

\subsection{Coherent bremsstrahlung}

For an electron crossing a family of atomic planes by a
near-straight trajectory at a small misalignment angle $\chi$, the
time in Eq.~(\ref{dw-lab}) expresses as $t=x/\chi$, where $x$ is the
coordinate transverse to the planes, and the oscillation period in
the lab equals
\begin{equation}\label{T-CB}
T=\frac{d}{\chi},
\end{equation}
$d$ being the plane spacing. Therewith, Eq.~(\ref{dw-lab}) becomes
\begin{equation}\label{dw-CB}
b\omega_0
=\frac{e^2}{\pi}\left|\int^{d/2}_{-d/2}\frac{dx}{\chi}\frac{F(x)}m
e^{-{2\pi i x}/{d}}\right|^2 \frac{L\chi}{d},
\end{equation}
where $F$ as a function of time was replaced by the corresponding
function of $x$. If $F(x)$ within a period describes roughly
as\footnote{That corresponds to a harmonic interplanar potential,
providing as a satisfactory approximation, e.g., for a (110)
inter-planar channel in crystalline silicon
\cite{parabolic-potential}.}
\begin{equation}\label{Fx}
F(x)\approx\frac{2F_{\max}}{d}x\qquad (-d/2<x<d/2),
\end{equation}
evaluation of the integral in Eq.~(\ref{dw-CB}) gives
\begin{equation}\label{dw-CB-Fmax}
b\omega_0 =\frac{e^2F_{\max}^2d}{\pi^3m^2\chi}L\approx
6\cdot10^{-5}\frac{L[\text{mm}]}{\chi},
\end{equation}
where we used numerical estimates $d\approx2\,\AA$, and
$|F_{\max}|\approx6\text{ GeV/cm}$.

Parameters $L$ and $\chi$ are not completely independent: they must
obey inequality (cf. \cite{PDG})
\begin{equation}\label{CB-ineq}
\delta\chi=\frac{13.6\text{ MeV}}{E_e}\sqrt{\frac{L}{X_0}}\ll\chi,
\end{equation}
in order for the multiple scattering spread $\delta\chi$ not to blur
the coherent peak. At the same time, $\chi$ correlates with $E_e$,
since under assumption $\omega_0\ll E_e$, Eqs.~(\ref{2gamma2},
\ref{T-CB}) yield
\begin{equation}\label{Echi-ll20}
E_e\chi\ll\frac{m^2d}{4\pi}=20\text{ MeV}.
\end{equation}
For practical misalignment angles $10^{-2}<\chi<10^{-4}$, this
constrains the electron energy to $1<E_e<100$ GeV.
Eqs.~(\ref{CB-ineq}) and (\ref{Echi-ll20}) are only compatible
provided $\sqrt{L/X_0}\ll 1$, which is essentially the condition
(\ref{all1}) adopted in our paper.

\subsection{Channeling radiation}

For the case of channeling, the oscillation period depends on the
particle energy:
\begin{equation}\label{T-chann}
T=2\pi\sqrt{\frac{E_e}{|\partial F/\partial x|}}.
\end{equation}
For a harmonic potential [corresponding to force (\ref{Fx})], the
period will be constant, but with the account of anharmonicity, it
may acquire also some dependence on the particle transverse energy
in the channel. For our estimates, it will suffice to adopt the
harmonic approximation, letting
\begin{equation}\label{F-chann}
F(t)=F_0\cos\frac{2\pi t}{T},
\end{equation}
where $F_0$ depends on the particle impact parameters, in the spirit
of Eq.~(\ref{Fx}). As an order of magnitude estimate, we may use
$F_0\sim F_{\max}/2$. Substituting (\ref{F-chann}) to
Eq.~(\ref{dw-lab}), we get
\begin{subequations}
\begin{eqnarray}
b\omega_0 &=&\frac{e^2}{4\pi m^2}F_0^2 TL\\
&\sim&\frac{e^2}{8
m^2}\sqrt{\frac{E_ed}{2F_{\max}}}F_{\max}^2L.\label{dw-chann}
\end{eqnarray}
\end{subequations}

At practice, the length of a crystal for channeling is limited by
dechanneling effects. For positrons, the dechanneling length may be
estimated crudely as \cite{Ld-for-positrons}
\begin{equation}\label{Ld}
L_{\text{d}}\sim0.5\frac{\text{mm}}{\text{GeV}}E_e,
\end{equation}
in terms of which Eq.~(\ref{dw-chann}) assumes the form
\begin{equation}\label{dw-chann-fin}
b\omega_0 \approx
8\left(E_e[\text{GeV}]\right)^{3/2}\frac{L}{L_{\text{d}}}.
\end{equation}

\subsection{Undulator radiation}

In an undulator, the particle is moving in a fixed channel,
similarly to channeling, but the motion period, determined by the
magnet spacing, is energy-independent, like in coherent
bremsstrahlung. Since $T$ is macroscopically large, according to
Eqs.~(\ref{dw-lab}--\ref{2gamma2}), the source must be bright, but
the photon energy be relatively low. Nonetheless, gamma-range
undulators were created recently, with periods $T\lesssim1$ cm, and
electron energies $\sim 10^2$ GeV \cite{Moortgat-Pick}.

For description of the undulator field strength, it is customary to
introduce parameter
\begin{equation}\label{K-def}
K=\frac{eB_0}{m}\frac{T}{2\pi}\approx 1 B_0[\text{T}]T[\text{cm}],
\end{equation}
where $B_0$ is the field amplitude, $B(t)=B_0\sin\frac{2\pi t}{T}$.
In terms of $K$, Eq.~(\ref{dw-lab}) reads
\begin{equation}\label{dw-undul}
b\omega_0 =\pi e^2K^2N\approx 0.02 K^2N.
\end{equation}
At practice, parameter $K$ can be $\sim1$ (somewhat violating the
dipole radiation condition $K\ll 1$), whilst the number of periods
can amount several hundreds or thousands. The latter number is
particularly large for undulator-based positron sources, where
$N\sim10^4$ \cite{Moortgat-Pick}.

\section{High-intensity multiphoton spectrum in peripheral regions}\label{app:periph}

As had been pointed out in Sec.~\ref{subsubsec:centr}, beyond the
central region specified by condition (\ref{central-region}), the
Gaussian approximation breaks down, and the multiphoton spectrum
becomes dependent on all the detail of the single-photon spectrum.
Even though in those regions the radiation spectrum is relatively
faint (and at high $\omega$ it may be overwhelmed by the incoherent
bremsstrahlung `tail'), it may still prove measurable under high
absolute radiation intensity. Therefore, it would be useful to find
some approximations for peripheral regions, as well.

Suitable approximations for coherent radiation in peripheral regions
can be built based on the same saddle-point approximation described
in Sec.~\ref{subsubsec:steepest-descent}, provided transcendental
saddle-point equation (\ref{saddle-point-eq}) is solved at least
approximately. That requires development of a special approach on
each side from $\overline{\omega_{1\!}}_{\text{c}}$.

\subsection{Trans-central region}

\begin{figure}
\includegraphics{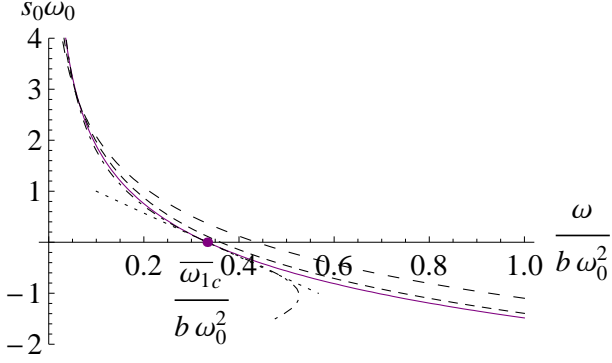}
\caption{\label{fig:s0c-vs-omega} Graphical real solution of
saddle-point equation (\ref{saddle-point-eq}), for
$\frac{dw_{1\text{c}}}{d\omega_1}$ given by
Eqs.~(\ref{dw1coh-real}--\ref{f-def}). Solid purple curve, exact
solution. Dotted black line, central region approximation
(\ref{saddle-point-eq-lin-soln}). Purple point, the central
solution. Long-dashed black curve, large-$\omega$ approximation
(\ref{s0c-log-1}). Short-dashed black curve, improved large-$\omega$
approximation (\ref{s0c-log-2}). Dot-dashed black curve,
low-$\omega$ approximation (\ref{quadr-eq}).}
\end{figure}

Let us consider the region
$\omega\gg\overline{\omega_{1\!}}_{\text{c}}$ first. From the
viewpoint of  Eq.~(\ref{saddle-point-eq}), this corresponds to large
and negative $s_0$ (see Fig.~\ref{fig:s0c-vs-omega}). It is
therefore beneficial to rewrite Eq.~(\ref{saddle-point-eq}) as
\begin{equation}\label{saddle-point-eq-large-omega}
\ln\frac{\omega}{b\omega_0^2}+s_0\omega_0=\ln\left[\int_0^1
dzzP(z)e^{-s_0\omega_0(z-1)}\right],
\end{equation}
where the r.h.s. is smaller than any of the terms on the l.h.s.
(granted that the exponential within the integration domain is
$\leq1$, achieving unity only at the endpoint $z=1$). Thence, the
r.h.s. can be treated as a perturbation.

To solve Eq.~(\ref{saddle-point-eq-large-omega}) iteratively, first,
we express $s_0\omega_0$ from equating the l.h.s. to zero (leading
log approximation):
\begin{equation}\label{s0c-log}
-s_0\omega_0=
\ln\frac{\omega}{b\omega_0^2}+\mathcal{O}\left(\ln\ln\frac{\omega}{b\omega_0^2}\right).
\end{equation}
Substituting this to the r.h.s. of
(\ref{saddle-point-eq-large-omega}) gives a NLL approximation for
$s_0\omega_0$:
\begin{equation}\label{s0c-log-1}
s_0\omega_0=\ln\left[\int_0^1
dzzP(z)\left(\frac{\omega}{b\omega_0^2}\right)^{z-2}\right]+\mathcal{O}\left(\ln\ln\ln\frac{\omega}{b\omega_0^2}\right).
\end{equation}
Its behavior is illustrated in Fig.~\ref{fig:s0c-vs-omega} by the
long-dashed black curve. It does not seem to be accurate enough,
which compels us to proceed to NNLL: substitute $s_0\omega_0$ in the
r.h.s. of Eq.~(\ref{saddle-point-eq-large-omega}) by that expressed
from the l.h.s., whereupon replace the remaining $s_0\omega_0$ in
the r.h.s. by LL expression (\ref{s0c-log}):
\begin{eqnarray}\label{s0c-log-2}
s_0\omega_0\simeq\ln\Bigg\{\int_0^1
dzzP(z)\left(\frac{\omega}{b\omega_0^2}\right)^{z-2}\qquad\quad\nonumber\\
\times\left[\int_0^1
dyyP(y)\left(\frac{\omega}{b\omega_0^2}\right)^{y-1}\right]^{1-z}\Bigg\}.
\end{eqnarray}
The latter solution is to be inserted to Eqs.~(\ref{H}--\ref{D}),
which are then inserted to (\ref{dwdomega-saddle}). The resulting
expression is rather bulky, and will be omitted. Its accuracy can be
visualized from Fig.~\ref{fig:s0c-vs-omega}, and may be regarded as
satisfactory. The qualitative corollary from Eq.~(\ref{s0c-log-2})
is that the multiphoton pure coherent spectrum decreases by the law
close to linear exponential [since in (\ref{dwdomega-saddle}) $A\sim
s_0\omega_0 e^{-s_0\omega_0}$].

%When ..., the growth of $\left\langle n \right\rangle(\omega)$ must
%slow down, too. To logarithmic, as at $a\neq0$, but with a greater
%coefficient.

\subsection{Sub-central region}

The opposite peripheral region
$\omega<\overline{\omega_{1\!}}_{\text{c}}$ corresponds to positive
$s_0$ (see Fig.~\ref{fig:s0c-vs-omega}). There, the main
contribution to the integral entering Eq.~(\ref{saddle-point-eq})
stems from the lower integration limit. Asymptotically, at large
$s_0$,
\begin{equation}\label{1/s2}
\omega=\int_0^{\omega_0} d\omega_1
\omega_1\frac{dw_{1\text{c}}}{d\omega_1}e^{-s_0\omega_1}=
\frac{bP(0)}{s_0^2}+\mathcal{O}\left(\frac{bP(0)}{s_0^3\omega_0}\right),
\end{equation}
but the latter asymptotic solution becomes accurate at
$\omega\ll\overline{\omega_{1\!}}_{\text{c}}$ only, whereas we need
to cover the whole region
$\overline{\omega_{1\!}}_{\text{c}}-\omega\gg\overline{\omega^2_{1\!}}_{\text{c}}$.
Therefore, Eq.~(\ref{saddle-point-eq}) needs to be solved accurately
on the whole interval $0<\omega<\overline{\omega_{1\!}}_{\text{c}}$,
which is challenging insofar as this equation is transcendental.

In the present situation, one of the simplest approaches may be to
interpolate the solution between asymptotes (\ref{1/s2}) and
(\ref{saddle-point-eq-lin}), e.g., writing it in form
\begin{equation}\label{quadr-eq}
\omega\approx\frac1{\frac1{bP(0)}
s_0^2+\frac{\overline{\omega^2_{1\!}}_{\text{c}}}{\overline{\omega_{1\!}}_{\text{c}}^2}s_0+\frac1{\overline{\omega_{1\!}}_{\text{c}}}}.
\end{equation}
The behavior of the latter interpolation is illustrated in
Fig.~\ref{fig:s0c-vs-omega} by the dot-dashed black curve, which is
rather close to the exact solution on the interval of interest.
Expressing $s_0$ (or $s_0^{-1}$) from quadratic equation
(\ref{quadr-eq}) and inserting to Eq.~(\ref{H}--\ref{D}), we get an
approximation holding up to the central region). Qualitatively, at
large $s_0$,
$A\sim\frac{2bP(0)}{s_0}-w_{1\text{c}}\sim2\sqrt{b\omega}-w_{1\text{c}}$,
which may be interpreted in the sense that the function rises faster
than a Gaussian.

%To evaluate $\bar n (\omega)$, it suffices to
%express $s_0$ from quadratic equation (\ref{quadr-eq}) and insert to
%(\ref{n-omega-through-s0}).

\end{document}